\documentclass[acmsmall,screen]{acmart}
\AtBeginDocument{%
  \providecommand\BibTeX{{%
    \normalfont B\kern-0.5em{\scshape i\kern-0.25em b}\kern-0.8em\TeX}}}

\renewcommand\footnotetextcopyrightpermission[1]{}
\acmConference{}
\setcopyright{acmcopyright}
\copyrightyear{2024}
\acmYear{2018}
\acmDOI{XXXXXXX.XXXXXXX}

\acmPrice{15.00}
\acmISBN{978-1-4503-XXXX-X/18/06}

\newtheorem{theorem}{Theorem}[section]
\newtheorem{assumption}{Assumption}[section]
\newtheorem{lemma}{Lemma}[section]

\newtheorem{definition}{Definition}[section]

\newcounter{numrellocal}
\renewcommand{\thenumrellocal}{\arabic{numrellocal}}
\newcounter{numrelglobal}

\makeatletter
\newcommand{\numrel}[2]{
  \stepcounter{numrellocal}
  \refstepcounter{numrelglobal}
  \ltx@label{#2}
  \overset{(\thenumrellocal)}{#1}
}
\makeatother

\usepackage{thmtools} 
\usepackage{bbm}
\usepackage{thm-restate}
\usepackage{subcaption}
\usepackage{multirow}

\newcommand{\argmax}{\mathop{\mathrm{argmax}}}

\newcommand{\cY}{\mathcal{Y}}

\newcommand{\cG}{\mathcal{G}}

\newcommand{\den}[1]{\llbracket #1 \rrbracket}

\newcommand{\ten}[1]{#1}

\begin{document}

\title[Uncertainty Quantification for Neurosymbolic Programs]{Uncertainty Quantification for Neurosymbolic Programs via Compositional Conformal Prediction}

\author{Ramya Ramalingam}
\author{Sangdon Park}
\author{Osbert Bastani}
\authorsaddresses{}

\begin{abstract}
Machine learning has become an effective tool for automatically annotating unstructured data (e.g., images) with structured labels (e.g., object detections). As a result, a new programming paradigm called \emph{neurosymbolic programming} has emerged where users write queries against these predicted annotations. However, due to the intrinsic fallibility of machine learning models, these programs currently lack any notion of correctness. In many domains, users may want some kind of conservative guarantee that the results of their queries contain all possibly relevant instances. Conformal prediction has emerged as a promising strategy for quantifying uncertainty in machine learning by modifying models to predict sets of labels instead of individual labels. A key feature is that it provides a probabilistic guarantee that the prediction set contains the true label with high probability (called \emph{coverage}). We propose a novel framework for adapting conformal prediction to neurosymbolic programs; our strategy is to represent prediction sets as abstract values in some abstract domain, and then to use abstract interpretation to propagate prediction sets through the program. Our strategy satisfies three key desiderata: (i) correctness (i.e., the program outputs a prediction set that contains the true output with high probability), (ii) compositionality (i.e., we can quantify uncertainty separately for different modules and then compose them together), and (iii) structured values (i.e., we can provide uncertainty quantification for structured values such as lists). When the full program is available ahead-of-time, we propose an optimization to our approach that incorporates conformal prediction at intermediate program points to reduce imprecision in abstract interpretation. We evaluate our approach on programs that take MNIST and MS-COCO images as input, demonstrating that it produces reasonably sized prediction sets while satisfying our coverage guarantee.
\end{abstract}



\maketitle
\pagestyle{plain}

\section{Introduction}

With the recent success of machine learning, there has been a great deal of interest in programs that incorporate machine learning components, called \emph{neurosymbolic programs}~\cite{chaudhuri2021neurosymbolic}. A prominent example stems from the use of deep neural networks (DNNs) to automatically annotate unstructured data---e.g., images may be annotated with object detections and text may be annotated with word embeddings and parts of speech. Then, a neurosymbolic program might query these annotated images~\cite{bastani2021skyquery,mell2023synthesizing,barnaby2023imageeye} or text~\cite{chen2021web,chen2023data} to identify instances matching a user-specified pattern. \ten{Thus far, the neurosymbolic programming literature has largely focused on expressing and synthesizing such programs, leaving aside important questions about the trustworthiness of the resulting programs.}

A key challenge facing neurosymbolic programs is that they inherit the fallibility of their machine learning components---if a DNN makes a mistake, then the program is likely to output the wrong value. For individual machine learning models, \emph{uncertainty quantification} has emerged as a promising strategy to improve the trustworthiness of the model by informing the user when predictions may be incorrect. Recent work has demonstrated that uncertainty quantification can improve the quality of human decisions based on model predictions~\cite{tschandl2020human,mcgrath2023does}. Thus, a key question is how to provide analogous uncertainty quantification for neurosymbolic programs. There are several desirable properties for any uncertainty quantification method in this setting:
\begin{itemize}
\item \textbf{Correctness:} First and foremost, it should satisfy some natural notion of correctness. Ideally, it would enable us to provide some kind of probabilistic guarantees on the uncertainty quantification for the program outputs.
\item \textbf{Compositional:} Second, it should be compositional, enabling uncertainty for different modules to be quantified independently and then composed together. Importantly, the composition should be computationally efficient.
\item \textbf{Structured values:} Third, it should extend beyond classification (i.e., categorical labels) or regression (i.e., real-valued labels) to structured values such as lists, which are frequently encountered in programs. Importantly,  it should retain interpretability in this setting.
\end{itemize}

At a high level, two paradigms have emerged for providing uncertainty quantification of machine learning models. The first approach is \emph{calibrated prediction}~\cite{dawid1982well,guo2017calibration}, which builds on the standard approach where the model predicts the probability of each label given the input, but asks for these probabilities to be \emph{calibrated}. Intuitively, calibration says that among inputs where a model predicts the label is correct with probability $p$, the accuracy of the model is $p$. However, mathematical issues make it impossible to provide theoretical guarantees for calibration (roughly speaking, the set of inputs where a model predicts probability $p$ usually has measure zero); to circumvent this issue, the notion of calibration must be discretized in some way~\cite{gupta2020distribution}. Perhaps more problematically, calibrated probabilities become complicated for structured prediction problems (e.g., object detection), where many different probabilities are needed for different pieces of the output (e.g., a separate probability for each bounding box height, width, and object category), and these probabilities can be highly correlated. Typically, in calibrated structured prediction, only limited information such as marginal probabilities are reported~\cite{kuleshov2015calibrated}, which discards valuable information about these correlations.



An alternative is \emph{conformal prediction}~\cite{vovk2005algorithmic,park2019pac,angelopoulos2020uncertainty}, which modifies machine learning components to predict sets of labels instead of individual labels in a way that comes with probabilistic guarantees. At a high level, a machine learning model $\widehat{f}:\mathcal{X}\to\mathcal{Y}$ is turned into a \emph{prediction set model} $\widetilde{f}:\mathcal{X}\to2^{\mathcal{Y}}$ that predicts a set of labels for each input instead of an individual label. Conformal prediction guarantees that the resulting prediction sets contain the true label with high probability:
\begin{align}
\label{eqn:introconformal}
\mathbb{P}_{(x,y^*)\sim D}\left[y^*\in\widetilde{f}(x)\right]\ge1-\epsilon,
\end{align}
where $\epsilon\in\mathbb{R}_{>0}$ is a user-provided error bound and $D$ is the underlying data distribution (there are some additional details which we describe later).
We could guarantee (\ref{eqn:introconformal}) by choosing $\widetilde{f}(x)=\mathcal{Y}$ to always include all labels; however, this choice may be unnecessarily conservative. Instead, conformal prediction aims to minimize the size of the prediction sets while still guaranteeing (\ref{eqn:introconformal}).

We propose a novel strategy for quantifying the uncertainty of neurosymbolic programs based on conformal prediction. Given a neurosymbolic program, our strategy proceeds in three steps. First, we convert each machine learning component in this program into a prediction set model. Second, we overapproximate the resulting prediction sets using abstract values in some abstract domain~\cite{cousot1977abstract}. Finally, we compose these abstract values together using abstract interpretation. This strategy translates guarantees of the form (\ref{eqn:introconformal}) for each prediction set model into a guarantee of the same form for the program outputs. Furthermore, our strategy is naturally compositional, and can handle structured values as long as we can design an abstract domain for those values.

One shortcoming of this approach is that for larger programs, composing prediction sets can lead to increasing conservativeness---i.e., while the prediction sets still satisfy (\ref{eqn:introconformal}), they may become very large. When the entire program is available ahead-of-time, we can mitigate this issue by applying conformal prediction not only to the output of a machine learning component, but also at intermediate program points that have simple structure (e.g., a real-valued prediction). Then, we can intersect prediction sets at intermediate program points with those obtained via abstract interpretation. Modules not available ahead-of-time can still be calibrated separately. This strategy reduces the size of the output prediction set at the cost of diminishing compositionality.

\begin{figure*}[t]
\begin{subfigure}[c]{0.9\linewidth}
\begin{align*}
p~=~(\textsf{foldr}~+~(\textsf{map}~(\lambda d\,.\,1)~(\textsf{filter}~(\lambda d\,.\,d_1=\textsf{person}\wedge d_2\le300)~(\widehat{f}~X)))~0)
\end{align*}
\caption{Example image query}
\label{fig:prog}
\end{subfigure}
\begin{subfigure}[c]{0.45\linewidth}
\includegraphics[width=\linewidth]{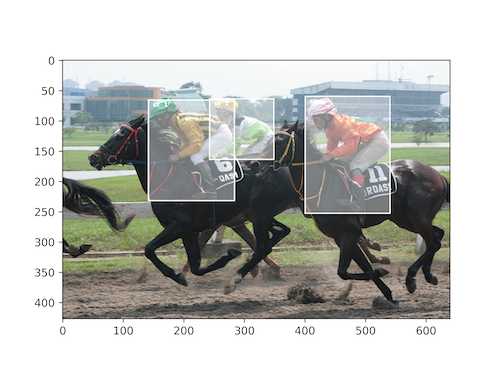}
\caption{Ground truth bounding boxes}
\label{fig:gt}
\end{subfigure}
\begin{subfigure}[c]{0.45\linewidth}
\includegraphics[width=\linewidth]{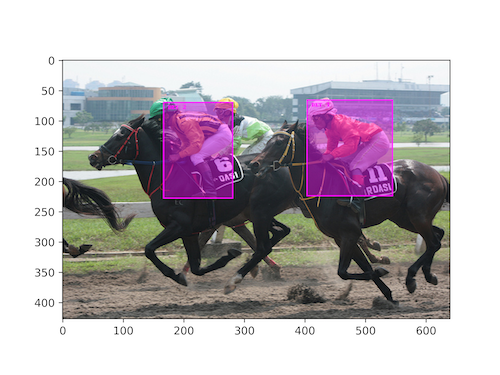}
\caption{Predicted detections}
\label{fig:pred}
\end{subfigure}
\begin{subfigure}[c]{0.45\linewidth}
\includegraphics[width=\linewidth]{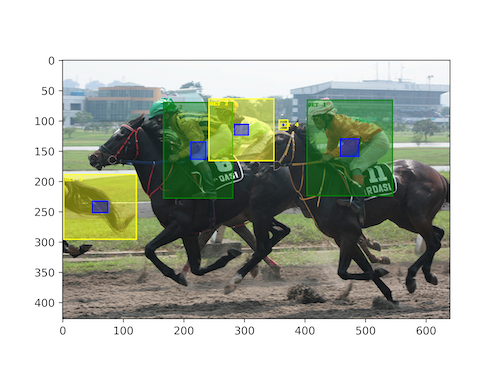}
\caption{Prediction set bounding boxes}
\label{fig:conf}
\end{subfigure}
\begin{subfigure}[c]{0.45\linewidth}
\begin{align*}
\den{p}^*(x)~&=~2 \\
\den{p}(x)~&=~1 \\
\den{p}^{C0}_{\epsilon_0}(x,Z)~&=~(1-1,1+1)~=~(0,2) \\
\den{p}^C_{\epsilon}(x,Z)~&=~(0,2)\sqcap(1,3)~=~(1,2)
\end{align*}
\caption{Ground truth, standard, direct conformal, and overall conformal semantics, respectively}
\label{fig:semantics}
\end{subfigure}
\caption{(a) A program in our image query DSL that counts the number of people in an image within 300 pixels of the left-hand side of the image; $\widehat{f}$ produces a list of detections, the program filters out people within 300 pixels of the left-hand side, maps each remaining element to $1$, and uses a fold to sum the list. (b) The ground truth bounding boxes. (c) The bounding boxes predicted by $\widehat{f}$. (d) The prediction sets output by the conformal predictor $\tilde{f}$. (e) The ground truth semantics $\den{p}^*(x)$ (evaluated using (b)) evaluates to 2. The standard semantics $\den{p}(x)$ (evaluated using (c)) evaluates to 1. The direct conformal semantics $\den{p}_{\epsilon_0}^{C0}(x,Z)$ evaluates to $(0, 2)$. The overall conformal semantics $\den{p}_{\epsilon}^{C}(x,Z)$ evaluates to $(1, 2)$.}
\label{fig:example}
\end{figure*}

As an example, Figure~\ref{fig:pred} shows the output of an object detector, which takes as input an image and outputs a list of \emph{detections} corresponding to objects predicted to be in the image; each detection consists of a \emph{bounding box} identifying the region of the image occupied by the object, and a category label for the object (e.g., person, car, etc.). Then, the image processing program in Figure~\ref{fig:prog} is designed to count the number of people who are within 300 pixels of the left border. Our framework converts this program into one that outputs a set of integers for each input image, which is guaranteed to contain the true number of such people in that image with high probability.

Finally, we construct a benchmark of image processing programs and empirically evaluate the effectiveness of our approach at solving programs in this benchmark. Example programs include selecting people in the image who are within a certain distance of a car, or computing the maximum distance between two people in the image. In summary, our contributions are:
\begin{itemize}
\item Formulating the conformal prediction problem for neurosymbolic programs (Section~\ref{sec:framework}).
\item A framework that combines conformal prediction and abstract interpretation to transform neurosymbolic programs into \emph{conformal neurosymbolic programs}, which come with coverage guarantees similar to (\ref{eqn:introconformal}) (Section~\ref{sec:framework}), with an extension to programs with loops (Section~\ref{sec:genframework}).
\item An evaluation of our framework, demonstrating that our framework constructs small prediction sets while achieving the desired coverage guarantee (Section~\ref{sec:experiments}).
\end{itemize}


\section{Motivating Example}


Consider a data scientist querying a database of images that are annotated using a deep neural network (DNN). For instance, recent work has shown how these kinds of queries can be used to identify frequent pedestrian crossings from drone images or where drivers routinely drive above the speed limit~\cite{bastani2021skyquery}. The results can be used to inform downstream interventions to improve safety, such as installing speed bumps. Na\"{i}vely, the data scientist may miss important instances due to mistakes made by the DNN. Thus, they may be interested in uncertainty quantification that identifies all but a small fraction of instances that satisfy their query.

\paragraph{Example program.}

Consider a program that takes as input an image $x\in\mathcal{X}$, applies an object detector $\widehat{f}:\mathcal{X}\to\mathcal{Y}$ to obtain a list of detections $\widehat{f}(x)\in\mathcal{Y}$, and then runs program logic over these detections. First, the object detector $\widehat{f}:\mathcal{X}\to\mathcal{Y}$ takes as input an image $x\in\mathcal{X}$, and outputs a list $y\in\mathcal{Y}=\mathcal{D}^*$ of predicted detections. Each detection $d\in\mathcal{D}=\mathcal{C}\times\mathbb{N}^2$ is a tuple, where $d_1\in\mathcal{C}$ is the object category (e.g., ``person'' or ``car'') and $(d_2,d_3)\in\mathbb{N}^2$ is the position of the center of the object (according to its predicted bounding box). Note that object detectors traditionally output bounding boxes; for simplicity, we focus on programs that only require the centers of the bounding boxes.

Figure~\ref{fig:prog} shows an example of a program $p$ that counts the number of people in an input image $x$ within 300 pixels of the left-hand side of the image. This program first applies the \textsf{filter} operator to filter out detections $d$ that are people ($d_1=\textsf{person}$) within 300 pixels of the left-hand side of this image ($d_2\le300$). Next, the \textsf{map} operator maps each remaining detection to $1$. Finally, the \textsf{foldr} operator sums the list, producing the desired count of the number of people.

We define two basic kinds of semantics for programs. First, the \emph{ground truth semantics} $\den{p}^*(x)$ evaluate the program when the object detector outputs the ground truth detections. For example, Figure~\ref{fig:gt} shows an example of an input image $x$ and the associated ground truth detections $y^*$. In this case, there are $2$ people within 300 pixels of the left edge the image, so $\den{p}^*(x)=2$.

However, in practice, we do not know $y^*$, so we cannot evaluate the ground truth semantics. Instead, we need to use an object detector $\widehat{f}$ to predict the set of detections, but $\widehat{f}$ may make mistakes. For example, in Figure~\ref{fig:pred}, we show the detections predicted by $\widehat{f}$; as can be seen, it misses a ground truth detection. The \emph{standard semantics} $\den{p}(x)$ evaluates $p$ using the detections produced by $\widehat{f}$; since $\widehat{f}$ may output the wrong detections, the standard semantics may output the incorrect output as well. Indeed, in our example, we have $\den{p}(x)=1\neq2=\den{p}^*(x)$.

\paragraph{Conformal prediction.}

Next, we describe how conformal prediction can be used to obtain prediction sets for the object detector $\widehat{f}:\mathcal{X}\to\mathcal{Y}$. In particular, the corresponding conformal predictor $\widetilde{f}:\mathcal{X}\to2^{\mathcal{Y}}$ outputs a set of possible labels. For object detectors, $\widehat{f}$ outputs a list of detections, so $\widetilde{f}$ outputs a set of lists of detections. Conformal prediction assumes that $\widehat{f}$ is derived from a \emph{scoring function} $g:\mathcal{X}\times\mathcal{Y}\to\mathbb{R}$---i.e., given image $x\in\mathcal{X}$ and label $y\in\mathcal{Y}$, $g(x,y)$ is a real-valued score indicating how likely it is that $y=y^*$, where $y^*$ is the true label. Neural networks typically already output scores (e.g., prediction probabilities) or can be modified to do so. No assumptions are made about the performance of $g$; if $g$ outputs low-quality scores, then conformal prediction will construct large prediction sets to compensate (with $\widetilde{f}(x)=\mathcal{Y}$ in the worst case).

At a high level, conformal prediction takes as input a held-out \emph{calibration set} $Z\subseteq\mathcal{X}\times\mathcal{Y}$ of labeled examples, and uses these labeled examples to evaluate the quality of the scoring function. In more detail, it considers a class of prediction sets of the form
\begin{align*}
\widetilde{f}_{\tau}(x)=\{y\in\mathcal{Y}\mid g(x,y)\ge\tau\},
\end{align*}
where $\tau\in\mathbb{R}$ is a real-valued parameter. In other words, $\widetilde{f}_{\tau}(x)$ is the $\tau$ level set of $g$. For a desired coverage rate of $1 - \epsilon$, the algorithm uses $Z$ to choose $\tau=\widehat{\tau}(Z,\epsilon)$ so that $\widetilde{f}_{\widehat{\tau}(Z,\epsilon)}$ satisfies a guarantee of the form (\ref{eqn:introconformal}):
\begin{align*}
\mathbb{P}_{Z\sim D^n,(x,y^*)\sim D}\left[y^*\in\widetilde{f}_{\widehat{\tau}(Z,\epsilon)}(x)\right]\ge1-\epsilon.
\end{align*}
Note that the conformal guarantee includes the randomness over $Z\sim D^n$. 

In Figure~\ref{fig:conf}, we show the output of the conformal predictor for the given image $x$. Note that because object detections are structured predictions, prediction sets over a collection of detections typically contain infinitely many elements. Thus, we use \emph{abstract values} to represent the possible prediction sets. We represent the integer coordinates of the detection using interval abstractions, and the object category using set abstractions (since the set of possible object categories is finite). Then, an abstract detection is a tuple $d^\#=(C,(\ell_1,u_1),(\ell_2,u_2))\in\mathcal{D}^\#$, where $C\in\mathcal{C}^\#=2^{\mathcal{C}}$ is a set of object categories, $(\ell_1,u_1)\in\mathbb{N}^\#=\mathbb{N}^2$ represents the set of possible horizontal positions $[\ell_1,u_1]$, and $(\ell_2,u_2)\in\mathbb{N}^\#$ represents the set of possible vertical positions $[\ell_2,u_2]$ of the center of the object. \ten{Here, $(\ell,u)$ is not an open interval, but an abstract value consisting of a pair of real numbers, which represents the closed interval $[\ell,u]=\gamma((\ell,u))$.}

Next, the conformal object detector outputs a list of abstract detections, but it may additionally be uncertain about which detections are even in the list. Thus, our domain of abstract lists of detections is $\mathcal{Y}^\#=(\mathcal{D}^\#\times\mathbb{B}^\#)$, where $\mathbb{B}^\#=\{\textsf{true},\textsf{false}, \top\}$ is the space of abstract Booleans. In particular, an element $(a^\#,b^\#)\in\mathcal{D}^\#\times\mathbb{B}^\#$ of an abstract list indicates that detection $a^\#$ is either definitely present in the list if $b^\#=\textsf{true}$, or possibly present in the list if $b^\#=\top$ (if it is definitely not present in the list, i.e. $b^\#=\textsf{false}$, we simply omit it). In Figure~\ref{fig:conf}, we visualize the Boolean flag for each detection by coloring the detection green if $b=\textsf{true}$ and yellow if $b=\top$. The blue box shows the prediction set $[\ell_1,u_1]\times[\ell_2,u_2]$ around the center of the corresponding detection.

\paragraph{Conformal semantics.}

Our goal is to construct a prediction set for the overall program, not just the object detector. We combine two strategies to do so. First, we use abstract interpretation~\cite{cousot1977abstract} to propagate prediction sets through the program.
\ten{In some cases, it may be possible to use abstractions that do not result in any overapproximation (e.g., set abstractions if all prediction sets are finite), but in other cases overapproximations are either unavoidable (e.g., prediction sets are real intervals) or beneficial (e.g., they significantly reduce running time).}
We assume we are given abstract transformers for each component. Since prediction sets are abstract values, we can evaluate the abstract transformer on the abstract values at each program point, and obtain an abstract output representing a set of possible outputs. We denote this abstract output by $\den{p}_{\epsilon_1}^{C1}(x,Z)$, where $\epsilon_1\in\mathbb{R}_{\ge0}$ is the desired coverage level and $Z=\{x_1,...,x_n\}\subseteq\mathcal{X}$ is a calibration set used to construct conformal predictors; rather than explicitly include labels in $Z$, we implicitly assume that we can evaluate the ground truth semantics for each $x\in Z$. If $y^*\in\widetilde{f}_{\widehat{\tau}(Z,\epsilon_1)}(x)$ with high probability, it follows that $\den{p}^*(x)\in\gamma(\den{p}_{\epsilon_1}^{C1}(x,Z))$ with high probability (specifically, probability at least $1-\epsilon_1$), where $\gamma$ is the concretization function mapping abstract values to sets of concrete values.

In our example image, we have $\den{p}^{C1}_{\epsilon_1}(x,Z)=(1,3)$. We call this the \emph{compositional conformal semantics}. Intuitively, the prediction set visualized in Figure~\ref{fig:conf} contains between 2 and 5 detections labeled ``person'', since there are 2 green boxes definitely labeled ``person'', and 3 yellow boxes definitely labeled ``person'' but that may not correspond to actual objects (one of the yellow boxes is very small). Out of these, one green box (a ``sure'' object) and two yellow boxes (an ``unsure'' object) are within 300 pixels of the left edge. 

While this strategy of constructing prediction sets for machine learning components and propagating them through the program provides the desired coverage guarantee on its own, we can also apply conformal prediction in other ways to improve performance. In our example program, the output is an integer value, so we can construct a prediction set (with conformal guarantees) directly around this value. In more detail, we can construct the modified calibration set
\begin{align*}
Z_p=\{(x,\den{p}^*(x))\mid x\in Z\},
\end{align*}
where $\den{p}^*(x)$ can be evaluated for examples in $Z$, since the ground truth labels $y$ corresponding to all $x \in Z$ are given.
In addition, we can construct a trivial scoring function of the following form:
\begin{align*}
g_p(x,y)=-|y-\den{p}(x)|.
\end{align*}
That is, the score for output $y$ is the distance of $y$ from the evaluated standard semantics. Next, we consider the prediction set model $\widetilde{p}_{\tau}:\mathcal{X}\to2^{\mathbb{N}}$ of the form
\begin{align}
\label{eqn:exprogconformal}
\widetilde{p}_{\tau}(x)=\{y\in\mathbb{N}\mid g_{p}(x,y)\ge\tau\}.
\end{align}
This output $\widetilde{p}_{\tau}$ is represented as an abstract value in our domain, e.g., if we use the interval domain to represent integers, $\widetilde{p}_{\tau}(x) = (\den{p}(x)-\tau,\den{p}(x)+\tau)$, with corresponding concrete values
\begin{align*}
\gamma((\den{p}(x)-\tau,\den{p}(x)+\tau))
=\{y\in\mathbb{N}\mid\den{p}(x)-\tau\le y\le\den{p}(x)+\tau\},
\end{align*}
which is equivalent to (\ref{eqn:exprogconformal}). For a desired coverage rate $1 - \epsilon_0$, using conformal prediction in conjunction with $Z_p$ and $g_p$ will produce a specific parameter $\widehat{\tau}(Z_p,\epsilon_0)$ (the scoring function $g_p$ is implicit) and corresponding prediction set model $\widetilde{p}_{\widehat{\tau}(Z_p,\epsilon_0)}$, which satisfies the coverage guarantee
\begin{align*}
\mathbb{P}_{Z\sim D^n,x\sim D}\left[\den{p}^*(x)\in\widetilde{p}_{\widehat{\tau}(Z_p,\epsilon_0)}(x)\right]\ge1-\epsilon_0.
\end{align*}
We call $\widetilde{p}_{\widehat{\tau}(Z_p,\epsilon_0)}$ the \emph{direct conformal semantics}, and denote it by
$\den{p}_{\epsilon_0}^{C0}(x,Z)=\widetilde{p}_{\widehat{\tau}(Z_p,\epsilon_0)}(x)$,
where $Z_p$ is the prediction set constructed from $Z$ and $p$, and the scoring function $g_p$ is implicit. In our example in Figure~\ref{fig:semantics}, $\den{p}(x)=1$. Suppose that conformal prediction chooses $\widehat{\tau}(Z_p,\epsilon_0)=1$; then, $\widetilde{p}_{\widehat{\tau}(Z_p,\epsilon_0)}(x)=(1-1,1+1)=(0,2)$, which is an interval containing the true output $\den{p}^*(x)=2$. 

\paragraph{Combining Approaches for Optimization.}
When the program $p$ is available ahead-of-time, we can combine both of these approaches; our algorithm constructs both the direct conformal semantics as well as the compositional conformal semantics obtained by propagating prediction sets throughout the program. This strategy produces two different abstract values, $\den{p}_{\epsilon_0}^{C0}(x,Z)$ obtained by the direct conformal semantics, and the abstract value $\den{p}_{\epsilon_1}^{C1}(x,Z)$ obtained by propagating $\widetilde{f}_{\widehat{\tau}(Z,\epsilon_1)}(x)$ using abstract interpretation. These two values have tradeoffs---abstract interpretation can be imprecise, leading to larger prediction sets, but the scoring function used in direct conformal prediction may not be as high quality as the one used to construct $\widetilde{f}_{\widehat{\tau}(Z,\epsilon_1)}$. They are both valid abstract values (i.e., both contain $\den{p}^*(x)$ with high probability); thus, we let
\begin{align*}
\den{p}_{\epsilon}^C(x,Z)=\den{p}_{\epsilon_0}^{C0}(x,Z)\sqcap\den{p}_{\epsilon_1}^{C1}(x,Z),
\end{align*}
where $\sqcap$ is the meet operator that overapproximates intersection. In our example, we have
\begin{align*}
\den{p}_{\epsilon}^C(x,Z)=\den{p}_{\epsilon_0}^{C0}(x,Z)\sqcap\den{p}_{\epsilon_1}^{C1}(x,Z)=(0,2)\sqcap(1,3)=(1,2),
\end{align*}
i.e., there are between 1 and 2 people within 300 pixels of the left side. Note that $(1,2)$ contains the true output $\den{p}^*(x)=2$, and is smaller than either $\den{p}_{\epsilon_0}^{C0}(x,Z)=(0,2)$ or $\den{p}_{\epsilon_1}^{C1}(x,Z)=(1,3)$.

In fact, our algorithm can do even better---it constructs direct conformal predictors at \emph{every} intermediate program point and immediately meets it with the abstract value obtained from abstract interpretation at that program point, and continues abstract interpretation with the resulting value. This strategy ensures that we keep the abstract values as precise as possible throughout execution.

Note that when combining prediction sets with probabilistic guarantees, we need to take a union bound, so $\epsilon_0$ and $\epsilon_1$ must be chosen to satisfy $\epsilon_0 + \epsilon_1 \leq \epsilon$.

\section{Background}
\label{sec:background}

We provide background on abstract interpretation~\cite{cousot1977abstract}, and on conformal prediction~\cite{vovk2005algorithmic}.

\paragraph{Abstract interpretation.}

For a concrete domain $\mathcal{X}$ and functions $f\in\mathcal{F}$ of type $f:\mathcal{X}^k\to\mathcal{X}$, we consider an abstract semantics $(\mathcal{X}^\#,\sqsubseteq,\sqcup,\sqcap,\alpha,\gamma,\mathcal{F}^\#)$, where the following hold:
\begin{itemize}
\item \textbf{Lattice:} $(\mathcal{X}^\#,\sqsubseteq,\sqcup,\sqcap)$ forms a lattice with partial order $x_\#\sqsubseteq x_\#'$, join $x_\#\sqcup x_\#'$, and meet $x_\#\sqcap x_\#'$. 
We assume it satisfies an \emph{ascending chain condition}, which says any sequence $x_\#^{(1)}\sqsubseteq x_\#^{(2)}\sqsubseteq\cdots$ \emph{stabilizes}---i.e., there exists $m\in\mathbb{N}$ such that $x_\#^{(m')}=x_\#^{(m)}$ for all $m'\ge m$. This condition ensures that our semantics for while loops are well defined.
\item \textbf{Abstraction/concretization functions:} The abstraction function $\alpha:2^{\mathcal{X}}\to\mathcal{X}^\#$ transforms a set of concrete values $X\subseteq\mathcal{X}$ into an abstract value $x_{\#}$ containing $X$. Given $x\in\mathcal{X}$, we let $\alpha(x)=\alpha(\{x\})$. The concretization function $\gamma:\mathcal{X}^\#\to2^{\mathcal{X}}$ transforms an abstract value into the set of concrete values it contains. We assume that $\alpha$ and $\gamma$ form a Galois connection:
\begin{align}
\label{eqn:abs1}
\alpha(X)\sqsubseteq x_\#\Leftrightarrow X\subseteq\gamma(x_\#)
\qquad(\forall X\in2^{\mathcal{X}},x_\#\in\mathcal{X}^\#).
\end{align}
Then, the join overapproximates union (i.e., $x\in\gamma(x_\#)\vee x\in\gamma(x_\#')\Rightarrow x\in\gamma(x_\#\sqcup x_\#')$) and the meet overapproximates intersection (i.e., $x\in\gamma(x_\#)\wedge x\in\gamma(x_\#')\Rightarrow x\in\gamma(x_\#\sqcap x_\#')$).
\item \textbf{Abstract transformers:} Each function $f\in\mathcal{F}$ corresponds to an abstract transformer $f^\#\in\mathcal{F}^\#$  that overapproximates the semantics of $f$:
\begin{align}
\label{eqn:abs4}
x_1\in\gamma(x_{1,\#})\wedge...\wedge x_k\in\gamma(x_{2,\#})\Rightarrow f(x_1,...,x_k)\in\gamma(f^\#(x_{1,\#},...,x_{k,\#})).
\end{align}
\end{itemize}

\paragraph{Conformal prediction.}

Next, we describe how we use conformal prediction to modify machine learning models to construct prediction sets that come with probabilistic guarantees. Consider a machine learning model $\widehat{f}:\mathcal{X}\to\mathcal{Y}$ that has already been trained; we assume $\widehat{f}$ is derived from a scoring function $g:\mathcal{X}\times\mathcal{Y}\to\mathbb{R}$, i.e., $\widehat{f}(x)=\operatorname*{\arg\max}_{y\in\mathcal{Y}}g(x,y)$. This assumption is standard in conformal prediction; typical machine learning models can be modified to predict probabilities of labels, which can be used as the scoring function. Then, given an error bound $\epsilon\in\mathbb{R}_{>0}$ and a calibration set $\{(x_i,y_i^*)\}_{i=1}^n$ of i.i.d. examples $(x_i,y_i^*)\sim D$, conformal prediction constructs a \emph{conformal predictor} $\widetilde{f}_{Z,\epsilon}:\mathcal{X}\to2^{\mathcal{Y}}$ satisfying the \emph{coverage guarantee}
\begin{align*}
\mathbb{P}_{Z\sim D^n,(x,y^*)\sim D}\left[y^*\in\widetilde{f}_{Z,\epsilon}(x)\right]\ge1-\epsilon.
\end{align*}
Note that in this case, the calibration dataset additionally includes the ground truth labels $y_i^*$ for each input $x_i$; our algorithm will construct these labels from $x_i$ using the ground truth semantics. We denote the conformal prediction subroutine by $\widetilde{f}_{Z,\epsilon}=C_{\epsilon}(g,Z)$. In our experiments, we use a slightly stronger variant of conformal prediction~\cite{vovk2012conditional,park2019pac}; see Appendix~\ref{sec:pac} for details.

We summarize the conformal prediction strategy; see~\citet{angelopoulos2023conformal} for details. The main idea is to consider conformal predictors of the form $\widetilde{f}_{Z,\epsilon}=\widetilde{f}_{\widehat{\tau}(Z,\epsilon)}$, where $\widehat{\tau}(Z,\epsilon)\in\mathbb{R}$ is a real-valued parameter chosen by the conformal prediction algorithm based on $Z$ and $\epsilon$, and
\begin{align*}
\widetilde{f}_{\tau}(x)=\{y\in\mathcal{Y}\mid g(x,y)\ge\tau\}.
\end{align*}
Intuitively, we could choose $\tau$ to empirically satisfy the coverage guarantee (\ref{eqn:compconformal}) for the calibration dataset $Z$; however, this strategy does not guarantee (\ref{eqn:compconformal}) due to generalization error. However, because we are only choosing a single parameter $\tau$, we can obtain explicit bounds on the generalization error, which can be used to adjust $\tau$ to obtain a theoretical guarantee.

Finally, conformal prediction sets are often infinitely large; thus, they are implicitly represented using alternative representations. For example, for regression, where the labels $\mathcal{Y}=\mathbb{R}$ are real valued, prediction sets are usual intervals $[\ell,u]\subseteq\mathbb{R}$. This strategy works well with our abstract interpretation framework, since we can reasonably assume that the conformal predictor $\widetilde{f}_{Z,\epsilon}$ outputs abstract values instead of sets---i.e., $\widetilde{f}_{Z,\epsilon}:\mathcal{X}\to\mathcal{X}^\#$. Then, the coverage guarantee becomes
\begin{align}
\label{eqn:compconformal}
\mathbb{P}_{Z\sim D^n,(x,y^*)\sim D}\left[y^*\in\gamma\left(\widetilde{f}_{Z,\epsilon}(x)\right)\right]\ge1-\epsilon.
\end{align}
We use this form of conformal prediction in our approach. Note that for this strategy to work, the abstract values we use must be compatible with the structure of the prediction sets. In particular, they must be able to overapproximate the level sets of the scoring function $g$ (since prediction sets have this form)---i.e., for any $\tau\in\mathbb{R}$ and $x\in\mathcal{X}$, there exists $y_{\tau,x}^\#\in\mathcal{Y}^\#$ such that
\begin{align*}
\gamma(y_{\tau,x}^\#)\supseteq\{y\in\mathcal{Y}\mid g(x,y)\ge\tau\}.
\end{align*}
In other words, $y_{\tau,x}^\#$ contains the $\tau$ level set of $g$; ideally, the inequality should be as tight as possible. We assume that $y_{\tau,x}^\#$ can be computed efficiently, which is typically the case for practical abstract domains. Once we have chosen $\tau$ using conformal prediction, we can define
$\widetilde{f}_\tau(x)=y_{\tau,x}^\#$.
For example, for real-valued predictions $\mathcal{Y}=\mathbb{R}$, a typical choice of scoring function is
\begin{align*}
g(x,y)=\frac{-|y-\widehat{\mu}(x)|}{\widehat{\sigma}(x)},
\end{align*}
where $\widehat{\mu}(x)$ is the predicted value of $y^*$ and $\widehat{\sigma}(x)$ is an estimate of the variance of this prediction. Then, the level sets of $g$ have the form $\{y\in\mathcal{Y}\mid|y-\widehat{\mu}(x)|\le\widehat{\sigma}(x)\cdot\tau\}$, so we can take
\begin{align*}
\widetilde{f}_\tau(x)=\left(\widehat{\mu}(x)-\widehat{\sigma}(x)\cdot\tau,\widehat{\mu}(x)+\widehat{\sigma}(x)\cdot\tau\right)
\end{align*}
where the abstract outputs are intervals---i.e.,
\begin{align*}
\gamma\left(\left(\widehat{\mu}(x)-\widehat{\sigma}(x)\cdot\tau,\widehat{\mu}(x)+\widehat{\sigma}(x)\cdot\tau\right)\right)=\left[\widehat{\mu}(x)-\widehat{\sigma}(x)\cdot\tau,\widehat{\mu}(x)+\widehat{\sigma}(x)\cdot\tau\right].
\end{align*}
Typically, $\widehat{\mu}$ is the prediction of a regression model (e.g., linear regression). Models such as deep neural networks can straightforwardly be modified to predict both $\widehat{\mu}$ and $\widehat{\sigma}$~\cite{park2019pac}. Alternatively, we can use $\widehat{\sigma}(x)=1$, with the caveat that prediction sets will all have the same size.

\section{Framework for Loop-Free Programs}
\label{sec:framework}

Now, we describe our framework for constructing \emph{conformal semantics} for programs that modify standard semantics to provide conformal guarantees. We focus on loop-free programs, and discuss an extension to programs with loops in Section~\ref{sec:genframework}. In particular, for each input, the conformal semantics produces a \emph{prediction set} of outputs (rather than a single output) that is guaranteed to cover the true output with high probability (Section~\ref{sec:problem}). Our framework uses two strategies:
\begin{itemize}
\item \textbf{Direct conformal semantics:} First, our framework directly uses conformal prediction to construct conformal semantics for each intermediate program point (Section~\ref{sec:direct}).
\item \textbf{Compositional conformal semantics:} A shortcoming of direct conformal semantics is that conformal prediction requires a \emph{scoring function} to work well, which is not available for all program points. Thus, our framework evaluates each program point on the prediction set of values outputted by previous program points (Section~\ref{sec:compositional}).
\end{itemize}
Finally, our full conformal semantics intersects these two at each program point (Section~\ref{sec:full}).

\subsection{Problem Formulation}
\label{sec:problem}

We consider a domain-specific language (DSL) represented by the following grammar $\cG$:
\begin{align}
\label{eqn:dsl}
P~::=~X\mid f(P,...,P)~(\forall f\in\mathcal{F}),
\end{align}
where $\mathcal{F}$ is a given set of functions (constants are functions with no arguments), and $X$ is a symbol representing the input $x\in\mathcal{X}$ of a program $p$ generated by these productions. In addition, we consider the following denotational semantics $\den{p}:\mathcal{X}\to\mathcal{X}$:
\begin{align*}
\den{X}(x)=x,
\qquad
\den{f(p_1,...,p_k)}(x)=f(\den{p_1}(x),...,\den{p_k}(x)).
\end{align*}
Here, $\mathcal{X}$ is the union of all possible program values (inputs, intermediate values, and outputs). 

We are interested in settings where some of the functions $\widehat{\mathcal{F}}\subseteq\mathcal{F}$ are machine learning models, which we refer to as \emph{machine learning components}. We additionally associate each machine learning component $\widehat{f}\in\widehat{\mathcal{F}}$ with a \emph{ground truth function} $f^*$, where $f^*(x_1,...,x_k)$ is the ground truth label associated with $\widehat{f}(x_1,...,x_k)$, and corresponding \emph{ground truth semantics} $\den{p}^*:\mathcal{X}\to\mathcal{X}$, given by
\begin{align*}
\den{X}^*(x)=x,
\qquad
\den{f(p_1,...,p_k)}^*(x)=f^*(\den{p_1}^*(x),...,\den{p_k}^*(x)),
\end{align*}
where we let $f^*=f$ if $f\not\in\widehat{\mathcal{F}}$. Importantly, we do not assume access to $f^*$ or $\den{\cdot}^*$ except on inputs $x$ in our calibration set (discussed below). Now, our goal is to modify $p$ to obtain a program that outputs \emph{sets} of values instead of individual values. In particular, we define a \emph{conformal semantics} $\den{p}_{\epsilon}^C:\mathcal{X}\times\mathcal{X}^n\to\mathcal{X}^\#$ that contains the true label with high probability:
\begin{align}
\label{eqn:progconformal}
\mathbb{P}_{Z\sim D^n,x\sim D}\left[\den{p}^*(x)\in\gamma\left(\den{p}_{\epsilon}^C(x,Z)\right)\right]\ge1-\epsilon,
\end{align}
where $D$ is a distribution over inputs $x\in\mathcal{X}$ and $\epsilon\in\mathbb{R}_{>0}$ is a user-provided error bound. We have dropped the true labels $y^*$ since they are captured in (unknown) ground truth functions $f^*$. In addition, $Z=(x_1,...,x_n)\subseteq\mathcal{X}$ is a given \emph{calibration set} of i.i.d. samples $x_1,...,x_n\sim D$ intended to help us construct all conformal predictors for all of the machine learning components in $p$. \ten{We assume that for each input $x\in Z$, we know the relevant output of the ground truth function $f^*$ for each call to a machine learning component $\widehat{f}$; thus, we can evaluate the ground truth semantics $\den{p}^*(x)$ for all $x\in Z$.} The calibration set should be a separate held-out set from the training and test sets. We also assume that each machine learning component $\widehat{f}\in\mathcal{F}$ (assumed to have form $\widehat{f}:\mathcal{X}_1\times...\times\mathcal{X}_k\to\mathcal{X}$, where the subscripts are to distinguish the $k$ inputs) is associated with a scoring function $g:\mathcal{X}_1\times...\times\mathcal{X}_k\times\mathcal{X}\to\mathbb{R}$ such that
\begin{align*}
\widehat{f}(x_1,...,x_k)=\operatorname*{\arg\max}_{y\in\mathcal{X}}g(x_1,...,x_k,y).
\end{align*}
In summary, given $\epsilon\in\mathbb{R}_{\ge0}$, our goal is to construct the conformal semantics $\den{p}_{\epsilon}^C$ from the calibration set and the scoring functions, such that the guarantee (\ref{eqn:progconformal}) holds.

\subsection{Direct Conformal Semantics}
\label{sec:direct}

Our direct conformal semantics applies conformal prediction directly to intermediate program points. It suffices to describe how to obtain conformal prediction guarantees for an arbitrary program $p$, since each subprogram is itself a program.
In particular, suppose we are given a program $p$ and a set of calibration inputs $Z=\{x_1,...,x_n\}\subseteq\mathcal{X}$, where $x_i\sim D$ are i.i.d. samples. Then, our goal is to construct conformal semantics $\den{\cdot}_{\epsilon}^{C0}$ that satisfy (\ref{eqn:progconformal}). For variables $p=X$, the standard semantics and the ground truth semantics coincide, so we can take $\den{X}_{\epsilon}^{C0}(x,Z)=\alpha(\den{X}(x))=\alpha(x)$ and satisfy (\ref{eqn:progconformal}).
The interesting case is $p=f(p_1,...,p_k)$; in this case, we use conformal prediction to construct $\den{p}_{\epsilon}^{C0}$. If $f\in\widehat{\mathcal{F}}$, we can use its conformal predictor---i.e., $\den{p}_{\epsilon}^{C0}(x,Z)=\widetilde{f}_{Z,\epsilon}(x)$. Otherwise, we need to construct a conformal predictor. To this end, we first construct the calibration set
\begin{align*}
Z_p=\left\{\left(x,\den{p}^*(x)\right)\mid x\in Z\right\}.
\end{align*}
To use conformal prediction, we still need to construct the scoring function $g_p:\mathcal{X}\times\mathcal{X}\to\mathbb{R}$, as well as an abstract domain that can be used to represent the prediction sets. Both of these objects must be specialized based on $\text{range}(p)$---i.e., the values that $\den{p}(x)$ may take. Our framework uses a few simple choices, but alternative choices can easily be incorporated. First, if $p=\widehat{f}(p_1,...,p_k)$, then letting $g$ be the scoring function for $\widehat{f}$, we take the scoring function for $p$ to be
\begin{align*}
g_p(x,y)=g(\den{p_1}(x),...,\den{p_k}(x),y)
\end{align*}
Note that we use the standard semantics to evaluate the inputs $p_i$ to $p$, and the ground truth semantics to evaluate the output $y$---i.e., $\den{p}_{\epsilon}^{C0}(x,Z)$ computes the inputs to $p$ using the standard semantics, and the conformal guarantee ensures $\den{p}^*(x)\in\gamma(\den{p}_{\epsilon}^{C0}(x,Z))$ with high probability.

In the second case, $p=f(p_1,...,p_k)$ for some $f\not\in\widehat{\mathcal{F}}$. We consider three sub-cases depending on $\text{range}(p)$, i.e., the possible outputs $\den{p}(x)$. First, if $\text{range}(p)\subseteq\mathbb{R}$, then we use
\begin{align*}
g_p(x,y)=-|y-\den{p}(x)|,
\end{align*}
i.e., higher scores for labels closer to the standard semantics $\den{p}(x)$. Intuitively, assuming $\den{p}(x)\approx\den{p}^*(x)$, this scoring function will assign relatively high scores to ground truth outputs.

Second, if $\text{range}(p)$ is any discrete set; in this case, we use
\begin{align*}
g_p(x,y)=\begin{cases}
1 &\text{if }\den{p}(x)=y \\
0 &\text{if }y\in\text{range}(p) \\
-1 &\text{otherwise},
\end{cases}
\end{align*}
i.e., the only label with a score of one is the standard semantics $\den{p}(x)$, and all other possible outputs have zero score (impossible outputs have a score of $-1$). Note that the scoring function for this case is very coarse-grained---it will only construct prediction sets that are either $\widetilde{p}(x)=\alpha(\den{p}(x))$ or $\widetilde{p}(x)=\top$ (where $\gamma(\top)=\text{range}(p)$ represents all possible values $p$ may output). However, we cannot do better without additional assumptions on its structure.

Third, if $\text{range}(p)$ contains both discrete and real components (i.e., $\text{range}(p)\subseteq\mathcal{X}_0\times\mathbb{R}^d$, where $\mathcal{X}_0$ is discrete), we can apply conformal prediction separately to each component (i.e., $\widetilde{p}_i$ for component $i\in\{0,1,...,d\}$) and take the product of the resulting conformal predictors
$\widetilde{p}(x)=\prod_{i=0}^d\widetilde{p}_i(x)$.
These three cases cover ranges that arise in typical applications.

Finally, once we have constructed the calibration set $Z_p\subseteq\mathcal{X}\times\mathcal{X}$ and the scoring function $g_p:\mathcal{X}\times\mathcal{X}\to\mathbb{R}$, we can construct the conformal predictor $\widetilde{p}=C_{\epsilon}(g_p,Z_p)$, which has type $\widetilde{p}:\mathcal{X}\to\mathcal{X}^\#$, and define the conformal semantics for $p$ to be
\begin{align*}
\den{p}_{\epsilon}^{C0}(x,Z)=\widetilde{p}(x),
\end{align*}
where $\widetilde{p}$ depends on $Z$ and $\epsilon$. We have the following result, which follows directly from (\ref{eqn:compconformal}):
\begin{lemma}
\label{lem:direct}
The direct conformal semantics $\den{p}_{\epsilon}^{C0}(x,Z)$ satisfies (\ref{eqn:progconformal}).
\end{lemma}

\subsection{Compositional Conformal Semantics}
\label{sec:compositional}

While $\den{p}_{\epsilon}^{C0}$ can provide conformal programs satisfying (\ref{eqn:progconformal}), they can produce prediction sets that are overly conservative, especially for programs involving discrete outputs. As a consequence, we augment these prediction sets with an alternative strategy that comes from the compositional structure of the program. Intuitively, for a program $p=f(p_1,...,p_k)$, we can combine an \emph{abstract transformer} $f^\#$ for $f$ with the direct conformal semantics $\den{p}_{\epsilon}^{C0}$ by running $f^\#$ on the conformal semantics for $p_1,...,p_k$ (obtained compositionally), and then \emph{intersecting} its output with $\den{p}_{\epsilon}^{C0}$.

To this end, we assume we are given an abstract transformer $f^\#:\mathcal{X}_1^\#\times...\times\mathcal{X}_k^\#\to\mathcal{X}^\#$ for the component $f:\mathcal{X}_1\times...\times\mathcal{X}_k\to\mathcal{X}$, where $\mathcal{X}_i^\#$ is the abstract domain for $\mathcal{X}_i$. Then, we define the \emph{compositional conformal semantics} $\den{p}_{\epsilon}^{C1}$ by $\den{X}^{C_1}_{\epsilon}(x,Z)=\alpha(x)$ and
\begin{align*}
\den{f(p_1,...,p_k)}_{\epsilon}^{C1}(x,Z)=f^\#\left(\den{p_1}_{\epsilon_1}^{C1}(x,Z),...,\den{p_k}_{\epsilon_k}^{C1}(x,Z)\right),
\end{align*}
where $\epsilon\ge\sum_{i = 1}^k \epsilon_i$; we discuss this constraint and the specific choices of $\epsilon_1,...,\epsilon_k$ below. By the standard property of abstract transformers, we can show $\den{p}^{C1}_{\epsilon}$ satisfies (\ref{eqn:progconformal}) by structural induction:
\begin{lemma}
\label{lem:compositional}
The compositional conformal semantics $\den{p}_{\epsilon}^{C1}(x,Z)$ satisfies (\ref{eqn:progconformal}).
\end{lemma}

\subsection{Full Conformal Semantics}
\label{sec:full}

In principle, we could define our \emph{full conformal semantics} $\den{p}_{\epsilon}^C:\mathcal{X}\times\mathcal{X}^n\to\mathcal{X}^\#$ to be
\begin{align*}
\den{p}^C_{\epsilon}(x,Z)=\den{p}^{C0}_{\epsilon_0}(x,Z)\sqcap\den{p}^{C1}_{\epsilon_1}(x,Z),
\end{align*}
for some $\epsilon_0,\epsilon_1$ satisfying $\epsilon\ge\epsilon_0+\epsilon_1$. However, we can actually take the intersection with $\den{p}^{C0}_{\epsilon}$ at every program point. In particular, we define $\den{X}_{\epsilon}^C(x,Z)=\alpha(x)$ and
\begin{align*}
\den{f(p_1,...,p_k)}_{\epsilon}^C(x,Z)=
\begin{cases}
f^\#\left(\den{p_1}_{\epsilon_1}^C(x,Z),...,\den{p_k}_{\epsilon_k}^C(x,Z)\right)\sqcap\den{p}_{\epsilon_0}^{C0}(x,Z) &\text{if }f\not\in\widehat{\mathcal{F}} \\
\den{p}_{\epsilon}^{C0}(x,Z) &\text{otherwise}.
\end{cases}
\end{align*}
Note that we do not use abstract interpretation for machine learning components $f\in\widehat{\mathcal{F}}$ since we typically do not have abstract transformers for these components.
In other words, we evaluate the abstract transformer on the abstract values of the arguments $p_1,...,p_k$ of $p$ (which are recursively evaluated using the same strategy), and then intersect this abstract value with the direct conformal semantics, where we need to specify the error bounds $\epsilon_i$, for both the recursive evaluation of $p_i$ and the evaluation of the direct conformal semantics.

To obtain an overall error bound of $\epsilon$, we need to take a union bound over all applications of machine learning components---i.e., we need $\{\epsilon_i\}_{i=0}^k$ to satisfy
\begin{align}
\label{eqn:deltaconstraint}
\epsilon\ge\epsilon_0+\sum_{i=1}^k\epsilon_i.
\end{align}
We can use any choice of $\{\epsilon_i\}_{i=0}^k$ satisfying (\ref{eqn:deltaconstraint}); for example, we can choose all the $\epsilon_i=\epsilon/(1+k)$ to all be equal.
In practice, we want to allocate more weight to $\epsilon_i$ corresponding to less conservative conformal predictors to minimize prediction set size. Typically, it suffices to optimize the tradeoff between the direct conformal semantics (i.e., $\epsilon_0$) and the compositional approach (i.e., $\epsilon_1=\epsilon_2=...=\epsilon_k$). For any $\{\epsilon_i\}_{i=0}^k$ satisfying (\ref{eqn:deltaconstraint}), we have the following result; see Appendix~\ref{sec:thm:loopfree:proof} for a proof:
\begin{theorem}
\label{thm:loopfree}
The full conformal semantics $\den{p}_{\epsilon}^C$ satisfies (\ref{eqn:progconformal}).
\end{theorem}
\section{General Framework}
\label{sec:genframework}

Next, we consider a simple imperative programming language that includes while loops. The main challenge is how to construct conformal predictors, which only satisfy a coverage guarantee if the calibration dataset consists of i.i.d. examples from the same distribution as the current input. For loop-free programs, all inputs take the same execution path, so we could use all examples to construct each conformal predictor.
However, for imperative programs, this strategy no longer works due to control flow---a given calibration example may not ``reach'' a given machine learning component. Instead, our conformal semantics dynamically constructs conformal predictors when needed based on the calibration examples that reach the current machine learning component.

\begin{figure}
\footnotesize
\begin{subfigure}[c]{\linewidth}
\begin{align*}
k\coloneqq 0;~
v\coloneqq 0;~
\textbf{while }
v\le5
\textbf{ do }
v\coloneqq\widehat{f}(x,k);~
k\coloneqq k+1
\textbf{ od}
\end{align*}
\caption{Program $p$; at the end of execution, $v$ is the value of the first element of $x$ such that $\widehat{f}(x,k)>5$.}
\end{subfigure}
\begin{subfigure}[c]{\linewidth}
\begin{alignat*}{3}
&\sigma_{\text{in}}&&=\{x=[I_1,I_2],k=0,v=0\}
\qquad&&\text{where}\qquad
f^*(I_1)=3,~f^*(I_2)=9 \\
&\sigma_{\text{in},1}&&=\{x=[I_{1,1},I_{1,2}],k=0,v=0\}
\qquad&&\text{where}\qquad
f^*(I_{1,1})=7,~f^*(I_{1,2})=2 \\
&\sigma_{\text{in},2}&&=\{x=[I_{2,1},I_{2,2},I_{2,3}],k=0,v=0\}
\qquad&&\text{where}\qquad
f^*(I_{2,1})=4,~f^*(I_{2,2})=8,~f^*(I_{2,3})=1
\end{alignat*}
\caption{New input $\sigma_{\text{in}}$ and calibration dataset $\{\sigma_{\text{in},1},\sigma_{\text{in},2}\}$.}
\end{subfigure}
\begin{subfigure}[c]{\linewidth}
\vspace{5pt}
\begin{enumerate}
\item $(\sigma_\#^{(0)},\{\sigma_i^{(0)}\}_{i=1}^2)=\den{k\coloneqq0;~v\coloneqq0}^C(\alpha(\sigma_{\text{in}}),\sigma_{\text{in}},\{\sigma_{\text{in},i}\}_{i=1}^2,\{\sigma_{\text{in},i}\}_{i=1}^2)$, where
\begin{align*}
\sigma_\#^{(0)}&=\{x=[I_1,I_2],k=(0,0),v=(0,0)\} \\
\sigma_1^{(0)}&=\{x=[I_{1,1},I_{1,2}],k=0,v=0]\} \\
\sigma_2^{(0)}&=\{x=[I_{2,1},I_{2,2},I_{2,3}],k=0,v=0]\}
\end{align*}
\item $(\sigma_\#^{(1)},\{\sigma_i^{(1)}\}_{i=1}^2)=\den{\textbf{if }v\le5\textbf{ then }v\coloneqq\widehat{f}(x,k);~k\coloneqq k+1\textbf{ fi}}^C(\sigma_\#^{(0)},\sigma_{\text{in}},\{\sigma_1^{(0)},\sigma_2^{(0)}\},\{\sigma_{\text{in},i}\}_{i=1}^2)$, where
\begin{align*}
C_{\epsilon/2}(g,\{(\sigma_{\text{in},1},f^*(\sigma_1^{(0)})),(\sigma_{\text{in},2},f^*(\sigma_2^{(0)}))\})(\sigma_{\text{in}})&=(3,6)
\end{align*}
\begin{align*}
\sigma_\#^{(1)}&=\{x=[I_1,I_2],k=(1,1),v=(3,6)\} \\
\sigma_1^{(1)}&=\{x=[I_{1,1},I_{1,2}],k=1,v=7]\} \\
\sigma_2^{(1)}&=\{x=[I_{2,1},I_{2,2},I_{2,3}],k=1,v=4]\}
\end{align*}
\item $(\sigma_\#^{(2)},\{\sigma_i^{(2)}\}_{i=1}^2)=\den{\textbf{if }v\le5\textbf{ then }v\coloneqq\widehat{f}(x,k);~k\coloneqq k+1\textbf{ fi}}^C(\sigma_\#^{(1)},\sigma_{\text{in}},\{\sigma_1^{(1)},\bot\},\{\sigma_{\text{in},i}\}_{i=1}^2)$, where
\begin{align*}
C_{\epsilon/4}(g,\{(\sigma_{\text{in},1},f^*(\sigma_1^{(1)}))\}(\sigma_{\text{in}})=(8,9)
\end{align*}
\begin{align*}
\sigma_\#^{(2)}&=\{x=[I_1,I_2],k=(2,2),v=(8,9)\} \\
\sigma_1^{(2)}&=\bot \\
\sigma_2^{(1)}&=\{x=[I_{2,1},I_{2,2},I_{2,3}],k=2,v=8]\}
\end{align*}
\item $(\sigma_\#',\{\sigma_i'\}_{i=1}^2)=\den{\textbf{while }v\le5\textbf{ do }v\coloneqq\widehat{f}(x,k);~k\coloneqq k+1\textbf{ od}}^C(\sigma_\#^{(0)},\sigma_{\text{in}},\{\sigma_1^{(0)},\sigma_2^{(0)}\},\{\sigma_{\text{in},i}\}_{i=1}^2)$, where
\begin{align*}
\sigma_\#'&=\{x=[I_1,I_2],k=(2,2),v=(3,9)\} \\
\sigma_1'&=\{x=[I_{1,1},I_{1,2}],k=1,v=7\} \\
\sigma_2'&=\{x=[I_{2,1},I_{2,2},I_{2,3}],k=2,v=8]\}
\end{align*}
\end{enumerate}
\caption{Relevant steps to computing our compositional conformal semantics for the loop.}
\end{subfigure}
\caption{Example of our general framework applied to a program with loops.}
\label{fig:whileloopexample}
\end{figure}

\subsection{Problem Formulation}

\paragraph{Programming language.}

We consider the following standard grammar of imperative programs:
\begin{align}
P&::=\vec{x}\coloneqq f(\vec{x})~(\forall f\in\mathcal{F})\mid P;P\mid\textbf{if }x\textbf{ then }P~(\forall x\in\mathcal{X})\mid\textbf{while }x\textbf{ do }P~(\forall x\in\mathcal{X}),
\label{eqn:implang}
\end{align}
where $\mathcal{X}=\{x_1,...,x_k\}$ is a finite set of variables, $\vec{x}=(x_1,...,x_k)$ enumerates the variables in $\mathcal{X}$, and $\mathcal{F}$ is a set of components $f:\mathbb{R}^k\to\mathbb{R}^k$;
\ten{without loss of generality, components directly map stores to stores, which means that each function can depend on all program variables.}

The ground truth semantics of our language, shown in Figure~\ref{fig:impgtsemantics}, are the standard denotational semantics. Here, $\Sigma=\mathbb{R}^\mathcal{X}=\{\sigma:\mathcal{X}\to\mathbb{R}\}$ is the space of stores; for simplicity, we assume variables are real-valued (Boolean values are implicitly encoded as $\mathbb{B}=\{0,1\}$), but any measurable space of values can be used. As before, we assume $f^*=f$ for $f\not\in\widehat{\mathcal{F}}$. The conditional and while loop semantics are equivalent to the usual semantics---e.g., for conditionals, exactly one of $\iota_{\text{true}}(\sigma,x)\neq\bot$ and $\iota_{\text{false}}(\sigma,x)\neq\bot$ holds depending on the value of $\sigma(x)$, so the join evaluates to the value on the branch taken. This notation simplifies our formalization of our conformal semantics. In the while loop semantics, ``lfp'' standards for ``least fixed point''; it is the usual fixed point semantics. 

As presented, the semantics for conditions and loops always execute both branches for every input; this notation extends more cleanly to conformal semantics. Thus, our semantics actually have type $\den{p}^*:(\Sigma\cup\{\bot\})\to(\Sigma\cup\{\bot\})$. We can handle inputs $\sigma=\bot$ simply by assuming that components return $\bot$ iff they are given $\bot$ as input---i.e., $f(\bot)=\bot$ and $\widehat{f}(\bot)=\bot$. Note that one of the branches is always evaluated using $\sigma=\bot$; in an implementation, this branch does not need to be executed, avoiding the possibility of nontermination.

As a running example, consider the program $p$ in Figure~\ref{fig:whileloopexample} (a); here, $x=[I_1,...,I_n]$ is an input list of images, where each $I_k$ is an image of a digit. Also, $\widehat{f}(x,k)$ predicts the value of the digit in image $I_k$; its ground truth counterpart is $f^*(x,k)=f^*(I_k)$. After executing $p$ with the ground truth semantics, the value of $v$ is $f^*(I_k)$, where $k$ is the index of the first element of $x$ with $f^*(I_k)>5$. 

\begin{figure}[t]
\begin{align*}
\den{\vec{x}\coloneqq f(\vec{x})}^*(\sigma)&=f^*(\sigma)
\qquad
\den{p;q}^*(\sigma)=\den{p}^*(\den{q}^*(\sigma))
\end{align*}
\begin{align*}
\den{\textbf{if }x\textbf{ then }p}^*(\sigma)
=\den{p}^*(\iota_{\text{true}}(\sigma,x))\sqcup(\iota_{\text{false}}(\sigma,x))
\quad\text{where}\quad
\iota_b(\sigma,x)=\begin{cases}
\sigma&\text{if }\sigma(x)=b \\
\bot&\text{otherwise}.
\end{cases}
\end{align*}
\begin{align*}
\den{\textbf{while }x\textbf{ do }p}^*(\sigma)=\text{lfp}(h=\phi_{p,x}^*(h))(\sigma)
~~\text{where}~~
\phi_{p,x}^*(h)(\sigma)=h(\den{p}^*(\iota_{\text{true}}(\sigma,x))\sqcup(\iota_{\text{false}}(\sigma,x))
\end{align*}
\caption{The ground truth semantics for our imperative programming language (\ref{eqn:implang}).}
\label{fig:impgtsemantics}
\end{figure}

\paragraph{Problem formulation.}

Our goal is to define a conformal semantics that contains the ground truth semantics with high probability. We assume given a (measurable) abstract domain $\mathbb{R}^\#$ for the reals (e.g., intervals), and let  $\Sigma^\#=(\mathbb{R}^\#)^{\mathcal{X}}=\{\sigma_\#:\mathcal{X}\to\mathbb{R}^\#\}$ be the space of abstract stores.
For each $f\in\mathcal{F}\setminus\widehat{\mathcal{F}}$, we assume given an abstract transformer $f^\#$ satisfying
$\sigma\in\gamma(\sigma_\#)\Rightarrow f(\sigma)\in\gamma(f^\#(\sigma_\#))$
 for all $\sigma\in\Sigma,\sigma_\#\in\Sigma^\#$. We assume all $f$, $f^*$, and $f^\#$ are measurable.
Given a program $p$ and a confidence level $\epsilon\in\mathbb{R}_{\ge0}$, our conformal semantics $\den{p}^C_{\epsilon}:\Sigma^\#\times\Sigma^n\to\Sigma^n$ are a function of the current input that is an abstract store $\sigma_\#\in\Sigma^\#$, and a calibration set of concrete stores $Z=\{\sigma_i\}_{i=1}^n\in\Sigma^n$, where $\{\sigma_i\}_{i=1}^n$ is interpreted as a length $n$ tuple. We have omitted labels from $Z$; we simply assume the ground truth semantics can be evaluated for examples $\sigma_i\in Z$.

Next, we formalize our desired coverage guarantee. Because this guarantee refers two different evaluations of the same program (i.e., the ground truth semantics and our conformal semantics), it is a 2-safety property~\citep{clarkson2010hyperproperties}; thus, we need to augment our conformal semantics with the ground truth semantics to define it. In particular, we define the \emph{joint semantics} to be $\den{p}_{\epsilon}^J=\den{p}_{\epsilon}^C\times\den{p}^*$---i.e.,
\begin{align*}
\den{p}_{\epsilon}^J(\sigma_\#,\{\sigma_i\}_{i=1}^n,\sigma_*)=(\den{p}^C_{\epsilon}(\sigma_\#,\{\sigma_i\}_{i=1}^n),\den{p}^*(\sigma_*))
\qquad(\forall\sigma_\#\in\Sigma^\#,\{\sigma_i\}_{i=1}^n\in\Sigma^n,\sigma_*\in\Sigma),
\end{align*}
where $\sigma_*$ is the true input. Intuitively, our coverage property says that if $\sigma_*\in\gamma(\sigma_\#)$, then we have $\den{p}^*(\sigma_*)\in\gamma(\den{p}^C_{\epsilon}(\sigma_\#))$ with probability at least $1-\epsilon$, where we have defined $\gamma(\den{p}^C_{\epsilon}(\sigma_\#))=\gamma(\sigma_\#')$ assuming $\den{p}^C_{\epsilon}(\sigma_\#)=(\sigma_\#',\{\sigma_i'\}_{i=1}^n)$. In this case, we assume the distribution over inputs to our joint semantics is given as a probability measure $\lambda\in\mathcal{M}(\Sigma^\#\times\Sigma^{n+1})$, where $\mathcal{M}\mathcal{X}$ is the space of probability measures over a measurable space $\mathcal{X}$. We need to ensure that the calibration set $Z=\{\sigma_i\}_{i=1}^n$ consists of i.i.d. examples from the same distribution as the true input $\sigma_*$. Formally, consider the projection $\pi:\Sigma^\#\times\Sigma^{n+1}\to\Sigma^{n+1}$that removes the abstract store $\sigma_\#$, i.e.,
\begin{align*}
\pi(\sigma_\#,\{\sigma_i\}_{i=1}^n,\sigma_*)=(\{\sigma_i\}_{i=1}^n,\sigma_*).
\end{align*}
Then, we assume that $\lambda$ satisfies
\begin{align}
\label{eqn:independencecond}
\pi_*(\lambda)=\otimes_{i=1}^{n+1}\mu
\end{align}
for some $\mu\in\mathcal{M}\Sigma$, where $g_*:\mathcal{M}\mathcal{X}\to\mathcal{M}\mathcal{Y}$ denotes the pushforward of $g:\mathcal{X}\to\mathcal{Y}$, and $\mu\otimes\nu\in\mathcal{M}(\mathcal{X}\times\mathcal{Y})$ denotes the product measure. Product measures encode probability distributions where the different components are independent, so (\ref{eqn:independencecond}) encodes that $(\{\sigma_i\}_{i=1}^n,\sigma_*)$ consists of i.i.d. random variables $\sigma_1,...,\sigma_n,\sigma_*$ with distribution $\mu$. Finally, given $\lambda\in\mathcal{M}(\Sigma^\#\times\Sigma^{n+1})$ satisfying $\pi_*(\lambda)=\otimes_{i=1}^{n+1}\mu$ for some $\mu\in\mathcal{M}\Sigma$, our desired coverage guarantee is
\begin{align}
\label{eqn:whilecoverage}
\mathbb{P}_{\lambda(\sigma_\#,\{\sigma_i\}_{i=1}^n,\sigma_*)}[\den{p}^*(\sigma_*)\in\gamma(\den{p}^C_{\epsilon}(\sigma_\#,\{\sigma_i\}_{i=1}^n))\mid\sigma_*\in\gamma(\sigma_\#)]\ge1-\epsilon,
\end{align}
i.e., assuming $\sigma_*\in\gamma(\sigma_\#)$, then our conformal semantics covers the ground truth semantics.

Continuing our running example, in Figure~\ref{fig:whileloopexample} (b), we show a new input $\sigma_{\text{in}}$ (for which $f^*$ is not known) and calibration dataset $\{\sigma_{\text{in},1},\sigma_{\text{in},2}\}$, where the input images have the given values. Then, our goal is for the following to hold with high probability over both $\sigma_{\text{in}}$ and $\{\sigma_{\text{in},i}\}_{i=1}^n$:
\begin{align*}
\den{p}^*(\sigma_{\text{in}},\{\sigma_{\text{in},i}\}_{i=1}^2)\in\gamma(\den{p}^C(\alpha(\sigma_{\text{in}}),\{\sigma_{\text{in},i}\}_{i=1}^2)).
\end{align*}

\subsection{Conformal Semantics}

\begin{figure}[t]
\begin{align*}
\den{\vec{x}\coloneqq f(\vec{x})}_{\epsilon}^C(\sigma_\#,\{\sigma_i\}_{i=1}^n)=(f^\#(\sigma_\#),\{f(\sigma_i)\}_{i=1}^n)
\quad
\den{\vec{x}\coloneqq\widehat{f}(\vec{x})}_{\epsilon}^C(\sigma_\#,\{\sigma_i\}_{i=1}^n)=(\widetilde{p}(\bot),\{f^*(\sigma_i)\}_{i=1}^n)
\end{align*}
\begin{align*}
\text{where}\qquad\widetilde{p}=C_{\epsilon}(g_0,\{(\bot,f^*(\sigma_i))\mid\sigma_i\neq\bot\}_{i=1}^n)
\qquad\text{and}\qquad
g_0(\bot,\sigma)=0
\end{align*}
\begin{align*}
\den{p;q}_{\epsilon}^C(\sigma_\#,\{\sigma_i\}_{i=1}^n)=\den{p}_{\epsilon/2}^C(\den{q}_{\epsilon/2}^C(\sigma_\#,\{\sigma_i\}_{i=1}^n))
\end{align*}
\begin{align*}
\den{\textbf{if }x\textbf{ then }p}_{\epsilon}^C(\sigma_\#,\{\sigma_i\}_{i=1}^n)
=\den{p}_{\epsilon}^C(\iota_{\text{true}}^\#(\sigma_\#,x),\{\iota_{\text{true}}(\sigma_i,x)\}_{i=1}^n)\sqcup(\iota_{\text{false}}^\#(\sigma_\#,x),\{\iota_{\text{false}}(\sigma_i,x)\}_{i=1}^n)
\end{align*}
\begin{align*}
\text{where}\qquad\iota_b^\#(\sigma_\#,x)&=\begin{cases}
\sigma_\#&\text{if }b\in\gamma(\sigma_\#) \\
\bot&\text{otherwise}
\end{cases}
\qquad\text{and}\qquad
\iota_b(\sigma,x)=\begin{cases}
\sigma&\text{if }\sigma(x)=b \\
\bot&\text{otherwise}.
\end{cases}
\end{align*}
\begin{align*}
\den{\textbf{while }x\textbf{ do }p}_{\epsilon}^C(\sigma_\#,\{\sigma_i\}_{i=1}^n)=\text{lfp}(h=\phi_{p,x}^C(h))(\sigma_\#,\{\sigma_i\}_{i=1}^n,\epsilon),
\end{align*}
\begin{align*}
&\text{where} \\
&\phi_{p,x}^C(h)(\sigma_\#,\{\sigma_i\}_{i=1}^n,\epsilon)=h(\den{p}_{\epsilon/2}^C(\iota_{\text{true}}(\sigma_\#,x),\{\iota_{\text{true}}(\sigma_i,x)\}_{i=1}^n),\epsilon/2)\sqcup(\iota_{\text{false}}(\sigma_\#,x),\{\iota_{\text{false}}(\sigma_i,x)\}_{i=1}^n).
\end{align*}
\caption{The conformal semantics for our imperative programming language (\ref{eqn:implang}).}
\label{fig:impconfsemantics}
\end{figure}

Our conformal semantics $\den{p}^C_{\epsilon}$ are shown in Figure~\ref{fig:impconfsemantics}. Note that they are deterministic, unlike the case of probabilistic programs~\cite{kozen1979semantics}---in our language, randomness only occurs at the input, so we can use the pushforward of our semantics to convert an input distribution to its output distribution.

\paragraph{Challenges.}

Before we describe our conformal semantics, we first provide some intuition on the main challenge they face. Returning to our running example, the key steps to computing the while loop semantics are shown in Figure~\ref{fig:whileloopexample} (c); in particular, we want to evaluate the semantics of the example program $p$ on input $\sigma_{\text{in}}$ given calibration dataset $\{\sigma_{\text{in},1},\sigma_{\text{in},2}\}$. First, step (1) evaluates the first two statements $k\coloneqq0$ and $v\coloneqq0$, which simply ensures that $k=v=0$. Also, note that we switch to the abstract store $\alpha(\sigma_{\text{in}})$, where $k=(0,0)$ and $v=(0,0)$ (so $\gamma(k)=\{0\}$ and $\gamma(v)=\{0\}$).

Now, we need to evaluate the while loop on the resulting abstract store $\sigma_\#^{(0)}$ and the (concrete) calibration stores $\{\sigma_1^{(0)},\sigma_2^{(0)}\}$. To do so, we essentially unroll the loop as a series of conditionals. Steps 2 and 3 show the unrolling, and step 4 shows the whole loop.

Compared to Section~\ref{sec:framework}, the main challenge is that the calibration examples $\{\sigma_1^{(0)},\sigma_2^{(0)}\}$ may take different control flow paths; further complicating matters is that the control flow path taken by $\sigma_\#^{(0)}$ depends on the conformal predictors encountered, which in turn depend on the control flow paths taken by the calibration examples $\{\sigma_1^{(0)},\sigma_2^{(0)}\}$.

Intuitively, our conformal semantics addresses these challenges by always executing both branches of any conditional. For a calibration example $\sigma_i$, we are using the ground truth semantics, so one of the branches is always executed with value $\sigma_i=\bot$. Then, when constructing conformal predictors, we simply filter out calibration examples such that $\sigma_i=\bot$, and construct the prediction set using the remaining examples. It is possible to show that using this strategy, the remaining examples are i.i.d. samples from the same distribution as the (unknown) store $\sigma$ obtained by evaluating the ground truth semantics on $\sigma_{\text{in}}$; thus, $\sigma\in\gamma(\sigma_\#)$ holds with high probability.

However, for an abstract store, we may be uncertain about which branch is taken. In our running example, if $v$ has abstract value $(3,6)$ (so $\gamma(v)=\{3,4,5,6\}$), then we do not know if the while loop terminates or continues execution. In this case, we actually need to execute both possibilities. In particular, we first obtain the value resulting from continuing execution of the loop; in our example, we get that $v$ has abstract value $(8,9)$. Then, we join that result with the result from terminating immediately, to get $(3,6)\sqcup(8,9)=(3,9)$. This strategy overapproximates the loop semantics.

Below, we formalize these intuitions in our conformal semantics, and then return to our running example to show how the computation works in more detail.

\paragraph{Conformal semantics.}

We describe our conformal semantics. Assignment for a standard component $f\in\mathcal{F}\setminus\widehat{\mathcal{F}}$ uses the abstract transformer $f^\#$ on the abstract store and the standard component on the calibration stores. We have omitted the direct conformal semantics for simplicity; it works as before, by using meet to intersect the direct and compositional conformal semantics:
\begin{align*}
\den{\vec{x}\coloneqq f(\vec{x})}_{\epsilon}^C(\sigma_\#,\{\sigma_i\}_{i=1}^n)=f^\#(\sigma_\#)\sqcap \widetilde{p}(\bot),
\end{align*}
where $\widetilde{p}$ is defined as in the semantics for $\widehat{f}$ (discussed next). As before, we use meet since both $f^\#(\sigma_\#)$ and $\widetilde{p}(\bot)$ are separately guaranteed to contain $f(\sigma_*)$ with high probability.

Assignment for a machine learning component $\widehat{f}\in\widehat{\mathcal{F}}$ constructs a conformal predictor $\widetilde{p}$ using the scoring function $g_0:\{\bot\}\times\Sigma^n\to\mathbb{R}$ defined by $g_0(\bot,f^*(\sigma_i))=0$. Unlike before, we have provided the scoring function with a trivial input $\bot$. This choice is merely to simplify notation, and our approach can easily be extended to the kinds of scoring functions discussed in Section~\ref{sec:framework}. In particular, we can allow dependence on an input store $\sigma_{\text{in}}\in\Sigma_{\text{in}}$ (e.g., representing input images). To do so, we would assume given an input store $\sigma_{\text{in}}$ for the current input, an input store $\sigma_{\text{in},i}$ for each calibration example $i\in[n]$, and a scoring function $g:\Sigma_{\text{in}}\times\Sigma\to\mathbb{R}$. Then, we would modify our conformal semantics to be $\den{p}^C_{\epsilon}:\Sigma^\#\times\Sigma_{\text{in}}\times\Sigma^{n+1}\times\Sigma_{\text{in}}^{n+1}\to\Sigma^\#\times\Sigma^{n+1}$, where
\begin{align}
&\den{\vec{x}\coloneqq\widehat{f}(\vec{x})}_{\epsilon}^C(\sigma_\#,\sigma_{\text{in}},\{\sigma_i\}_{i=1}^n,\{\sigma_{\text{in},i}\}_{i=1}^n) \nonumber \\
&=(C_{\epsilon}(g,\{(\sigma_{\text{in},i},f^*(\sigma_i))\mid\sigma_i\neq\bot\}_{i=1}^n)(\sigma_{\text{in}}),\{f^*(\sigma_i)\}_{i=1}^n,\{\sigma_{\text{in},i}\}_{i=1}^n).
\label{eqn:altconformalsemantics}
\end{align}
Here, we replaced inputs $\bot$ to the scoring function with $\sigma_{\text{in},i}$ (for the calibration examples) and $\sigma_{\text{in}}$ (for the current input). Our guarantees hold as before, with extra bookkeeping to track input stores. Note that the conformal semantics in Figure~\ref{fig:whileloopexample} (c) use the form (\ref{eqn:altconformalsemantics}). In particular, they carry around the input stores $\sigma_{\text{in}}$ and $\{\sigma_{\text{in},i}\}_{i=1}^2$ throughout the computation.

In addition, when constructing our conformal predictor, we filter out calibration stores $\sigma_i=\bot$ for $i\in[n]$; these are stores that do not take the same control flow path in conditionals or while loops as $\sigma_*$ (as discussed below). This strategy ensures $C_{\epsilon}$ is called using i.i.d. samples from the distribution of $\sigma_*$ conditioned on $\sigma_*\neq\bot$, which is needed for our coverage guarantee to hold.

Next, our semantics for sequences $p;q$ is standard; it simply executes $p$ followed by $q$. One minor point is that they distribute the confidence level $\epsilon$ evenly over the two sub-programs (i.e., $\epsilon/2$ each); in practice, any division works as long as it sums to $\epsilon$. For instance, if we can statically check that one branch never executes a conformal predictor, we could assign it $\epsilon=0$.

For conditional statements, as indicated in the ground truth semantics, we use a slightly different notation, but it is equivalent to the standard semantics for conditional statements applied independently to the abstract store $\sigma_\#$ and each calibration example $\sigma_i$ (for $i\in[n]$). While the calibration examples are only executed along one branch, the abstract semantics may execute $\sigma_\#$ along both branches if $\gamma(\sigma_\#)=\{\text{true},\text{false}\}$, and then join together the results. This strategy soundly overapproximates the concrete semantics of conditional statements. As with the ground truth semantics, conditionals can evaluate one of the branches using stores $\sigma_\#=\bot$ or $\sigma_i=\bot$ (for some $i\in[n]$). In this case, multiple computations are being run in parallel, so a subset of inputs may equal $\bot$. Defining $f(\bot)=\bot$ for all $f\in\mathcal{F}$ naturally accounts for different components separately.

For while loops, our presentation differs slightly from the standard denotational semantics. The reason is that we are effectively evaluating the semantics on multiple stores in parallel (where the execution of $\sigma_\#$ depends on the execution of $\{\sigma_i\}_{i=1}^n$, but not vice versa), and different stores may require different numbers of loop iterations. Our use of the join operator naturally combines values across all possible loop iteration counts, since $z\sqcup\bot=\bot\sqcup z=z$ for any value $z$. The least fixed point (lfp) is slightly different than usual; we provide a derivation in Appendix~\ref{sec:whileloopsemantics}.

Our conformal semantics satisfies the desired coverage guarantee; see Appendix~\ref{sec:thm:impmain:proof} for a proof:
\begin{theorem}
\label{thm:impmain}
Given a program $p$ guaranteed to terminate (see Appendix~\ref{sec:whileloopsemantics}), $\epsilon\in\mathbb{R}_{\ge0}$, and $\lambda\in\mathcal{M}(\Sigma^\#\times\Sigma^{n+1})$ satisfying $\pi_*(\lambda)=\otimes_{i=1}^{n+1}\mu$ for some $\mu\in\mathcal{M}\Sigma$, then $\den{p}^C_{\epsilon}$ satisfies (\ref{eqn:whilecoverage}).
\end{theorem}

\paragraph{Example.}

Returning to our running example in Figure~\ref{fig:whileloopexample}, the while loop semantics are evaluated by unrolling the loop as a series of conditionals. For the example inputs, the while loop is executed at most twice, so it suffices to unroll it twice:
\begin{align}
\label{eqn:unrolling}
\textbf{if }v\le5\textbf{ then }v\coloneqq\widehat{f}(x,k);~k\coloneqq k+1;~\textbf{if }v\le5\textbf{ then }v\coloneqq\widehat{f}(x,k);~k\coloneqq k+1\textbf{ fi fi}
\end{align}
Step 2 shows the first execution of the loop body; since $v\le5$, the input $\sigma_\#^{(0)}$ and both calibration examples $\sigma_1^{(0)}$ and $\sigma_2^{(0)}$ take this branch. As a consequence, the conformal predictor in this loop body is constructed with both calibration examples. Our example assumes that the output of the conformal predictor on this step is $(3,6)$; thus, the value of $v$ in $\sigma_\#^{(1)}$ is $(3,6)$.

Next, note that for the second calibration example $\sigma_2^{(0)}$, the while loop terminates after one iteration since $v>5$. Thus, the inner conditional in (\ref{eqn:unrolling}) is executed with $\sigma_2^{(1)}=\bot$. In contrast, the while loop does not terminate for $\sigma_1^{(0)}$. As a consequence, the conformal predictor in this loop body is constructed with just the calibration example $\sigma_1^{(1)}$. Our example assumes that the output of the conformal predictor on this step is $(8,9)$; thus, the value of $v$ in $\sigma_\#^{(2)}$ is $(8,9)$.

Finally, note that it is uncertain whether the input $\sigma_\#^{(0)}$ takes one or two loop iterations, since the $(3,6)\widehat{\le}5=\{\text{true},\text{false}\}$ can be either true or false. If we execute the while loop just once, then the abstract value of $v$ is $(3,6)$; if we execute it twice, then it is $(8,9)$. Thus, the final output of $v$ is obtained by joining these two values to get $(3,6)\sqcup(8,9)=(3,9)$. Because $(8,9)\widehat{\le}5=\{\text{false}\}$, the loop is guaranteed to terminate after two iterations, so we are done.

\section{Experiments}
\label{sec:experiments}


\ten{We describe the basic experimental setup in Section~\ref{sec:setup}. Then, we describe our evaluation on programs processing lists of MNIST images in Section~\ref{sec:mnist}. These results demonstrate that our compositional semantics achieves the desired coverage; when the entire program can be calibrated at once, the full conformal semantics leverages both the direct and compositional semantics to achieve on average the smallest prediction set sizes. In addition, we show that in most settings, using interval semantics only introduces minimal size increase compared to the set semantics (which are exact), demonstrating that approximation is not an issue. We also report the dependence of our approach on various problem parameters. Finally, we report results on programs processing lists of detected objects in MS-COCO images in Section~\ref{sec:obj-detect}, demonstrating that our approach extends to realistic problem settings.}

\subsection{Experimental Setup}
\label{sec:setup}

\paragraph{Programs.}

We evaluate the efficacy of our approach on example programs in two domains. First, as a simple example, we consider queries that use image detection models as components to label images of handwritten digits. We make use of image data from the MNIST dataset, synthetically altering it to introduce variation in model uncertainty across images. Second, as a more realistic example, we consider queries on images from the MS-COCO dataset, using an object detector as a subroutine to obtain information (label and position) about relevant objects in the image.

\paragraph{Datasets.}

For each program $p = f(p_1, \cdots, p_k)$ in our benchmarks, we construct a dataset of inputs $Z = \{x_i\}_{i=1}^m$ such that $x_i \sim D$ are i.i.d. samples. We divide $Z$ into a calibration dataset $Z_{\text{cal}}$ and a test dataset $Z_{\text{test}}$. We use $Z_{\text{cal}}$ to construct a conformal predictor with coverage guarantees for each machine learning component used in $p$, and evaluate our metrics on $Z_{\text{test}}$.

\paragraph{Baselines.}

We compare our full conformal semantics $\den{p}_{\epsilon}^C$ to
both the direct conformal semantics $\den{p}_{\epsilon}^{C0}$ and the compositional conformal semantics $\den{p}_{\epsilon}^{C1}$. For the latter, we compare the performance of two different abstract semantics for propagating prediction sets. 

\paragraph{Metrics.}

We fix an overall miscoverage rate $\epsilon$. Then, for each approach, we compute the coverage and average prediction set size on $Z_{\text{test}}$. Coverage is defined as
\begin{align*}
\text{coverage}=\frac{1}{|Z_{\text{test}}|}\sum_{x\in Z_{\text{test}}}\mathbbm{1}(\den{p}^*(x)\in\gamma(\den{p}_{\epsilon}^C(x,Z)),
\end{align*}
and the average prediction set size is
\begin{align*}
\text{size}=\frac{1}{|Z_{\text{test}}|}\sum_{x\in Z_{\text{test}}}|\gamma(\den{p}_{\epsilon}^C(x,Z))|,
\end{align*}
where $|\gamma((\ell,u))|=u-\ell+1$. For the ablations, $\den{\cdot}^C$ is replaced with $\den{\cdot}_{\epsilon}^{C0}$ or $\den{\cdot}_{\epsilon}^{C1}$. Each approach is theoretically guaranteed to satisfy the coverage guarantee, but may result in different prediction set sizes. Thus, our primary goal is to minimize the average prediction set size; we report coverage only to validate that our theoretical guarantees hold.

For average prediction set sizes, we additionally report standard deviations over the examples in our test set $x\in Z_{\text{test}}$. For coverage rates, we cannot do so since they are a function of the entire test set; instead, we perform our experiments using 25 random calibration sets $Z_{\text{cal}}$ (obtained by using different calibration-test set splits), and report mean and standard deviation over these 25 random runs. Then, for average prediction set sizes, we report the standard deviation over both examples in our test set and the 25 random runs (except in our plots, where we use a single run). This strategy is in line with prior work on conformal prediction~\cite{park2019pac}.

When comparing different abstract semantics, we report prediction set size as well as run-time. 

\paragraph{Conformal Prediction Technique.} In all of our experiments, we use PAC prediction set algorithms ~\cite{vovk2012conditional,park2019pac} to produce prediction sets. This is a variant on the classical conformal prediction algorithm which gives stronger guarantees. See Appendix~\ref{sec:pac} for details.


\subsection{List Processing Programs Over MNIST Images}
\label{sec:mnist}

First, we apply our framework to obtain prediction sets around the output of list processing programs, where the input is a list of handwritten digits. Each program applies a machine learning component to label each image, and then performs list operations over the resulting integers.

\paragraph{Dataset.}

The MNIST dataset contains grayscale images of individual handwritten digits. The training set contains 60000 images, and the test set contains 10000 images. We train $\widehat{f}$ on the training data. To construct the conformal model $\widetilde{f}$ for $\widehat{f}$, we require a calibration set $Z$ of images, and to train the models using direct conformal semantics for any program $p$ that takes a list of images as input, we require a calibration set $Z^{\text{list}}$ of lists of images. Therefore, we first split the original test set into a calibration set $Z_{\text{cal}}$ of size 2000 and a test set $Z_\text{test}$ of size 8000. Before using this data, we artificially inject $\approx80\%$ of the data-samples (in both $Z_{\text{cal}}$ and $Z_{\text{test}}$) with salt-and-pepper noise to increase the uncertainty of $\widehat{f}$. We use the altered $Z_{\text{cal}}$ directly to train the conformal model $\widetilde{f}$. Then, we randomly sample elements from $Z_\text{cal}$ to generate image lists of length between four and ten, each of which is added as a calibration point for $Z_\text{cal}^\text{list}$; we use $|Z_\text{cal}^\text{list}| = 2000$. Similarly, we sample uniformly from $Z_{\text{test}}$ to generate list of images to create $Z_\text{test}^\text{list}$; we use $|Z_\text{test}^\text{list}| = 5000$.

\begin{table}[t]
\centering
\tiny
\begin{tabular}{p{40mm}rrrrrr}
\toprule
\multicolumn{1}{c}{\multirow{2}{*}{\textbf{Program description}}}  & \multicolumn{3}{c}{\textbf{Average Prediction Set Size}} & \multicolumn{3}{c}{\textbf{Coverage}} \\
& \multicolumn{1}{c}{Direct}
& \multicolumn{1}{c}{Compositional}
& \multicolumn{1}{c}{Full}
& \multicolumn{1}{c}{Direct}
& \multicolumn{1}{c}{Compositional}
& \multicolumn{1}{c}{Full} \\
\midrule
sum of list elements & \textbf{16.52 $\pm$ 0.85} & 31.44 $\pm$ 2.45 & 16.55 $\pm$ 0.94 & 0.93 $\pm$  0.010 & 0.99 $\pm$ 0.000 &  0.94 $\pm$ 0.010 \\
sum of list elements less than 7 & 13.0 $\pm$ 0.00 & 34.46 $\pm$ 1.85 & \textbf{12.49 $\pm$ 1.22} & 0.94 $\pm$ 0.003 & 1.00 $\pm$ 0.000 & 0.94 $\pm$ 0.003 \\
max of list elements & 7.88 $\pm$ 0.99 & 8.06 $\pm$ 1.04 & \textbf{6.16 $\pm$ 1.41} & 0.94 $\pm$ 0.021 & 1.00 $\pm$ 0.000 & 0.95 $\pm$ 0.020 \\
\# of list elements less than 6 & 3.32 $\pm$ 0.73 & 4.90 $\pm$ 0.31 & \textbf{3.11 $\pm$ 0.74} & 0.94 $\pm$ 0.023 & 1.00 $\pm$ 0.000 & 0.95 $\pm$ 0.026 \\
\# of list elements equal to 2 & 3.00 $\pm$ 0.00 & 4.68 $\pm$ 0.55 & \textbf{2.36 $\pm$ 0.48} & 0.99 $\pm$ 0.001 & 1.00 $\pm$ 0.000 & 0.99 $\pm$ 0.001 \\
\# of list elements between 3 and 8 & 5.00 $\pm$ 0.00 & 4.98 $\pm$ 0.14 & \textbf{4.05 $\pm$ 0.71} & 0.98 $\pm$ 0.001 & 1.00 $\pm$ 0.000 & 0.99 $\pm$ 0.001 \\
max sum of any two list elements & 15.72 $\pm$ 1.87 & 15.12 $\pm$ 2.07 & \textbf{11.23 $\pm$ 2.81} & 0.95 $\pm$ 0.020 & 1.00 $\pm$ 0.000 & 0.95 $\pm$ 0.020 \\
max difference between two list elements & 9.00 $\pm$ 0.00 & 9.91 $\pm$ 0.29 & \textbf{7.60 $\pm$ 1.26}
&  0.95 $\pm$ 0.003 & 1.00 $\pm$ 0.000 &  0.95 $\pm$ 0.003 \\
sum of first $k$ list elements & 50.32 $\pm$ 1.77 & 78.33 $\pm$ 0.63 & \textbf{42.18 $\pm$ 1.11} & 0.93 $\pm$ 0.008 & 1.00 $\pm$ 0.000 & 0.94 $\pm$ 0.008 \\
sum list elements until one is $>5$ & 81.80 $\pm$ 4.00 & 99.15 $\pm$ 0.42 & \textbf{49.20 $\pm$ 2.40} & 0.99 $\pm$ 0.000 & 1.00 $\pm$ 0.000 & 0.99 $\pm$ 0.000 \\
\midrule
$\text{set size}/\text{our set size}$ & 1.23$\times$ & 1.52$\times$ & \textbf{1.00$\times$} & \multicolumn{1}{c}{--} & \multicolumn{1}{c}{--} & \multicolumn{1}{c}{--} \\
\bottomrule
\end{tabular}
\caption{Prediction set sizes and coverages for MNIST. The desired coverage rate is $1 - \epsilon = 0.9$. Coverage is rounded to 1.00 when it is $>$ 0.999, and standard deviation is rounded to 0.000 when it is $<10^{-3}$. In the second-to-last program, $k$ is a program input. The last row shows the average prediction set size for each of the three approaches divided by the average prediction set size of the full approach, averaged over programs.}
\label{table:MNIST}
\end{table}

\paragraph{Conformal predictor.}

We use a model for classifying individual MNIST digits~\cite{lecun1998gradient}, which has type $\widehat{f}: \textsf{image}\to\textsf{int}$. It takes an image as input and outputs a digit value between 0 and 9. This model is defined using a scoring function $g: \textsf{image}~\times~\textsf{int} \to \mathbb{R}$:
\begin{equation*}
\widehat{f}(x) = \argmax_{y \in \cY} g(x, y),
\end{equation*}
with $g(x, y)$ being a measure of the model's confidence that $x$ is an image of the digit $y$. 
The corresponding conformal model $\widetilde{f}: \textsf{image}\to\textsf{int}^{\#}$ outputs prediction sets over integer values---i.e., for each input $x$, $\widetilde{f}_{\widehat{\tau}(Z,\epsilon)}(x) = y_{\tau, x}^{\#}$, where $y_{\tau, x}^{\#}$ is an element of $\textsf{int}^{\#}$ such that
\begin{equation*}
\gamma(y_{\tau, x}^{\#} ) \supseteq \{y \in \mathbb{N} \mid g(x, y) \geq \tau\}.
\end{equation*}
Since our DSL represents prediction sets over integers as closed continuous intervals, this means $y_{\tau, x}^{\#}$ will be the smallest such interval enclosing the prediction set of integers.


\paragraph{Programs.}

Our programs process the integers represented by the input images in different ways to obtain an integer output---as an example, the program
\begin{equation*}
p(x)=(\textsf{foldr}~+~(\textsf{map}~\widehat{f}~x)~0),
\end{equation*}
takes as input a list $x$ of images, and outputs the sum of the numbers represented by those images. \ten{Importantly, the last two programs in our set include control flow involving uncertain values (e.g., sum the first $k$ elements of $x$, where $k$ is also an image whose value must be predicted). Therefore, these programs require the general framework described in Section~\ref{sec:genframework}.} We test a set of 10 programs and for each one, set the desired overall miscoverage $\epsilon = 0.1$.

\paragraph{Comparing different conformal semantics.}

Results comparing direct conformal semantics, compositional conformal semantics, and their combination are in Table~\ref{table:MNIST}. In Figure~\ref{fig:MNIST-Set-Size} in Appendix~\ref{sec:additionalresults}, we show box plots illustrating the distribution of prediction set size over different examples in our test set. Note that for the direct approach, all prediction sets have the same size. These plots show that in addition to the full approach having the smallest average prediction set sizes, its prediction sets are almost always smaller than those of the direct approach, and even the higher quantiles of prediction set sizes are consistently lower than those for the compositional approach. We also show box plots illustrating the distribution of coverage over 25 random trials in Figure~\ref{fig:MNIST-coverage} in Appendix~\ref{sec:additionalresults}. All three approaches consistently achieve coverage higher than the desired rate of 0.9.


\ten{The size of a prediction set generated using conformal prediction is strongly dependent on the accuracy of the machine learning model as well as on the desired coverage rate. Though most of our results correspond to noise $\eta = 0.2$ (recall that adding noise makes the model less accurate) and error bound $\epsilon = 0.1$, we also consider varying these parameters. In particular, Tables~\ref{table:MNIST-noise} \&~\ref{table:MNIST-epsilon} show results (averaged across all programs) for varying noise $\eta$ and error bound $\epsilon$, respectively.}



\paragraph{Comparing different abstract semantics.}

For MNIST, the prediction sets produced by $\widetilde{f}$ are always finite subsets of the set of possible digits $\{0, 1, \cdots, 9\}$. There are different ways to define abstract semantics for such prediction sets. One is set semantics (i.e., enumerate through all possible values); an alternative is interval semantics. While the former may produce smaller prediction sets on average, it may increase running time, especially as the prediction set size and the program complexity increase. We compare the performance of these two abstract semantics. Results are shown in Table~\ref{table:MNIST-2}. As expected, the average prediction set size is smaller for set abstractions, and the average runtime is smaller for interval abstractions. Notably, however, the increase in set size when using intervals is minimal (with intervals performing equally well as sets for some programs) while the runtime speed-up is significant, generally an order of magnitude.  Thus, using the interval semantics is an effective strategy for balancing running time and prediction set size.

\begin{figure}[t]
\centering
\begin{subfigure}[b]{0.4\textwidth}
\centering
\includegraphics[width=\textwidth]{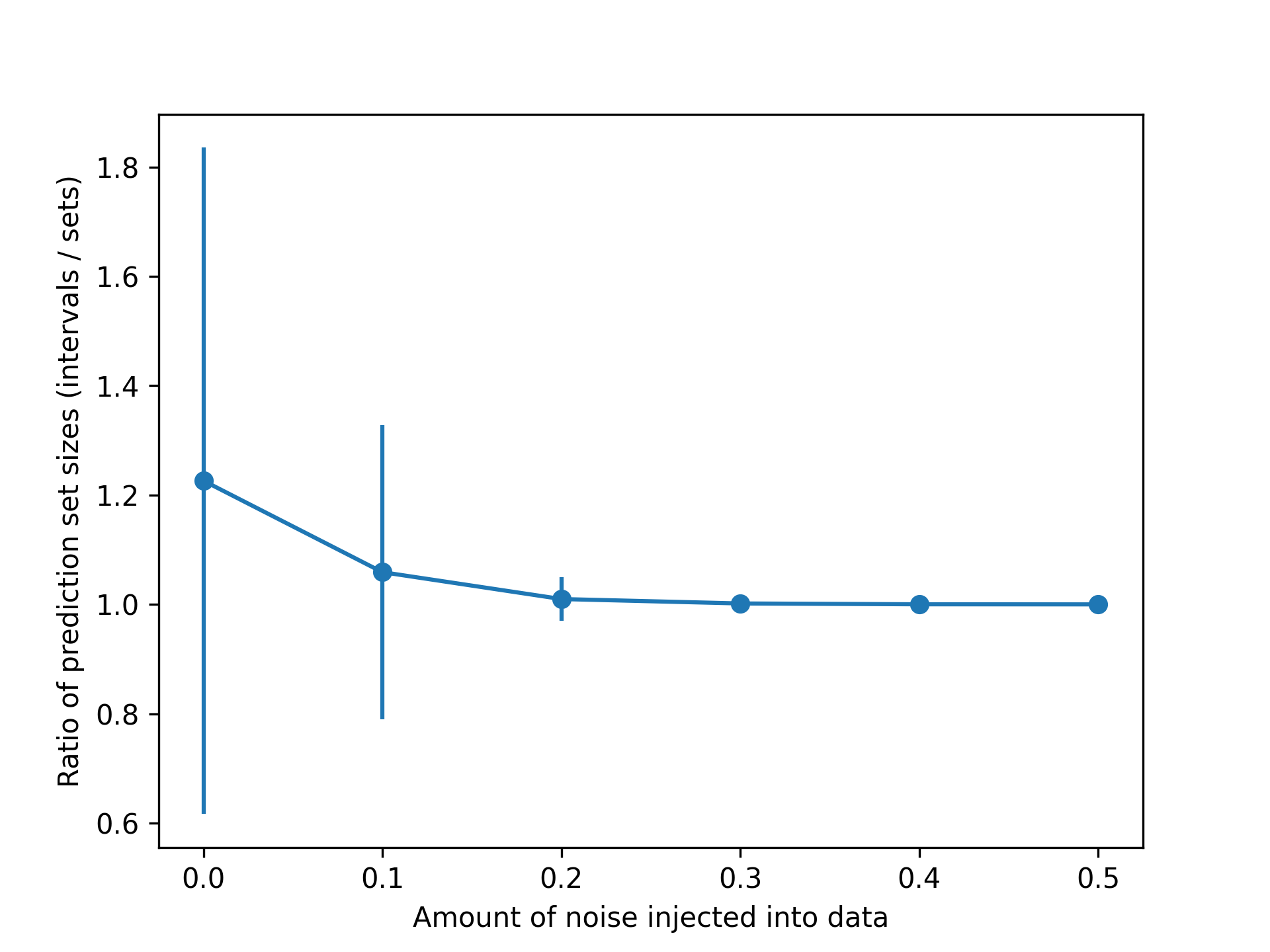}
\caption{Prediction set size vs. noise}
\label{fig:prog-1}
\end{subfigure}
\qquad
\begin{subfigure}[b]{0.4\textwidth}
\centering
\includegraphics[width=\textwidth]{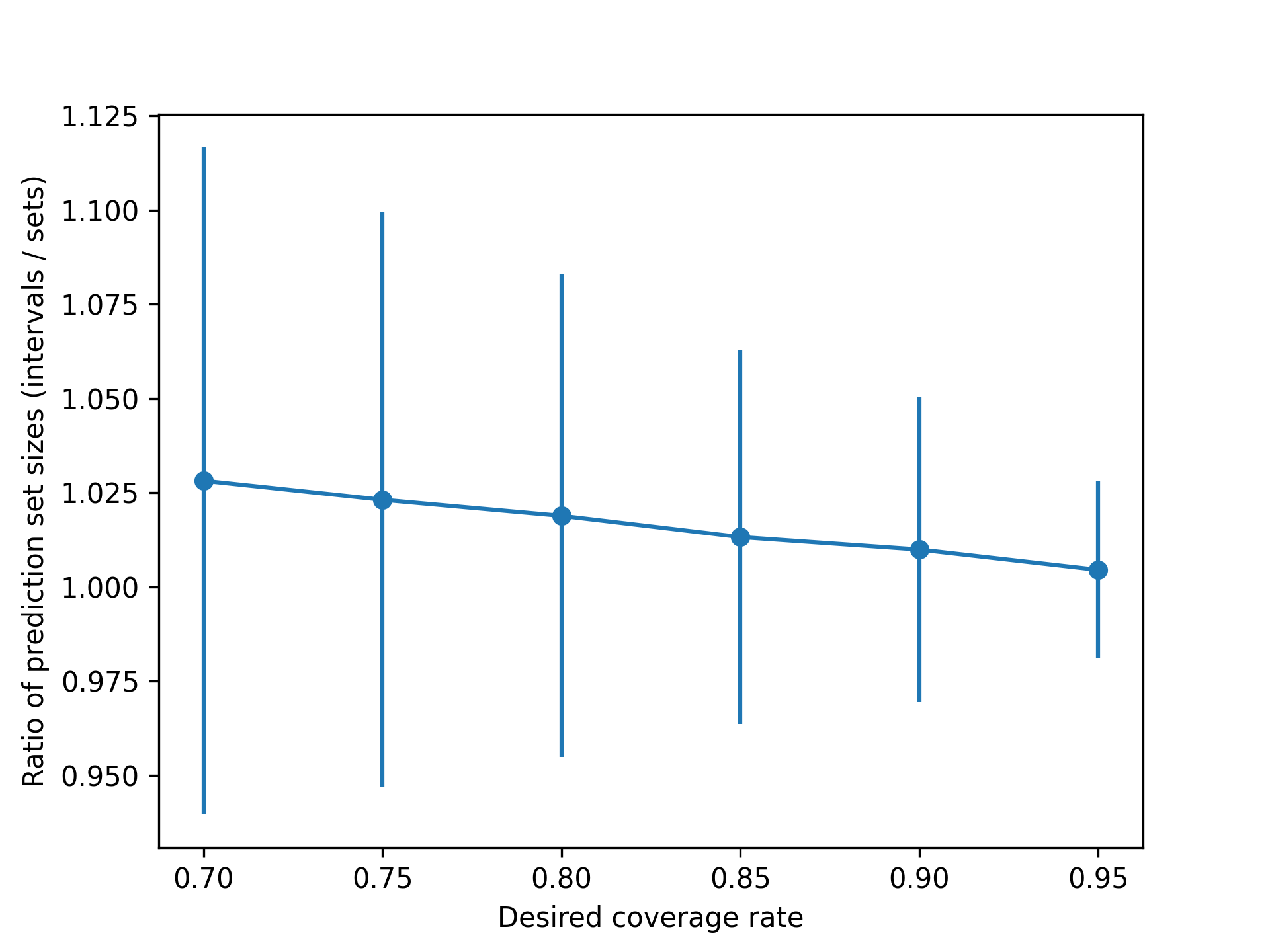}
\caption{Prediction set size vs. $1-\epsilon$.}
\label{fig:prog-3}
\end{subfigure}
\caption{Dependence of relationship between set and interval abstractions on level of added noise and on $\epsilon$.}
\label{fig:mnistablations}
\end{figure}

\ten{Furthermore, in Figure~\ref{fig:mnistablations}, we show how the relationship between these two semantics is affected by various problem parameters. First, we add more noise to the images; Figure~\ref{fig:mnistablations} (a) shows that the gap between the prediction set sizes becomes smaller; intuitively, with more noise, the prediction sets become larger overall, which increases the average prediction set size by more for the set semantics than the interval semantics. Similarly, Figure~\ref{fig:mnistablations} (b) shows that the gap between prediction set sizes become smaller when $1-\epsilon$ becomes larger (again, the prediction sets become larger).}

\begin{table}[t]
\centering
\tiny
\begin{tabular}{p{20mm}rrrrrr}
\toprule
\multicolumn{1}{c}{\multirow{2}{*}{\textbf{Noise $\eta$}}}  & \multicolumn{3}{c}{\textbf{Average Prediction Set Size}} & \multicolumn{3}{c}{\textbf{Coverage}} \\
& \multicolumn{1}{c}{Direct}
& \multicolumn{1}{c}{Compositional}
& \multicolumn{1}{c}{Full}
& \multicolumn{1}{c}{Direct}
& \multicolumn{1}{c}{Compositional}
& \multicolumn{1}{c}{Full} \\
\midrule
$0.0$ & 4.48 $\pm$ 2.79 & 4.20 $\pm$ 4.70 & 4.37 $\pm$ 2.90 & 0.96 $\pm$ 0.03 & 0.95 $\pm$ 0.10 & 0.96 $\pm$ 0.02 \\
0.1 & 6.08 $\pm$ 3.25 & 9.21 $\pm$ 8.29 & 5.64 $\pm$ 3.38 & 0.96 $\pm$ 0.02 & 0.97 $\pm$ 0.07 & 0.96 $\pm$ 0.02 \\
0.2 & 9.18 $\pm$ 5.10 & 14.19 $\pm$ 11.41 & 7.94 $\pm$ 4.92 & 0.95 $\pm$ 0.02 & 0.99 $\pm$ 0.00 & 0.96 $\pm$ 0.02 \\
0.3 & 11.25 $\pm$ 6.48 & 15.57 $\pm$ 12.37 & 9.87 $\pm$ 6.71 & 0.95 $\pm$ 0.01 & 0.99 $\pm$ 0.00 & 0.95 $\pm$ 0.01 \\
0.4 & 13.30 $\pm$ 7.35 & 15.86 $\pm$ 12.71 & 11.31 $\pm$ 7.71 & 0.95 $\pm$ 0.02 & 0.99 $\pm$ 0.00 & 0.96 $\pm$ 0.02 \\
0.5 & 13.74 $\pm$ 7.41 & 15.97 $\pm$ 12.85 & 11.55 $\pm$ 7.93 & 0.96 $\pm$ 0.02 & 0.99 $\pm$ 0.00 & 0.96 $\pm$ 0.02 \\
\bottomrule
\end{tabular}
\caption{Performance for different levels of noise $\eta$.}
\label{table:MNIST-noise}
\end{table}

\begin{table}[t]
\centering
\tiny
\begin{tabular}{p{20mm}rrrrrr}
\toprule
\multicolumn{1}{c}{\multirow{2}{*}{\textbf{Coverage $1 - \epsilon$}}}  & \multicolumn{3}{c}{\textbf{Average Prediction Set Size}} & \multicolumn{3}{c}{\textbf{Coverage}} \\
& \multicolumn{1}{c}{Direct}
& \multicolumn{1}{c}{Compositional}
& \multicolumn{1}{c}{Full}
& \multicolumn{1}{c}{Direct}
& \multicolumn{1}{c}{Compositional}
& \multicolumn{1}{c}{Full} \\
\midrule
0.95 & 11.5 $\pm$ 6.61 & 14.94 $\pm$ 12.00 & 9.51 $\pm$ 6.34 & 0.98 $\pm$ 0.01 & 0.99 $\pm$ 0.00 & 0.98 $\pm$ 0.00  \\
0.90 & 9.18 $\pm$ 5.10 & 14.19 $\pm$ 11.41 & 7.94 $\pm$ 4.92 & 0.95 $\pm$ 0.02 & 0.99 $\pm$ 0.00 & 0.96 $\pm$ 0.02 \\
0.85 & 7.75 $\pm$ 4.46 & 13.75 $\pm$ 11.05 & 6.87 $\pm$ 4.38 & 0.93 $\pm$ 0.03 & 0.99 $\pm$ 0.00 & 0.93 $\pm$ 0.03 \\
0.80 & 6.75 $\pm$ 3.66 & 13.10 $\pm$ 10.56 & 6.15 $\pm$ 3.80 & 0.91 $\pm$ 0.03 & 0.99 $\pm$ 0.00 & 0.91 $\pm$ 0.03\\
0.75 & 5.5 $\pm$ 3.42 & 12.72 $\pm$ 10.27 & 5.29 $\pm$ 3.28 & 0.84 $\pm$ 0.04 & 0.99 $\pm$ 0.00 & 0.85 $\pm$ 0.05 \\
0.70 & 5.0 $\pm$ 3.16 & 12.32 $\pm$ 9.97 & 4.76 $\pm$ 3.24 & 0.83 $\pm$ 0.05 & 0.99 $\pm$ 0.03 & 0.83 $\pm$ 0.05 \\
\bottomrule
\end{tabular}
\caption{Performance for different values of error bound $\epsilon$.}
\label{table:MNIST-epsilon}
\end{table}

\begin{table}[t]
\centering
\tiny
\begin{tabular}{p{40mm}rrrrrr}
\toprule
\multicolumn{1}{c}{\multirow{2}{*}{\textbf{Program description}}}  & \multicolumn{2}{c}{\textbf{Average Prediction Set Size}} & \multicolumn{2}{c}{\textbf{Runtime (in seconds)}} & \multicolumn{2}{c}{\textbf{Coverage}} \\
& \multicolumn{1}{c}{Intervals}
& \multicolumn{1}{c}{Sets}
& \multicolumn{1}{c}{Intervals}
& \multicolumn{1}{c}{Sets}
& \multicolumn{1}{c}{Intervals}
& \multicolumn{1}{c}{Sets} \\
\midrule
sum of list elements & 31.44 $\pm$ 2.45 & \textbf{31.33 $\pm$ 2.40} & \textbf{4.78e-05 $\pm$ 0.0002} & 1.82e-04 $\pm$ 0.0006 & 0.99 $\pm$ 0.000 &  0.99 $\pm$ 0.000 \\
sum of list elements less than 7 & 34.46 $\pm$ 1.85 & \textbf{34.16 $\pm$ 1.87} & \textbf{9.66e-05 $\pm$ 0.0003} & 3.16e-04 $\pm$ 0.0005 & 0.99 $\pm$ 0.000 & 0.99 $\pm$ 0.000 \\
max of list elements &  8.06 $\pm$ 1.04 & \textbf{7.63 $\pm$ 1.04} & \textbf{3.83e-05 $\pm$ 0.0001} & 1.39e-04 $\pm$ 0.0001 & 0.99 $\pm$ 0.000 &  0.99 $\pm$ 0.000 \\
\# of list elements less than 6 & \textbf{4.90 $\pm$ 0.31} & \textbf{4.90 $\pm$ 0.31} & \textbf{9.75e-05 $\pm$ 0.0003} & 1.79e-04 $\pm$ 0.0003 & 0.99 $\pm$ 0.000 & 0.99 $\pm$ 0.000 \\
\# of list elements equal to 2 & \textbf{4.68 $\pm$ 0.54} &\textbf{ 4.68 $\pm$ 0.55} &\textbf{ 1.19e-04 $\pm$ 0.0003 }& 2.45e-04 $\pm$ 0.0004 & 1.00 $\pm$ 0.000 & 1.00 $\pm$ 0.000 \\
\# of list elements between 3 and 8 & \textbf{4.98 $\pm$ 0.14} & \textbf{4.98 $\pm$ 0.14} & \textbf{1.26e-04 $\pm$ 0.0003 }& 2.66e-04 $\pm$ 0.0004 & 1.00 $\pm$ 0.000 & 1.00 $\pm$ 0.000 \\
max sum of any two list elements & 15.12 $\pm$ 2.07 & \textbf{15.06 $\pm$ 0.99} & \textbf{1.52e-04 $\pm$ 0.0003} & 2.81e-03 $\pm$ 0.0008 & 0.99 $\pm$ 0.000 & 0.99 $\pm$ 0.000 \\
max difference between two list elements & 9.919 $\pm$ 0.29 & \textbf{9.917 $\pm$ 0.30} & \textbf{3.16e-04 $\pm$ 0.0004} & 3.69e-03 $\pm$ 0.0009 & 1.00 $\pm$ 0.000 & 1.00 $\pm$ 0.000 \\
\midrule
$\text{sets performance}/\text{intervals performance}$ & 1.00$\times$ & \textbf{0.99$\times$} & \textbf{1.00$\times$} & 5.86$\times$ & \multicolumn{1}{c}{--} & \multicolumn{1}{c}{--} \\
\bottomrule
\end{tabular}
\caption{Prediction set sizes, coverages, and runtimes for two abstract semantics (set semantics and interval semantics). The desired coverage rate is $1 - \epsilon = 0.9$. In the second-to-last program, $k$ is a program input.}
\label{table:MNIST-2}
\end{table}


\begin{table}[t]
\centering
\tiny
\begin{tabular}{p{40mm}rrrrrr}
\toprule
\multicolumn{1}{c}{\multirow{2}{*}{\textbf{Program description}}}  & \multicolumn{3}{c}{\textbf{Average Prediction Set Size}} & \multicolumn{3}{c}{\textbf{Coverage}} \\
& \multicolumn{1}{c}{Direct}
& \multicolumn{1}{c}{Compositional}
& \multicolumn{1}{c}{Full}
& \multicolumn{1}{c}{Direct}
& \multicolumn{1}{c}{Compositional}
& \multicolumn{1}{c}{Full} \\
\midrule
\# objects in image  & 64.92 $\pm$  1.20  & \textbf{6.91 $\pm$ 10.52} & 7.04 $\pm$ 10.48 & 0.93 $\pm$ 0.008 & 0.99 $\pm$ 0.001 & 0.99 $\pm$ 0.001 \\
\# objects within 100 pixels of left & 16.84 $\pm$ 0.54 & \textbf{4.94 $\pm$ 5.39} & 5.10 $\pm$ 5.55 & 0.94 $\pm$ 0.007 & 1.00 $\pm$ 0.000 & 0.99 $\pm$ 0.001 \\
\# objects within 100 pixels of right & 13.00 $\pm$ 0.00 & \textbf{4.21 $\pm$ 4.56} & 4.34 $\pm$ 4.69 & 0.94 $\pm$ 0.003 & 1.00 $\pm$ 0.000 & 0.99 $\pm$ 0.001 \\
\# objects within 100 pixels of top & 10.92 $\pm$ 0.69 & \textbf{2.52 $\pm$ 4.03} & 2.58 $\pm$ 3.93 & 0.93 $\pm$ 0.007 & 1.00 $\pm$ 0.000 & 0.99 $\pm$ 0.000 \\
\# objects within 100 pixels of bottom & 15.08 $\pm$ 0.69 & \textbf{3.03 $\pm$ 5.57} & 3.12 $\pm$ 5.50 & 0.93 $\pm$ 0.005 & 1.00 $\pm$ 0.000 & 0.99 $\pm$ 0.000 \\
\# people within 100 pixels of a car & 132.36 $\pm$ 5.82 & 21.93 $\pm$ 69.90 & \textbf{21.85 $\pm$ 64.51} & 0.93 $\pm$ 0.007 & 1.00 $\pm$ 0.000 & 1.00 $\pm$ 0.001 \\
\# people left of a car within 100 pixels & 66.60 $\pm$ 3.30 & 21.21 $\pm$ 67.33 & \textbf{17.96 $\pm$ 45.37} & 0.93 $\pm$ 0.006 & 1.00 $\pm$ 0.000 & 1.00 $\pm$ 0.001 \\
\# people right of a car within 100 pixels & 67.40 $\pm$ 3.05 & 21.06 $\pm$ 66.75 & \textbf{18.39 $\pm$ 47.51} &  0.93 $\pm$ 0.005 & 1.00 $\pm$ 0.000 & 0.99 $\pm$ 0.001 \\
\# people below a car within 100 pixels & 72.36 $\pm$ 3.14 & 20.45 $\pm$ 64.56 & \textbf{18.16 $\pm$ 47.77}
& 0.93 $\pm$ 0.006 & 1.00 $\pm$ 0.000 & 1.00 $\pm$ 0.000 \\
\# people above a car within 100 pixels & 61.00 $\pm$ 3.53 & 20.70 $\pm$ 66.63 & \textbf{17.98 $\pm$ 46.86} & 0.92 $\pm$ 0.009 & 1.00 $\pm$ 0.000 & 1.00 $\pm$ 0.001 \\
max distance between two people & 1138.83 $\pm$ 7.43
& 2586.68 $\pm$ 2725.82 & \textbf{808.78 $\pm$ 500.00} & 0.93 $\pm$ 0.006 & 1.00 $\pm$ 0.000 & 0.99 $\pm$ 0.001 \\
max distance between a car and a person & 1179.01 $\pm$ 4.88
& 851.72 $\pm$ 1663.28 & \textbf{354.66 $\pm$ 535.28} & 0.92 $\pm$ 0.007 & 1.00 $\pm$ 0.000 & 1.00 $\pm$ 0.000 \\
\midrule
$\text{set size}/\text{our set size}$ & 4.13$\times$ & 1.34$\times$ & \textbf{1.00$\times$} & \multicolumn{1}{c}{--} & \multicolumn{1}{c}{--} & \multicolumn{1}{c}{--} \\
\bottomrule
\end{tabular}
\caption{Prediction set sizes and coverages for object detection. Here, ``object'' is a detection in $\{\text{car}, \text{person}\}$. The desired coverage rate is $1 - \epsilon = 0.9$. Coverage is rounded to 1.00 when it is $>$ 0.999; standard deviation is rounded to 0.000 when it is $<10^{-3}$. The last row shows the average prediction set size for each of the three approaches divided by the average prediction set size of the full approach, averaged over programs.}
\label{table:obj-det-1}
\end{table}

\begin{table}[t]
\centering
\tiny
\begin{tabular}{p{40mm}rrrrrr}
\toprule
\multicolumn{1}{c}{\multirow{2}{*}{\textbf{Program description}}}  & \multicolumn{3}{c}{\textbf{Fraction of Uncertain Prediction Sets}} & \multicolumn{3}{c}{\textbf{Total Coverage}} \\
& \multicolumn{1}{c}{Direct}
& \multicolumn{1}{c}{Compositional}
& \multicolumn{1}{c}{Full}
& \multicolumn{1}{c}{Direct}
& \multicolumn{1}{c}{Compositional}
& \multicolumn{1}{c}{Full} \\
\midrule
\# objects in image $\ge 3$ & 0.886 $\pm$ 0.01 & \textbf{0.385 $\pm$ 0.01} & 0.398 $\pm$ 0.01 & 0.983 $\pm$ 0.01 &  0.999 $\pm$ 0.00 & 0.999 $\pm$ 0.00 \\
\# objects within 100 pixels of left $\ge 3$ & 0.949 $\pm$ 0.01 & \textbf{0.440 $\pm$ 0.01} & 0.450 $\pm$ 0.01 & 0.967 $\pm$ 0.01 & 0.999 $\pm$ 0.00 & 0.999 $\pm$ 0.00 \\
\# objects within 100 pixels of right $\ge 3$ & 0.951 $\pm$ 0.01 & \textbf{0.383 $\pm$ 0.01} & 0.396 $\pm$ 0.01 & 0.966 $\pm$ 0.00 & 0.999 $\pm$ 0.00 & 0.999 $\pm$ 0.00 \\
\# objects within 100 pixels of top $\ge 3$ & 0.943 $\pm$ 0.01 & \textbf{0.168 $\pm$ 0.01} & 0.177 $\pm$ 0.01 & 0.968 $\pm$ 0.01 & 0.999 $\pm $ 0.00 & 0.999 $\pm$ 0.00 \\
\# objects within 100 pixels of bottom $\ge 3$ & 0.940 $\pm$ 0.00 & \textbf{0.179 $\pm$ 0.01} & 0.189 $\pm$ 0.01 & 0.970 $\pm$ 0.03 & 0.999 $\pm$ 0.00 & 0.999 $\pm$ 0.00 \\
\# pairs of (people, car) within 100 pixels $\ge 3$ & 0.926 $\pm$ 0.01 & \textbf{0.277 $\pm$ 0.01} & 0.281 $\pm$ 0.01 & 0.945 $\pm$ 0.01 & 0.999 $\pm$ 0.00 & 0.999 $\pm$ 0.00 \\
\# people left of a car within 100 pixels $\ge 3$ & 0.928 $\pm$ 0.01 & \textbf{0.277 $\pm$ 0.01} & 0.281 $\pm$ 0.01 & 0.942 $\pm$ 0.01 & 1.000 $\pm$ 0.00 & 0.999 $\pm$ 0.00 \\
\# people right of a car within 100 pixels $\ge 3$ & 0.928 $\pm$ 0.01 & \textbf{0.277 $\pm$ 0.01} & 0.281 $\pm$ 0.01 & 0.943 $\pm$ 0.01 & 1.000 $\pm$ 0.00 & 0.999 $\pm$ 0.00 \\
\# people below a car within 100 pixels $\ge 3$ &  0.930 $\pm$ 0.01 & \textbf{0.273 $\pm$ 0.01} & 0.278 $\pm$ 0.01
& 0.943 $\pm$ 0.01 & 1.000 $\pm$ 0.00 & 0.990 $\pm$ 0.00 \\
\# people above a car within 100 pixels $\ge 3$ & 0.928 $\pm$ 0.01 & \textbf{0.274 $\pm$ 0.01} & 0.278 $\pm$ 0.01 & 0.941 $\pm$ 0.01 & 1.000 $\pm$ 0.00 & 0.999 $\pm$ 0.00 \\
max dist between two people $\ge 500$ pixels & 1.000 $\pm$ 0.00
& \textbf{0.718 $\pm$ 0.01} & 0.725 $\pm$ 0.01 & 1.000 $\pm$ 0.00 & 1.000 $\pm$ 0.00 & 0.999 $\pm$ 0.01 \\
max dist between car and person $\ge 500$ pixels & 1.000 $\pm$ 0.00 
&\textbf{ 0.305 $\pm$ 0.01 }& 0.310 $\pm$ 0.01 & 1.000 $\pm$ 0.00 & 1.000 $\pm$ 0.00 & 1.000 $\pm$ 0.00 \\
\bottomrule
\end{tabular}
\caption{Fraction of "uncertain" prediction sets and total coverage for binarized programs. Prediction set sizes and coverages for object detection. The desired coverage rate is $1 - \epsilon = 0.9$. Coverage is rounded to 1.00 when it is $>$ 0.999; standard deviation is rounded to 0.000 when it is $<10^{-3}$.}
\label{table:obj-det-2}
\end{table}

\subsection{List Processing Programs Over MS-COCO Object Detections}
\label{sec:obj-detect}


Next, we consider programs that use an object detector to convert images into a list of \emph{detections} (a detection is an object in the image described by an identification label and position), and then use list processing components (e.g., map, filter, and fold) on this list. Examples include counting the number of cars in an image, or counting the number of people within 100 pixels of a car.

\paragraph{Dataset.}
The COCO dataset \cite{lin2014microsoft} consists of color images along with object detection labels. In particular, 
we use the 2017 dataset, which contains training images and  validation images. As before, we randomly split the validation set into our calibration set $Z_{\text{cal}}$ and test set $Z_{\text{test}}$. We only consider \textsf{person} and \textsf{car} labels, resulting in $|Z_{\text{cal}}| = 2476$ and $|Z_{\text{test}}|=2476$.

\paragraph{Conformal predictor.}

As before, we train the detector $\widehat{f}$ on the training set, construct the corresponding conformal detector using the calibration set $Z_{\text{cal}}$, and evaluate using the test set $Z_{\text{test}}$. 

\paragraph{Programs.}

The programs in this section perform various tasks such as counting the number of people in an image, counting the number of people within 100 pixels of a car, computing the maximum distance between two people, etc. \ten{For each program, we also consider a variant that binarizes the output (e.g., ``Does an image contain more than two people?''); intuitively, this variant captures image search problems where a user wants to select images matching some behavior.} We use error bound $\epsilon = 0.1$. For our approach, we use the single split $\epsilon_0 = 0.005$ and $\epsilon_1 = 0.095$ when intersecting the results of the direct and compositional approaches (since $\widetilde{f}$ is used only once per program to construct the prediction set around detections, only one split is required).


\paragraph{Results.}

We show results in Tables~\ref{table:obj-det-1} \&~\ref{table:obj-det-2}. The results in Table~\ref{table:obj-det-1} compare the sizes of the prediction sets across all three approaches (direct, compositional, and full conformal semantics). \ten{Table~\ref{table:obj-det-2} shows results for the binarized programs; for these programs, we report the fraction of ``uncertain'' prediction sets (i.e., output is $\{\text{true},\text{false}\}$) and the coverage. The number of uncertain prediction sets essentially captures the number of potentially spurious images the user needs to examine, whereas the miscoverage rate captures the number of true positive images that the user may have missed. Ideally, we want as few uncertain prediction sets as possible while ensuring coverage.}

\ten{As can be seen from Table~\ref{table:obj-det-1}, the compositional approach performs well in most cases, but the full approach can sometimes provide a significant advantage. Interestingly, in the binarized setting, the full conformal semantics actually performs worse than the compositional ones (the gap is due to the need to split $\epsilon$ in the full conformal semantics). This effect is likely because binary outcomes are very coarse-grained, so the direct conformal semantics is likely to almost always be uncertain, which eliminates the advantage of the full semantics. Thus, full conformal semantics are most useful when outputs have more complex structure.}

In addition, for each program in our benchmark, we provide box plots comparing the average prediction set sizes over test examples in Figure~\ref{fig:obj-det-set-size} in Appendix~\ref{sec:additionalresults}, and comparing coverage over 25 random calibration-test splits in Figure~\ref{fig:obj-det-coverage} in Appendix~\ref{sec:additionalresults}. In terms of average prediction set size, the full semantics approach performs significantly better in more than half of the programs; for the remaining programs, its performance is nearly identical to that of the slightly better performing ``compositional'' ablation. The desired coverage is achieved by all approaches across all calibration-test splits for all programs.

\section{Related work}

\paragraph{Neurosymbolic programs.}

There has been recent interest in neurosymbolic programs~\cite{gaunt2016terpret,chaudhuri2021neurosymbolic}, which are
programs that incorporate machine learning components. Much of the work has focused on synthesizing neurosymbolic programs from input-output examples, including 
functional programs~\cite{valkov2018houdini,shah2020learning}, reinforcement learning~\cite{anderson2020neurosymbolic,inala2020neurosymbolic},
data extraction from unstructured text~\cite{chen2021web,chen2023data}, and video queries~\cite{fu2019rekall,bastani2021skyquery,mell2023synthesizing}. However, most of these algorithms attempt to maximize some performance heuristic such as accuracy or $F_1$ score; thus, they are solving the same problem as a traditional machine learning strategy, but come with benefits such as improved interpretability and robustness. Outside of the work on safe reinforcement learning (which require strong assumptions as described below), none of these approaches provide any guarantees for the programs they synthesize. In contrast, our goal is to provide uncertainty quantification for neurosymbolic programs that come with statistical guarantees. Our approach complements existing neurosymbolic synthesis techniques.

\paragraph{Conformal prediction.}

Conformal prediction~\cite{vovk2005algorithmic,shafer2008tutorial,balasubramanian2014conformal,angelopoulos2023conformal} has recently gained prominence due to its ability to quantify uncertainty of deep neural networks with provable guarantees~\cite{park2019pac,angelopoulos2020uncertainty,park2020pac,bates2021distribution}. One shortcoming is that it only guarantees high-probability coverage for future examples that come from the same data distribution as the calibration examples. As a consequence, there has been a great deal of recent interest in adapting conformal prediction to other learning settings, including covariate shift~\cite{tibshirani2019conformal,park2021pac,barber2023conformal}, meta learning~\cite{fisch2021few,park2022pac}
and online learning~\cite{gibbs2022conformal,bastani2022practical}; our approach can straightforwardly be adapted to these settings by choosing $\widehat{\tau}$ using the respective algorithms.

\paragraph{Verification of machine learning models.}

There has been a great deal of interest in verifying machine learning models. For instance, robustness of neural networks says that norm-bounded input perturbations should not affect the predicted label in the classification setting~\cite{szegedy2013intriguing,goodfellow2014explaining}; there has been work on verifying this property for a given neural network~\cite{bastani2016measuring,katz2017reluplex,huang2017safety}. One particular strategy is to use abstract interpretation~\cite{gehr2018ai2}; however, this strategy focuses on propagating abstract values representing perturbed inputs through the neural network, whereas the abstract values in our setting are designed to quantify statistical uncertainty. Another property of interest is fairness~\cite{dwork2012fairness}, especially group fairness~\cite{hardt2016equality,kleinberg2018inherent}, which says that predictions should not discriminate against certain subgroups. There has been work on verifying fairness~\cite{albarghouthi2017fairsquare}, including a strategy based on statistical verification~\cite{bastani2019probabilistic}; however, this strategy cannot repair machine learning models that are not fair. Finally, in the context of reinforcement learning, where a neural network policy is used to control a system, there has been interest in proving that the corresponding closed-loop system satisfies a given safety constraint (e.g., avoid an obstacle)~\cite{bastani2018verifiable,ivanov2019verisig,anderson2020neurosymbolic,ivanov2021compositional}. However, these approaches assume that the environment is known, so the correctness property can be fully specified as a logical formula. In contrast, we are interested in uncertainty quantification of neurosymbolic programs, which provides an alternative strategy for improving trustworthiness.

\section{Conclusion}

We have proposed a novel framework for applying conformal prediction to obtain prediction sets for neurosymbolic programs that are guaranteed to cover the ground truth output with high probability. Our framework combines conformal prediction with abstract interpretation to do so, intersecting conformal predictions with abstract values at intermediate program points to minimize imprecision. These prediction sets can be used in downstream applications to guarantee probabilistic correctness by acting conservatively with respect to all possibilities present in the prediction set. There are a number of important directions for future work, such as leveraging our techniques in downstream applications to provide end-to-end correctness guarantees for neurosymbolic programs.

\bibliographystyle{ACM-Reference-Format}
\bibliography{ref}

\clearpage
\appendix

\section{Theoretical Guarantees}

\subsection{Probably Approximately Correct (PAC) Prediction Sets}
\label{sec:pac}

One shortcoming of the traditional conformal prediction guarantee (\ref{eqn:compconformal}) is that it ``mixes'' the probability over the current input $x\in\mathcal{X}$ and the calibration dataset $Z=\{(x_i,y_i^*)\}_{i=1}^n\subseteq\mathcal{X}\times\mathcal{Y}$. Recent work~\cite{vovk2012conditional,park2019pac} has demonstrated how to construct prediction set models $\widetilde{f}_{\widehat{\tau}(Z,\epsilon,\delta)}$ that satisfy a \emph{probably approximately correct (PAC)} guarantee of the form
\begin{align*}
\mathbb{P}_{Z\sim D^n}[\mathbb{P}_{(x,y^*)\sim D}[y^*\in\widetilde{f}_{\widehat{\tau}(Z,\epsilon,\delta)}(x)]\ge1-\epsilon]\ge1-\delta,
\end{align*}
where $\epsilon,\delta\in\mathbb{R}_{>0}$ are given. These guarantees are useful in ensuring that for most calibration datasets $Z$ (more precisely, with $1-\delta$ probability over $Z\sim D^n$), we obtain the $1-\epsilon$ coverage guarantee. In contrast, if we randomly construct a conformal predictor using traditional conformal prediction, there is a high probability that the resulting conformal predictor does not satisfy the conformal guarantee; instead, it only satisfies the guarantee on average over $Z\sim D^n$. The drawback is that PAC approaches are more conservative than traditional conformal prediction (i.e., they result in larger prediction set sizes on average), but we find that the differences are small in practice.

Incorporating PAC prediction set models into our framework is straightforward; we simply need to union bound over both $\epsilon$ and $\delta$ instead of just $\epsilon$. It is straightforward to show that we correspondingly obtain PAC guarantees for our approach---i.e., guarantees of the form
\begin{align*}
\mathbb{P}_{Z\sim D^n}\left[\mathbb{P}_{x\sim D}\left[\den{p}^*(x)\in\gamma\left(\den{p}_{\epsilon,\delta}^C(x,Z)\right)\right]\ge1-\epsilon\right]\ge1-\delta.
\end{align*}

\subsection{Proof of Theorem~\ref{thm:loopfree}}
\label{sec:thm:loopfree:proof}

We prove by structural induction on $p$---i.e., we prove the inductive hypothesis
\begin{align*}
\mathbb{P}_{Z\sim D^n,x\sim D}\left[\den{p}^*(x)\in \gamma\left(\den{p}_{\epsilon}^C(x,Z)\right)\right]\ge1-\epsilon.
\end{align*}
For the case $p=X$, we have
\begin{align*}
\den{X}^*(x)=x\in\gamma(\alpha(x))=\gamma(\den{X}_{\epsilon}^C(x,Z)),
\end{align*}
where the set inclusion follows by (\ref{eqn:abs1}), as claimed. For the case $p=f(p_1,...,p_k)$, there are two subcases. If $f \in \widehat{\mathcal{F}}$, then by Lemma \ref{lem:direct},
\begin{equation*}
\mathbb{P}_{Z\sim D^n,x\sim D}\left[\den{p}^*(x)\in\gamma\left(\den{p}_{\epsilon}^C(x,Z)\right)\right] = 
\mathbb{P}_{Z\sim D^n,x\sim D}\left[\den{p}^*(x)\in\gamma\left(\den{p}_{\epsilon}^{C0}(x,Z)\right)\right]\ge1-\epsilon
\end{equation*}
as claimed. If  $f \notin \widehat{\mathcal{F}}$, then by induction we have
\begin{align}
\label{eqn:thm:main:1}
\mathbb{P}_{Z\sim D^n,x\sim D}\left[\den{p_i}^*(x)\in\gamma\left(\den{p_i}_{\epsilon_i}^C(x,Z)\right)\right]\ge1-\epsilon_i.
\qquad(\forall i\in[k]),
\end{align}
and by Lemma~\ref{lem:direct}, we have
\begin{align}
\label{eqn:thm:main:2}
\mathbb{P}_{Z\sim D^n,x\sim D}\left[\den{p}^*(x)\in\gamma\left(\den{p}_{\epsilon_0}^{C0}(x,Z)\right)\right]\ge1-\epsilon_0.
\end{align}
Let $E(Z,x)$ be the event that $\den{p_i}^*(x)\in\gamma\left(\den{p_i}_{\epsilon_i}^C(x,Z)\right)$ for all $i\in[k]$ and $\den{p}^*(x)\in\gamma\left(\den{p}_{\epsilon_0}^{C0}(x,Z)\right)$; by a union bound, the assumption that the $\{\epsilon_i\}_{i=0}^k$ satisfy (\ref{eqn:deltaconstraint}), and lines (\ref{eqn:thm:main:1}) and (\ref{eqn:thm:main:2}), we have
\begin{align}
\label{eqn:thm:main:3}
\mathbb{P}_{Z\sim D^n,x\sim D}[E(Z,x)]\ge1-\epsilon.
\end{align}
On the event $E(Z,x)$, we know $\den{p}^*(x)\in\gamma(\den{p}_{\epsilon_0}^{C0}(x,Z))$. Further, by the definition of $f^\#$, we have
\begin{align*}
\den{p}^*(x)\in \gamma\left(f^\#(\den{p_1}_{\epsilon_1}^C(x,Z),...,\den{p_k}_{\epsilon_k}^C(x,Z))\right),
\end{align*}
Thus, by the definition of $\sqcap$, it follows that $\den{p}^*(x)\in\gamma\left(\den{p}_{\epsilon}^C(x,Z)\right)$. Together with (\ref{eqn:thm:main:3}), we have
\begin{align*}
\mathbb{P}_{Z\sim D^n,x\sim D}\left[\den{p}^*(x)\in\gamma\left(\den{p}_{\epsilon}^C(x,Z)\right)\right]\ge\mathbb{P}_{Z\sim D^n,x\sim D}[E(Z,x)]\ge1-\epsilon.
\end{align*}
as claimed. \hfill $\qed$

\subsection{Fixed Point Semantics for While Loops}
\label{sec:whileloopsemantics}

We formalize the solution to the least fixed point used to define while loops in our ground truth and conformal semantics for our imperative language. Starting with our conformal semantics, note that $\phi_{p,x}^C$ has type $\phi_{p,x}^C:\mathcal{H}\to\mathcal{H}$, where $\mathcal{H}=\mathcal{Z}\times\mathbb{R}_{\ge0}\to\mathcal{Z}$ and $\mathcal{Z}=(\Sigma^\#\cup\{\bot\})\times(\Sigma\cup\{\bot\})^n$. For all applicable sets, we assume the partial order induced the basic relation $a\ge\bot$ for all $a$. Define
\begin{align*}
h_\bot(\sigma_\#,\{\sigma_i\}_{i=1}^n,\epsilon)=\bot\qquad(\forall\sigma_\#\in\Sigma^\#,\sigma_1,...,\sigma_n\in\Sigma,\epsilon\in\mathbb{R}_{\ge0}),
\end{align*}
and define
\begin{align*}
h_m=(\phi_{p,x}^C)^{(m)}(h_\bot)=\begin{cases}
\phi_{p,x}^C((\phi_{p,x}^C)^{(m-1)}(h_\bot))&\text{if }m>0 \\
h_\bot&\text{if }m=0.
\end{cases}
\end{align*}
Given $\sigma_\#\in\Sigma^\#$, $\sigma_1,...,\sigma_n\in\Sigma$, and $\epsilon\in\mathbb{R}_{\ge0}$, let $(\sigma_\#^{(m)},\{\sigma_i^{(m)}\}_{i=1}^n)=h_m(\sigma_\#,\{\sigma_i\}_{i=1}^n,\epsilon)$, and define $h_{\omega}(\sigma_\#,\{\sigma_i\}_{i=1}^n,\epsilon)=(\sigma_\#^{(\omega)},\{\sigma_i^{(\omega)}\}_{i=1}^n)$, where
\begin{align*}
\sigma_\#^{(\omega)}
&=\lim_{m\to\infty}\sigma_\#^{(m)} \\
\sigma_i^{(\omega)}
&=\lim_{m\to\infty}\sigma_i^{(m)}
\qquad(\forall i\in[n]).
\end{align*}
The limits are well defined since the sequences of values form ascending chains: $\sigma_\#^{(0)}\sqsubseteq\sigma_\#^{(1)}\sqsubseteq\cdots$ and $\sigma_i^{(0)}\sqsubseteq\sigma_i^{(1)}\sqsubseteq...$ (for $i\in[n]$), which follows straightforwardly from the definition of $\phi_{p,x}^C$, and therefore stabilize (by our ascending chain condition assumption for the former case and because the partial order only has ascending chains of length two in the latter case). Then, we define $\text{lfp}(h=\phi_{p,x}^C(h))=h_{\omega}$. Similarly, for the ground truth semantics, define $h_\bot^*(\sigma_*)=\bot$ for all $\sigma\in\Sigma$ and $h_m^*=(\phi_{p,x}^*)^{(m)}(h_\bot^*)$. Given $\sigma_*\in\Sigma$, let $\sigma_*^{(m)}=h_m(\sigma_*)$, and define
\begin{align*}
h_{\omega}^*(\sigma_*)
=\sigma_*^{(m)}=\lim_{m\to\infty}\sigma_*^{(m)}.
\end{align*}
As before, the sequence $\sigma_*^{(0)}\sqsubseteq\sigma_*^{(1)}\sqsubseteq\cdots$ stabilizes, so the limit is well defined, and we define $\text{lfp}(h^*=\phi_{p,x}^*(h^*))=h_{\omega}^*$. Next, by a standard argument, we have:
\begin{lemma}
\label{lem:while}
For all $m\in\mathbb{N}$, we have $\sigma_\#^{(m)}\in\{\sigma_\#^{(\omega)},\bot\}$, $\sigma_i^{(m)}\in\{\sigma_i^{(\omega)},\bot\}$ for all $i\in[n]$, and $\sigma_*^{(m)}\in\sigma_*^{(\omega)}$. Furthermore, if $\sigma_\#^{(m)}=\bot$ for all $m\in\mathbb{N}$, then $\sigma_\#^{(\omega)}=\bot$; for all $i\in[n]$, if $\sigma_i^{(m)}=\bot$ for all $m\in\mathbb{N}$, then $\sigma_i^{(\omega)}=\bot$; and if $\sigma_*^{(m)}=\bot$ for all $m\in\mathbb{N}$, then $\sigma_*^{(\omega)}=\bot$.
\end{lemma}
Finally, we formalize the notion of loop termination as follows:
\begin{definition}
\label{def:whileterminate}
\rm
A while loop \emph{terminates} under our conformal semantics if $\sigma_\#^{(\omega)}=\bot$ iff $\sigma_\#=\bot$, and for all $i\in[n]$, $\sigma_i^{(\omega)}=\bot$ iff $\sigma_i=\bot$, and under our ground truth semantics if $\sigma_*^{(\omega)}=\bot$ iff $\sigma_*=\bot$.
\end{definition}
\begin{assumption}
\label{assump:terminate}
All while loops in $p$ terminate on all possible inputs.
\end{assumption}
Note that our requirements is that $p$ terminates for all inputs, not just with probability one. However, our guarantees can be straightforwardly extended to termination with probability one.

\subsection{Proof of Theorem~\ref{thm:impmain}}
\label{sec:thm:impmain:proof}

First, we show that our conformal predictors satisfy our coverage guarantee:
\begin{lemma}
\label{lem:impconformal}
For any $\lambda\in\mathcal{M}(\Sigma^\#\times\Sigma^{n+1})$ satisfying $\pi_*(\lambda)=\otimes_{i=1}^{n+1}\mu$, we have
\begin{align*}
\mathbb{P}_{\pi_*(\lambda)(\{\sigma_i\}_{i=1}^n,\sigma_*)}[f^*(\sigma_*)\in C_{\epsilon}(g_0,\{(\bot,f^*(\sigma_i))\mid\sigma_i\neq\bot\}_{i=1}^n)(\bot)\mid\sigma_*\neq\bot]\ge1-\epsilon.
\end{align*}
\end{lemma}
\begin{proof}
The claim follows from the conformal guarantee (\ref{eqn:compconformal}), assuming the calibration examples $\{(\bot,\sigma_i)\}_{i=1}^n$ and the current example $(\bot,\sigma_*)$ are i.i.d. random variables. The main issue is handling the conditional on $\sigma_*\neq\bot$. This step follows since the algorithm filters examples $\sigma_i\neq\bot$ from the calibration set, which is equivalent to performing rejection sampling~\citep{neumann1951various}, so the remaining calibration examples are i.i.d. samples from the probability distribution $\mu$ conditioned on $\sigma\neq\bot$.
\end{proof}
Next, one of the key challenges is demonstrating that the calibration set $\{\sigma_i\}_{i=1}^n$ and the true store $\sigma_*$ for the current input are i.i.d. random variables. To this end, we first show that the (deterministic) semantics decomposes into a product when we use $\pi$ to project out $\sigma_\#$ from the joint semantics:
\begin{lemma}
\label{lem:calibrationsemantics}
We have $\pi\circ\den{p}_{\epsilon}^J(\sigma_\#,\{\sigma_i\}_{i=1}^n,\sigma_*)=(\{\den{p}^*(\sigma_i)\}_{i=1}^n,\den{p}^*(\sigma_*))$.
\end{lemma}
\begin{proof}
The assignment semantics for both standard and machine learning components follow from the definition of our conformal semantics. The sequencing case follows by structural induction. The conditional case follows since exactly one of $\iota_b(\sigma,x)\neq\top$ for $b\in\{\text{true},\text{false}\}$, corresponding to the branch where $\sigma_i(x)=b$; thus, the join evaluates to the value of $\den{p}^*$ along the taken branch. Finally, for while loops, based on the semantics in Appendix~\ref{sec:whileloopsemantics}, it is easy to see by induction that if $\sigma_i=\sigma_*$, then $\sigma_i^{(m)}=\sigma_*^{(m)}$ for all $m\in\mathbb{N}$; thus, by definition, $\sigma_i^{(\omega)}=\sigma_*^{(\omega)}$ as well.
\end{proof}
Next, we have two standard technical lemmas from measure theory:
\begin{lemma}
\label{lem:measureseq}
Given measurable $g:\mathcal{X}\to\mathcal{Y}$ and $h:\mathcal{Y}\to\mathcal{Z}$, we have $(h\circ g)_*=h_*\circ g_*$.
\end{lemma}
\begin{proof}
By definition, given $\mu\in\mathcal{M}\mathcal{X}$ and measurable $A\subseteq\mathcal{Z}$, we have
\begin{align*}
(h\circ g)_*(\mu)(A)=\mu(\{x\in\mathcal{X}\mid h(g(x))\in A\},
\end{align*}
whereas
\begin{align*}
h_*(g_*(\mu))(A)=g_*(\mu)(\{y\in\mathcal{Y}\mid h(y)\in A\})&=\mu(\{x\in\mathcal{X}\mid g(x)\in\{y\in\mathcal{Y}\mid h(y)\in A\}\}) \\
&=\mu(\{x\in\mathcal{X}\mid h(g(x))\in A\} \\
&=(h\circ g)_*(\mu)(A),
\end{align*}
so the claim follows.
\end{proof}
\begin{lemma}
\label{lem:measureprod}
Given measurable $g:\mathcal{X}\to\mathcal{Y}$ and $g':\mathcal{X}'\to\mathcal{Y}'$, define $g\times g':\mathcal{X}\times\mathcal{X}'\to\mathcal{Y}\times\mathcal{Y}'$ by $(g\times g')(x,x')=(g(x),g'(x'))$. Given $\mu,\mu'\in\mathcal{M}\mathcal{X}$, we have $(g\times g')_*(\mu\otimes\mu')=g_*(\mu)\otimes g'_*(\mu')$.
\end{lemma}
\begin{proof}
Note that for any $\mu\in\mathcal{M}\mathcal{X}$, $\mu'\in\mathcal{M}\mathcal{X}'$, $A\subseteq\mathcal{Y}$, and $A'\times\mathcal{Y}'$, we have
\begin{align*}
(g\times g')_*(\mu\otimes\mu')(A\times A')
&=(\mu\otimes\mu')(\{(y,y')\in\mathcal{Y}\times\mathcal{Y}'\mid(g(y),g'(y'))\in A\times A'\}) \\
&=(\mu\otimes\mu')(\{y\in\mathcal{Y}\mid g(y)\in A\}\times\{y'\in\mathcal{Y}'\mid g'(y')\in A'\}) \\
&=\mu(\{y\in\mathcal{Y}\mid g(y)\in A\})\mu'(\{y'\in\mathcal{Y}'\mid g'(y')\in A'\}) \\
&=g_*(\mu)(A)g'_*(\mu')(A') \\
&=(g_*(\mu)\otimes g'_*(\mu'))(A\times A').
\end{align*}
Note that both $g_*(\mu)\otimes g'_*(\mu')$ and $(g\times g')_*(\mu\otimes\mu')$ are $\sigma$-finite. Then, by the Hahn–Kolmogorov theorem, they are uniquely defined by the values they take on sets of the form $A\times A'$, which generate the $\sigma$-algebra on $\mathcal{Y}\times\mathcal{Y}'$. Thus, $(g\times g')_*(\mu\otimes\mu')=g_*(\mu)\otimes g'_*(\mu')$, as claimed.
\end{proof}
Using these two results, we can extend Lemma~\ref{lem:calibrationsemantics} to the probabilistic semantics given by the pushforward of our joint semantics, which we achieve using our next two lemmas:
\begin{lemma}
\label{lem:gendecomp}
Define $G:\Sigma^\#\times\Sigma^{n+1}\to\Sigma^\#\times\Sigma^{n+1}$ by $G(\sigma_\#,\{\sigma_i\}_{i=1}^n,\sigma_*)=(g^\#(\sigma_\#,\{\sigma_i\}_{i=1}^n),\{g(\sigma_i)\}_{i=1}^n,g(\sigma_*))$ for measurable $g^\#:\Sigma^\#\times\Sigma^n\to\Sigma^\#$ and $g:\Sigma\to\Sigma$. If $\pi_*(\lambda)=\otimes_{i=1}^{n+1}\mu$, then $\pi_*\circ G_*(\lambda)=\otimes_{i=1}^{n+1}g_*(\mu)$.
\end{lemma}
\begin{proof}
Note that $\pi_*\circ G_*
=(\pi\circ G)_*
=(\times_{i=1}^{n+1}g)_*
=\times_{i=1}^{n+1}g_*$,
where the first equality follows by Lemma~\ref{lem:measureseq}, the second by definition of $G$, and the third by Lemma~\ref{lem:measureprod}, as claimed.
\end{proof}
\begin{lemma}
\label{lem:ind}
Given program $p$ and $\lambda\in\mathcal{M}(\Sigma^\#\times\Sigma^{n+1})$ such that $\pi_*(\lambda)=\otimes_{i=1}^{n+1}\mu$ for some $\mu\in\mathcal{M}\Sigma$,
\begin{align*}
\den{p}^J_{\epsilon,*}(\lambda)=\otimes_{i=1}^{n+1}\den{p}^*(\mu).
\end{align*}
\end{lemma}
\begin{proof}
This result follows by taking $g^\#=\den{p}^C_{\epsilon}$ and $g=\den{p}^*$ in Lemma~\ref{lem:gendecomp}, where the required structure on $G$ follows by Lemma~\ref{lem:calibrationsemantics}.
\end{proof}
Next, we have a straightforward technical lemma regarding evaluating to $\bot$:
\begin{lemma}
\label{lem:bot}
Given $\lambda\in\mathcal{M}((\Sigma\cup\{\bot\})^\#\times(\Sigma\cup\{\bot\})^{n+1})$ and $p$, we have $\sigma_*=\bot\Leftrightarrow\den{p}^*(\sigma_*)=\bot$.
\end{lemma}
\begin{proof}
We prove by structural induction. The case of assignments follows since by definition, components $f\in\mathcal{F}$ have type $f:\Sigma^n\to\Sigma^n$, which we have extended to components $f:(\Sigma\cup\{\bot\})^n\to(\Sigma\cup\{\bot\})^n$ by defining $f(\bot)=\bot$. The sequencing case follows by induction. The conditional case follows since if $\sigma\neq\bot$, then $\iota_b(\sigma)\neq\bot$ for some $b\in\mathbb{B}$, so by induction, one of the two inputs to the join are not $\bot$. The while loop case follows by Assumption~\ref{assump:terminate}.
\end{proof}
Finally, we combine the other results to form our key guarantee; as we discuss at the end, Theorem~\ref{thm:impmain} follows straightforwardly from this result:
\begin{lemma}
\label{lem:impmain}
Given $\lambda\in\mathcal{M}((\Sigma\cup\{\bot\})^\#\times(\Sigma\cup\{\bot\})^{n+1})$ satisfying (i) $\pi_*(\lambda)=\otimes_{i=1}^{n+1}\mu$ for some $\mu\in\mathcal{M}\Sigma$, (ii) $\mathbb{P}_{\lambda(\sigma_\#,\{\sigma_i\}_{i=1}^n,\sigma_*)}[\sigma_*\neq\bot]>0$, and (iii) $\mathbb{P}_{\lambda(\sigma_\#,\{\sigma_i\}_{i=1}^n,\sigma_*)}[\sigma_*\not\in\gamma(\sigma_\#)\mid\sigma_*\neq\bot]\le\epsilon_0$, then
\begin{align*}
\mathbb{P}_{\lambda(\sigma_\#,\{\sigma_i\}_{i=1}^n,\sigma_*)}[\den{p}^*(\sigma_*)\not\in\gamma(\den{p}^C_{\epsilon}(\sigma_\#,\{\sigma_i\}_{i=1}^n))\mid\sigma_*\neq\bot]\le\epsilon_0+\epsilon.
\end{align*}
\end{lemma}

\begin{proof}
We prove by structural induction on $p$. 

\paragraph{Assignment $p=\vec{x}\coloneqq f(\vec{x})$ (with $f\in\mathcal{F}$).}

On the event $\sigma_*\neq\bot\wedge\sigma_*\in\gamma(\sigma_\#)$, we have
\begin{align*}
\den{p}^*(\sigma_*)=f(\sigma_*)\in\gamma(f^\#(\sigma_\#))=\gamma(\den{p}^C_{\epsilon}(\sigma_\#,\{\sigma_i\}_{i=1}^n)),
\end{align*}
so
\begin{align*}
&\mathbb{P}_{\lambda(\sigma_\#,\{\sigma_i\}_{i=1}^n,\sigma_*)}[\den{p}^*(\sigma_*)\not\in\gamma(\den{p}^C_{\epsilon}(\sigma_\#,\{\sigma_i\}_{i=1}^n))\mid\sigma_*\neq\bot] \\
&\le\mathbb{P}_{\lambda(\sigma_\#,\{\sigma_i\}_{i=1}^n,\sigma_*)}[\sigma_*\not\in\gamma(\sigma_\#)\mid\sigma_*\neq\bot] \\
&\le\epsilon_0,
\end{align*}
so the claim follows.

\paragraph{Assignment $p=\vec{x}\coloneqq\widehat{f}(\vec{x})$ (with $\widehat{f}\in\widehat{\mathcal{F}}$).}

We have
\begin{align*}
&\mathbb{P}_{\lambda(\sigma_\#,\{\sigma_i\}_{i=1}^n,\sigma_*)}[f^*(\sigma_*)\not\in C_{\epsilon}(g_0,\{(\bot,f^*(\sigma_i))\mid\sigma_i\neq\bot\}_{i=1}^n)(\bot)\mid\sigma_*\neq\bot] \\
&=\mathbb{P}_{\pi_*(\lambda)(\{\sigma_i\}_{i=1}^n,\sigma_*)}[f^*(\sigma_*)\not\in C_{\epsilon}(g_0,\{(\bot,f^*(\sigma_i))\mid\sigma_i\neq\bot\}_{i=1}^n)(\bot)\mid\sigma_*\neq\bot] \\
&=\mathbb{P}_{\mu(\sigma_1),...,\mu(\sigma_n),\mu(\sigma_*)}[f^*(\sigma_*)\not\in C_{\epsilon}(g_0,\{(\bot,f^*(\sigma_i))\mid\sigma_i\neq\bot\}_{i=1}^n)(\bot)\mid\sigma_*\neq\bot] \\
&\le\epsilon,
\end{align*}
where the first equality follows since the event and condition do not depend on $\sigma_\#$ and by change of measure, the second equality follows by our assumption on $\lambda$, and the first inequality follows by our conformal guarantee in Lemma~\ref{lem:impconformal}. The claim follows.

\paragraph{Sequence $p=q;r$.} Letting $(\sigma_\#',\{\sigma_i'\}_{i=1}^n,\sigma_*')=\den{q}^J_{\epsilon/2}(\sigma_\#,\{\sigma_i\}_{i=1}^n,\sigma_*)$, by induction, we have
\begin{align*}
\mathbb{P}_{\lambda(\sigma_\#,\{\sigma_i\}_{i=1}^n,\sigma_*)}[\sigma_*'\not\in\gamma(\sigma_\#')\mid\sigma_*\neq\bot]\le\epsilon_0+\frac{\epsilon}{2}.
\end{align*}
By Lemma~\ref{lem:bot}, we equivalently have
\begin{align*}
\mathbb{P}_{\lambda(\sigma_\#,\{\sigma_i\}_{i=1}^n,\sigma_*)}[\sigma_*'\not\in\gamma(\sigma_\#')\mid\sigma_*'\neq\bot]\le\epsilon_0+\frac{\epsilon}{2}.
\end{align*}
Letting $\lambda'=\den{q}^J_{\epsilon/2,*}(\lambda)$ be the pushforward measure, by a change of measure, we have
\begin{align*}
\mathbb{P}_{\lambda'(\sigma_\#',\{\sigma_i'\}_{i=1}^n,\sigma_*')}[\sigma_*'\not\in\gamma(\sigma_\#')\mid\sigma_*'\neq\bot]\le\epsilon_0+\frac{\epsilon}{2}.
\end{align*}
By Lemma~\ref{lem:ind}, $\pi_*(\lambda')=\otimes_{i=1}^{n+1}\mu'$ with $\mu'=\den{q}^*_*(\mu)$, and by Lemma~\ref{lem:bot}, $\mathbb{P}_{\lambda'(\sigma_\#',\{\sigma_i'\}_{i=1}^n,\sigma_*')}[\sigma_*'\neq\bot]>0$. Then, letting $(\sigma_\#'',\{\sigma_i''\}_{i=1}^n,\sigma_*'')=\den{r}^J_{\epsilon/2}(\sigma_\#',\{\sigma_i'\}_{i=1}^n,\sigma_*')$, by induction, we have
\begin{align*}
\mathbb{P}_{\lambda'(\sigma_\#',\{\sigma_i'\}_{i=1}^n,\sigma_*')}[\sigma_*''\not\in\gamma(\sigma_\#'')\mid\sigma_*'\neq\bot]\le\epsilon_0+\epsilon.
\end{align*}
By another change of measure, we equivalently have
\begin{align*}
\mathbb{P}_{\lambda(\sigma_\#,\{\sigma_i\}_{i=1}^n,\sigma_*)}[\sigma_*''\not\in\gamma(\sigma_\#'')\mid\sigma_*'\neq\bot]\le\epsilon_0+\epsilon.
\end{align*}
By Lemma~\ref{lem:bot}, we equivalently have
\begin{align*}
\mathbb{P}_{\lambda(\sigma_\#,\{\sigma_i\}_{i=1}^n,\sigma_*)}[\sigma_*''\not\in\gamma(\sigma_\#'')\mid\sigma_*\neq\bot]\le\epsilon_0+\epsilon.
\end{align*}
Note that by definition, $(\sigma_\#'',\{\sigma_i''\}_{i=1}^n,\sigma_*'')=\den{q;r}^C_{\epsilon}(\sigma_\#,\{\sigma_i\}_{i=1}^n,\sigma_*)$, so the claim follows.

\paragraph{Condition $p=\textbf{if }x\textbf{ then }q$.}

First, letting $\mathbb{B}=\{\text{true},\text{false}\}$, by our assumption, we have
\begin{align}
\epsilon_0
>\mathbb{P}_{\lambda(\sigma_\#,\{\sigma_i\}_{i=1}^n,\sigma_*)}[\sigma_*\in\gamma(\sigma_\#)\mid\sigma_*\neq\bot]
=\sum_{b\in\mathbb{B}}\epsilon_b\cdot p_b,
\label{eqn:impthm:cond:0}
\end{align}
where we have defined
\begin{align*}
\epsilon_b&=\mathbb{P}_{\lambda(\sigma_\#,\{\sigma_i\}_{i=1}^n,\sigma_*)}[\sigma_*\in\gamma(\sigma_\#)\mid\sigma_*\neq\bot\wedge\sigma_*(x)=b] \\
p_b&=\mathbb{P}_{\lambda(\sigma_\#,\{\sigma_i\}_{i=1}^n,\sigma_*)}[\sigma_*(x)=b\mid\sigma_*\neq\bot].
\end{align*}
In addition, note that
\begin{align}
\epsilon_b
&=\mathbb{P}_{\lambda(\sigma_\#,\{\sigma_i\}_{i=1}^n,\sigma_*)}[\iota_b(\sigma_*)\in\gamma(\iota_b^\#(\sigma_\#))\mid\iota_b(\sigma_*)\neq\bot] \nonumber \\
&=\mathbb{P}_{\lambda_b(\sigma_\#',\{\sigma_i'\}_{i=1}^n,\sigma_*')}[\sigma_*'\in\gamma(\sigma_\#'))\mid\sigma_*'\neq\bot] \label{eqn:impthm:cond:1}
\end{align}
where the first equality holds since $\sigma_*\neq\bot\wedge\sigma_*(x)=b$ holds iff $\iota_b(\sigma_*)\neq\bot$ holds, and since $\sigma_*\in\gamma(\sigma_\#)$ iff $\iota_b(\sigma_*)\in\gamma(\iota_b^\#(\sigma_\#))$ on the event $\sigma_*(x)=b$, and the second equality holds by defining $\lambda_b=G_*(\lambda)$, where $G_*=\iota_b^\#\times(\times_{i=1}^{n+1}\iota_b)$. Now, by the law of total probability, we have
\begin{align}
&\mathbb{P}_{\lambda(\sigma_\#,\{\sigma_i\}_{i=1}^n,\sigma_*)}[\den{p}^*(\sigma_*)\not\in\gamma(\den{p}^C_{\epsilon}(\sigma_\#,\{\sigma_i\}_{i=1}^n))\mid\sigma_*\neq\bot] \nonumber \\
&=\sum_{b\in\mathbb{B}}\mathbb{P}_{\lambda(\sigma_\#,\{\sigma_i\}_{i=1}^n,\sigma_*)}[\den{p}^*(\sigma_*)\not\in\gamma(\den{p}^C_{\epsilon}(\sigma_\#,\{\sigma_i\}_{i=1}^n))\mid\sigma_*\neq\bot\wedge\sigma_*(x)=b]\cdot p_b. \label{eqn:impthm:cond:2}
\end{align}
For the term $b=\text{true}$ in (\ref{eqn:impthm:cond:2}), we have
\begin{align}
&\mathbb{P}_{\lambda(\sigma_\#,\{\sigma_i\}_{i=1}^n,\sigma_*)}[\den{p}^*(\sigma_*)\not\in\gamma(\den{p}^C_{\epsilon}(\sigma_\#,\{\sigma_i\}_{i=1}^n))\mid\sigma_*\neq\bot\wedge\sigma_*(x)=\text{true}] \nonumber \\
&=\mathbb{P}_{\lambda(\sigma_\#,\{\sigma_i\}_{i=1}^n,\sigma_*)}[\den{p}^*(\sigma_*)\not\in\gamma(\den{p}^C_{\epsilon}(\sigma_\#,\{\sigma_i\}_{i=1}^n))\mid\iota_{\text{true}}(\sigma_*)\neq\bot] \nonumber \\
&=\mathbb{P}_{\lambda(\sigma_\#,\{\sigma_i\}_{i=1}^n,\sigma_*)}[\den{q}^*(\iota_{\text{true}}(\sigma_*))\not\in\gamma(\den{p}^C_{\epsilon}(\sigma_\#,\{\sigma_i\}_{i=1}^n))\mid\iota_{\text{true}}(\sigma_*)\neq\bot] \nonumber \\
&\le\mathbb{P}_{\lambda(\sigma_\#,\{\sigma_i\}_{i=1}^n,\sigma_*)}[\den{q}^*(\iota_{\text{true}}(\sigma_*))\not\in\gamma(\den{q}^C_{\epsilon}(\iota_{\text{true}}^\#(\sigma_\#),\{\iota_{\text{true}}(\sigma_i)\}_{i=1}^n))\mid\iota_{\text{true}}(\sigma_*)\neq\bot] \nonumber \\
&=\mathbb{P}_{\lambda_{\text{true}}(\sigma_\#',\{\sigma_i'\}_{i=1}^n,\sigma_*')}[\den{q}^*(\sigma_*')\not\in\gamma(\den{q}^C_{\epsilon}(\sigma_\#',\{\sigma_i'\}_{i=1}^n))\mid\sigma_*'\neq\bot], \label{eqn:impthm:cond:3}
\end{align}
where the first equality follows since $\sigma_*\neq\bot\wedge\sigma_*(x)=\text{true}$ iff $\iota_{\text{true}}(\sigma_*)\neq\bot$, the second equality follows by definition of $\den{p}^*$ under the event that $\sigma_*(x)=\text{true}$, the first inequality follows since $\gamma(\den{q}^C_{\epsilon}(\iota_{\text{true}}^\#(\sigma_\#),\{\iota_{\text{true}}(\sigma_i)\}_{i=1}^n))\subseteq\gamma(\den{p}^C_{\epsilon}(\sigma_\#,\{\sigma_i\}_{i=1}^n))$ by definition of the join, and the third equality follows by a change of measure.

Now, by Lemma~\ref{lem:gendecomp} with $g=\iota_{\text{true}}$ and $g^\#=\iota_{\text{true}}^\#$, we have $\lambda_{\text{true}}=G_*(\lambda)=\otimes_{i=1}^ng(\mu)$. Furthermore, by (\ref{eqn:impthm:cond:1}), we have $\mathbb{P}_{\lambda_{\text{true}}(\sigma_\#',\{\sigma_i'\}_{i=1}^n,\sigma_*')}[\sigma_*'\in\gamma(\sigma_\#')\mid\sigma_*'\neq\bot]\le\epsilon_{\text{true}}$. Thus, by induction, we have
\begin{align*}
\mathbb{P}_{\lambda_{\text{true}}(\sigma_\#',\{\sigma_i'\}_{i=1}^n,\sigma_*')}[\den{q}^*(\sigma_*')\not\in\gamma(\den{q}^C_{\epsilon}(\sigma_\#',\{\sigma_i'\}_{i=1}^n))\mid\sigma_*'\neq\bot]\le\epsilon_{\text{true}}+\epsilon.
\end{align*}
Combining this inequality with (\ref{eqn:impthm:cond:3}), we have
\begin{align*}
\mathbb{P}_{\lambda(\sigma_\#,\{\sigma_i\}_{i=1}^n,\sigma_*)}[\den{p}^*(\sigma_*)\not\in\gamma(\den{p}^C_{\epsilon}(\sigma_\#,\{\sigma_i\}_{i=1}^n))\mid\sigma_*\neq\bot\wedge\sigma_*(x)=\text{true}]\le\epsilon_b+\epsilon.
\end{align*}
For the term $b=\text{false}$ in (\ref{eqn:impthm:cond:2}), we have
\begin{align}
&\mathbb{P}_{\lambda(\sigma_\#,\{\sigma_i\}_{i=1}^n,\sigma_*)}[\den{p}^*(\sigma_*)\not\in\gamma(\den{p}^C_{\epsilon}(\sigma_\#,\{\sigma_i\}_{i=1}^n))\mid\sigma_*\neq\bot\wedge\sigma_*(x)=\text{false}] \nonumber \\
&=\mathbb{P}_{\lambda(\sigma_\#,\{\sigma_i\}_{i=1}^n,\sigma_*)}[\den{p}^*(\sigma_*)\not\in\gamma(\den{p}^C_{\epsilon}(\sigma_\#,\{\sigma_i\}_{i=1}^n))\mid\iota_{\text{false}}(\sigma_*)\neq\bot] \nonumber \\
&=\mathbb{P}_{\lambda(\sigma_\#,\{\sigma_i\}_{i=1}^n,\sigma_*)}[\iota_{\text{false}}(\sigma_*)\not\in\gamma(\den{p}^C_{\epsilon}(\sigma_\#,\{\sigma_i\}_{i=1}^n))\mid\iota_{\text{false}}(\sigma_*)\neq\bot] \nonumber \\
&\le\mathbb{P}_{\lambda(\sigma_\#,\{\sigma_i\}_{i=1}^n,\sigma_*)}[\iota_{\text{false}}(\sigma_*)\not\in\gamma(\iota_{\text{false}}^\#(\sigma_\#))\mid\iota_{\text{false}}(\sigma_*)\neq\bot] \nonumber \\
&=\mathbb{P}_{\lambda_{\text{false}}(\sigma_\#',\{\sigma_i'\}_{i=1}^n,\sigma_*')}[\sigma_*'\not\in\gamma(\sigma_\#')\mid\sigma_*'\neq\bot] \nonumber \\
&=\epsilon_{\text{false}}, \label{eqn:impthm:cond:4}
\end{align}
where the first equality follows since $\sigma_*\neq\bot\wedge\sigma_*(x)=\text{false}$ iff $\iota_{\text{false}}(\sigma_*)\neq\bot$, the second equality follows by definition of $\den{p}^*$ under the event that $\sigma_*(x)=\text{false}$, the first inequality follows since $\gamma(\iota_{\text{false}}^\#(\sigma_\#))\subseteq\gamma(\den{p}^C_{\epsilon}(\sigma_\#,\{\sigma_i\}_{i=1}^n))$ by definition of the join, the third equality follows by a change of measure, and the fourth equality follows by (\ref{eqn:impthm:cond:1}).

Finally, we have
\begin{align*}
\mathbb{P}_{\lambda(\sigma_\#,\{\sigma_i\}_{i=1}^n,\sigma_*)}[\den{p}^*(\sigma_*)\not\in\gamma(\den{p}^C_{\epsilon}(\sigma_\#,\{\sigma_i\}_{i=1}^n))\mid\sigma_*\neq\bot]
&\le(\epsilon_{\text{true}}+\epsilon)\cdot p_{\text{true}}+\epsilon_{\text{false}}\cdot p_{\text{false}} \\
&\le\epsilon_0+\epsilon\cdot p_{\text{true}} \\
&\le\epsilon_0+\epsilon,
\end{align*}
where the first inequality follows by combining (\ref{eqn:impthm:cond:2}) with (\ref{eqn:impthm:cond:3}) \& (\ref{eqn:impthm:cond:4}), the second inequality follows from (\ref{eqn:impthm:cond:0}), and the last inequality follows since $p_{\text{true}}\le1$. For this case, we have implicitly assumed that $p_b>0$ for each $b\in\mathbb{B}$, which is required for certain conditional probabilities to be well defined and also for the inductive hypothesis to hold. If $p_b=0$ for either $b\in\mathbb{B}$, we can simply ignore that branch in our analysis since the corresponding term equals zero in (\ref{eqn:impthm:cond:2}). The claim follows.

\paragraph{While loop $p=\textbf{while }x\text{ do }q$.}

We modify $\gamma$ so $\gamma(\bot)=\{\bot\}$. This change does not affect our semantics since it never checks whether $\bot\in\gamma(\sigma_\#)$ for any $\sigma_\#\in(\Sigma^\#\cup\{\bot\})$; it also does not affect our above results since we always condition on $\sigma_*\neq\bot$. Now, let
\begin{align*}
p_m=\mathbb{P}_{\lambda(\sigma_\#,\{\sigma_i\}_{i=1}^n,\sigma_*)}[\sigma_*^{(m)}\not\in\gamma(\sigma_\#^{(m)})\mid\sigma_*\neq\bot]
\qquad(\forall m\in\mathbb{N}\cup\{\omega\}).
\end{align*}
Then, it suffices to show $p_{\omega}\le\epsilon_0+\epsilon$, since
\begin{align*}
p_{\omega}=\mathbb{P}_{\lambda(\sigma_\#,\{\sigma_i\}_{i=1}^n,\sigma_*)}[\sigma^{(\omega)}_*\not\in\gamma(\sigma^{(\omega)}_\#)\mid\sigma_*\neq\bot]
\end{align*}
is the probability we need to bound. By Lemma~\ref{lem:while}, $\sigma_*^{(m)}\in\{\sigma_*^{(\omega)},\bot\}$ and $\sigma_\#^{(m)}\in\{\sigma_\#^{(\omega)},\bot\}$, so
\begin{align*}
\sigma_*^{(\omega)}\not\in\gamma(\sigma_\#^{(\omega)})\Rightarrow\sigma_*^{(m)}=\bot\vee\sigma_\#^{(m)}=\bot\vee\sigma_*^{(m)}\not\in\gamma(\sigma_\#^{(m)}).
\end{align*}
Thus, we have
\begin{align*}
p_{\omega}\le p_m+e_m+e_m^*
\qquad(\forall m\in\mathbb{N}),
\end{align*}
where for all $m\in\mathbb{N}$, we have defined
\begin{align*}
e_m&=\mathbb{P}_{\lambda(\sigma_\#,\{\sigma_i\}_{i=1}^n,\sigma_*)}[\sigma^{(m)}_\#=\bot\mid\sigma_*\neq\bot] \\
e_m^*&=\mathbb{P}_{\lambda(\sigma_\#,\{\sigma_i\}_{i=1}^n,\sigma_*)}[\sigma^{(m)}_*=\bot\mid\sigma_*\neq\bot].
\end{align*}
Then, we have
\begin{align*}
p_{\omega}\le\lim_{m\to\infty}(p_m+e_m+e_m^*)=\lim_{m\to\infty}p_m,
\end{align*}
where the equality holds since it is easy to check that $e_0\ge e_1\ge\cdots$ and $e_0^*\ge e_1^*\ge\cdots$, so by Assumption~\ref{assump:terminate}, we have $\lim_{m\to\infty}e_m=\lim_{m\to\infty}e_m^*=0$. Thus, it suffices to show that $\lim_{m\to\infty}p_m\le\epsilon_0+\epsilon$. Specifically, we show that for any $\lambda'\in\mathcal{M}(\Sigma^\#\times\Sigma^{n+1})$ satisfying $\pi_*(\lambda')=\otimes_{i=1}^{n+1}\mu'$ for some $\mu'\in\mathcal{M}\Sigma$, and satisfying
\begin{align*}
\mathbb{P}_{\lambda'(\sigma_\#,\{\sigma_i\}_{i=1}^n,\sigma_*)}[\sigma_*\in\gamma(\sigma_\#)\mid\sigma_*\neq\bot]\le\epsilon_0'
\end{align*}
for some $\epsilon_0'\in\mathbb{R}_{\ge0}$, and for any $\epsilon'\in\mathbb{R}_{\ge0}$, we have
\begin{align}
\label{eqn:impthm:while:1}
p_m(\lambda',\epsilon_0',\epsilon')\coloneqq\mathbb{P}_{\lambda'(\sigma_\#,\{\sigma_i\}_{i=1}^n,\sigma_*)}[\sigma_*^{(m)}\not\in\gamma(\sigma_\#^{(m)})\mid\sigma_*\neq\bot]\le\epsilon_0'+\epsilon'
\qquad(\forall m\in\mathbb{N}),
\end{align}
where $(\sigma_\#^{(m)},\{\sigma_i^{(m)}\}_{i=1}^n,\sigma_*^{(m)})=(\phi^J_{p,x})^{(m)}(h_\bot)(\sigma_\#,\{\sigma_i\}_{i=1}^n,\sigma_*)$. Then, the claim follows, since we have $p_m=p_m(\lambda,\epsilon_0,\epsilon)\le\epsilon_0+\epsilon$ for all $m\in\mathbb{N}$. We prove by induction on $m$. First, note that the base case $m=0$ follows trivially, since letting $(\sigma_\#^{(0)},\{\sigma_i^{(0)}\}_{i=1}^n,\sigma_*^{(0)})=h_\bot(\sigma_\#,\{\sigma_i\}_{i=1}^n,\sigma_*)$, then by definition, we have $\sigma_\#^{(0)}=\sigma_*^{(0)}=\bot$, so
\begin{align*}
p_0=\mathbb{P}_{\lambda'(\sigma_\#,\{\sigma_i\}_{i=1}^n,\sigma_*)}[\bot\not\in\gamma(\bot)\mid\sigma_*\neq\bot]=0\le\epsilon_0'.
\end{align*}
For the inductive case, we introduce the production $\vec{x}\coloneqq h_m(\vec{x})$, and define its semantics to be
\begin{align*}
\den{\vec{x}\coloneqq h_m(\vec{x})}^J_{\epsilon'}(\sigma_\#,\{\sigma_i\}_{i=1}^n,\sigma_*)
=h_m(\sigma_\#,\{\sigma_i\}_{i=1}^n,\epsilon',\sigma_*).
\end{align*}
By the inductive hypothesis (\ref{eqn:impthm:while:1}), this component satisfies the statement of Theorem~\ref{thm:impmain}. With this new component, $\phi_{q,x}^J=\phi_{q,x}^C\times\phi_{q,x}^*$ can equivalently be defined by
\begin{align*}
h_{m+1}(\sigma_\#,\{\sigma_i\}_{i=1}^n,\epsilon',\sigma_*)=\phi_{q,x}^J(h)(\sigma_\#,\{\sigma_i\}_{i=1}^n,\epsilon',\sigma_*)=\den{r}^J_{\epsilon'}(\sigma_\#,\{\sigma_i\}_{i=1}^n,\sigma_*),
\end{align*}
where
\begin{align*}
r=\textbf{if }x\textbf{ then }(q;\vec{x}\coloneqq h_m(\vec{x})).
\end{align*}
We can use structural induction on $r$ since it does not refer to $p$ except for its child $q$, and we are assuming the statement of Theorem~\ref{thm:impmain} holds for $q$. The conditional in $r$ can be handled directly using the argument above for conditionals, and the assignment $\vec{x}\coloneqq h(\vec{x})$, is handled by (\ref{eqn:impthm:while:1}) as discussed above. Then, letting $(\sigma_\#^{(m+1)},\{\sigma_i^{(m+1)}\}_{i=1}^m,\sigma_*^{(m+1)})=h_{m+1}(\sigma_\#,\{\sigma_i\}_{i=1}^n,\epsilon,\sigma_*)$, we can apply structural induction to $r$ to obtain
\begin{align*}
\mathbb{P}_{\lambda'(\sigma_\#,\{\sigma_i\}_{i=1}^n,\sigma_*)}[\sigma_*^{(m+1)}\in\gamma(\sigma_\#^{(m+1)})\mid\sigma_*\neq\bot]\le\epsilon_0'+\epsilon',
\end{align*}
which is exactly the induction hypothesis (\ref{eqn:impthm:while:1}) we need to prove for $m+1$. The claim follows.
\end{proof}
Now, we prove Theorem~\ref{thm:impmain}. Given $\lambda\in\mathcal{M}(\Sigma^\#\times\Sigma^{n+1})$ as in the statement of Theorem~\ref{thm:impmain}, we can trivially extend it to a measure $\lambda\in\mathcal{M}((\Sigma\bot\{\bot\})^\#\times(\Sigma\cup\{\bot\})^{n+1})$ (by placing zero measure on $\bot$ in each component). Then, $\lambda$ satisfies the requirements of Lemma~\ref{lem:impmain} with $\mathbb{P}_{\lambda(\sigma_\#,\{\sigma_i\}_{i=1}^n,\sigma_*)}[\sigma_*\neq\bot]=1>0$ and taking $\epsilon_0$. Thus, by Lemma~\ref{lem:impmain}, we have
\begin{align*}
\mathbb{P}_{\lambda(\sigma_\#,\{\sigma_i\}_{i=1}^n,\sigma_*)}[\den{p}^*(\sigma_*)\not\in\gamma(\den{p}^C_{\epsilon}(\sigma_\#,\{\sigma_i\}_{i=1}^n))\mid\sigma_*\neq\bot]\le\epsilon,
\end{align*}
which implies
\begin{align*}
\mathbb{P}_{\lambda(\sigma_\#,\{\sigma_i\}_{i=1}^n,\sigma_*)}[\den{p}^*(\sigma_*)\in\gamma(\den{p}^C_{\epsilon}(\sigma_\#,\{\sigma_i\}_{i=1}^n))\mid\sigma_*\neq\bot]\ge1-\epsilon.
\end{align*}
Finally, we have $\mathbb{P}_{\lambda(\sigma_\#,\{\sigma_i\}_{i=1}^n,\sigma_*)}[\sigma_*=\bot]=0$, so
\begin{align*}
&\mathbb{P}_{\lambda(\sigma_\#,\{\sigma_i\}_{i=1}^n,\sigma_*)}[\den{p}^*(\sigma_*)\in\gamma(\den{p}^C_{\epsilon}(\sigma_\#,\{\sigma_i\}_{i=1}^n))] \\
&=\mathbb{P}_{\lambda(\sigma_\#,\{\sigma_i\}_{i=1}^n,\sigma_*)}[\den{p}^*(\sigma_*)\in\gamma(\den{p}^C_{\epsilon}(\sigma_\#,\{\sigma_i\}_{i=1}^n))\mid\sigma_*\neq\bot] \\
&\le1-\epsilon
\end{align*}
as claimed. \hfill $\qed$

\section{Additional Results}
\label{sec:additionalresults}

\begin{figure}[t]
\centering
\begin{subfigure}[b]{0.3\textwidth}
\centering
\includegraphics[width=\textwidth]{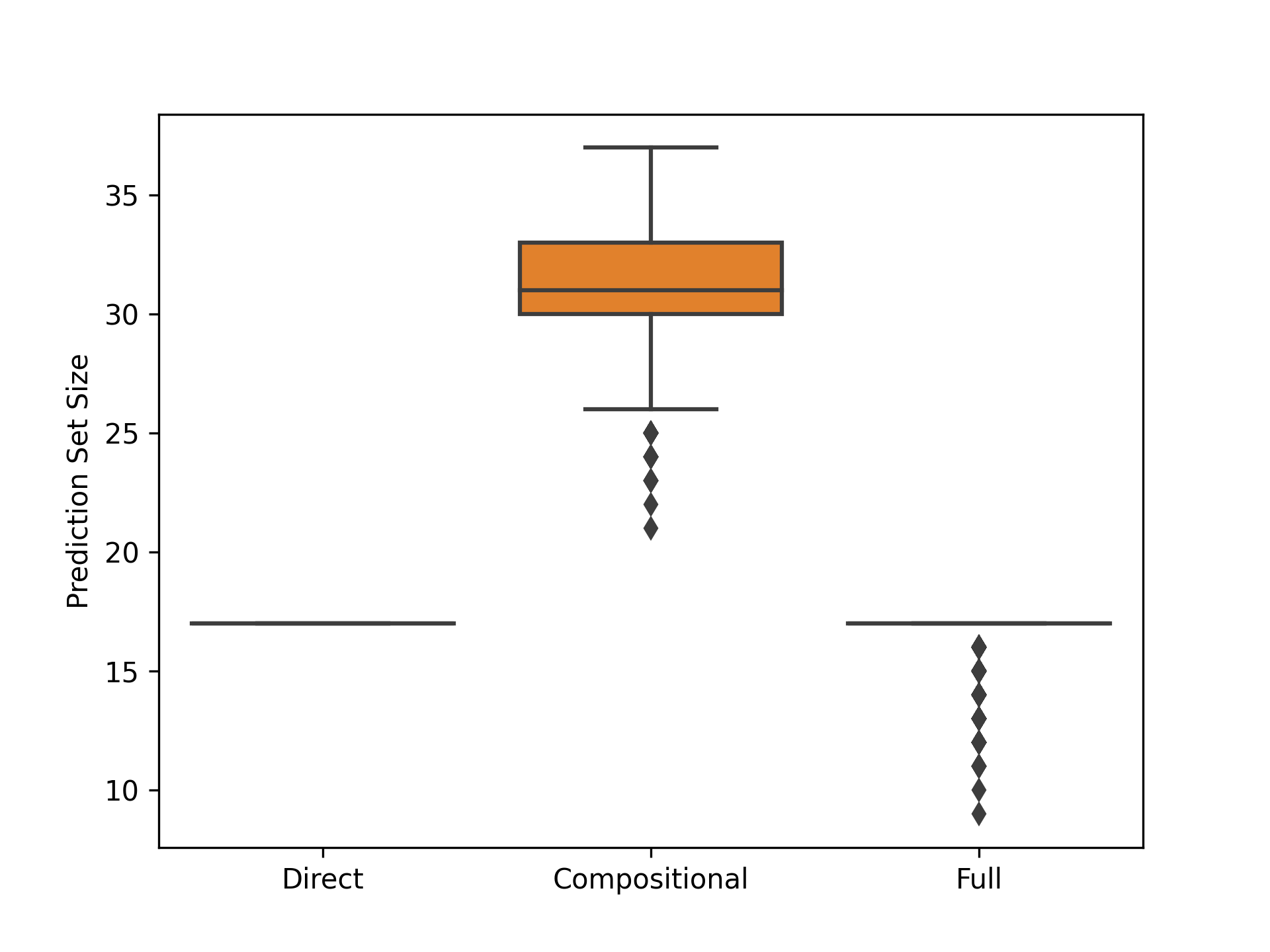}
\caption{Program 1}
\label{fig:prog-1}
\end{subfigure}
\hfill
\begin{subfigure}[b]{0.3\textwidth}
\centering
\includegraphics[width=\textwidth]{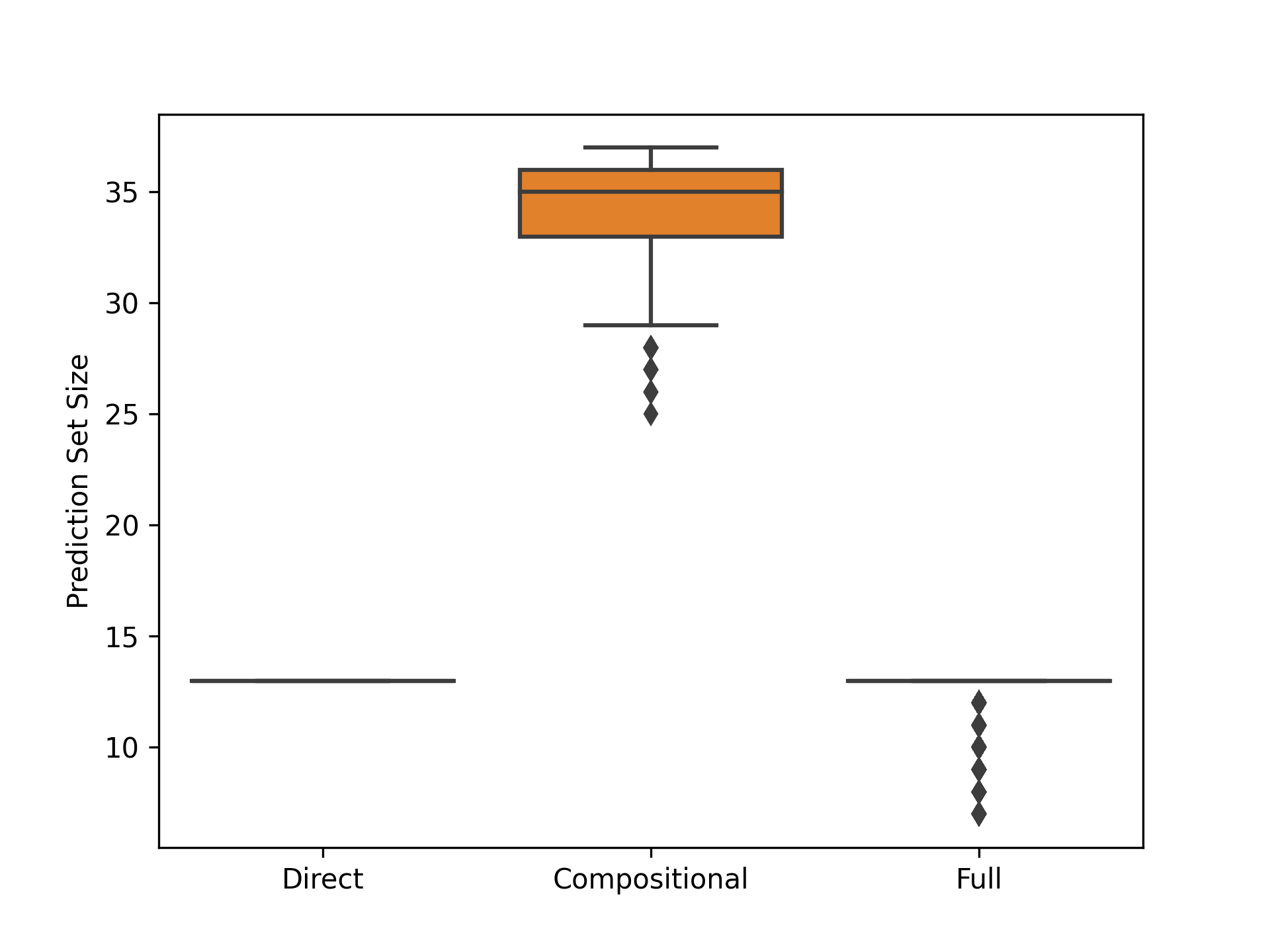}
\caption{Program 2}
\label{fig:prog-2}
\end{subfigure}
\hfill
\begin{subfigure}[b]{0.3\textwidth}
\centering
\includegraphics[width=\textwidth]{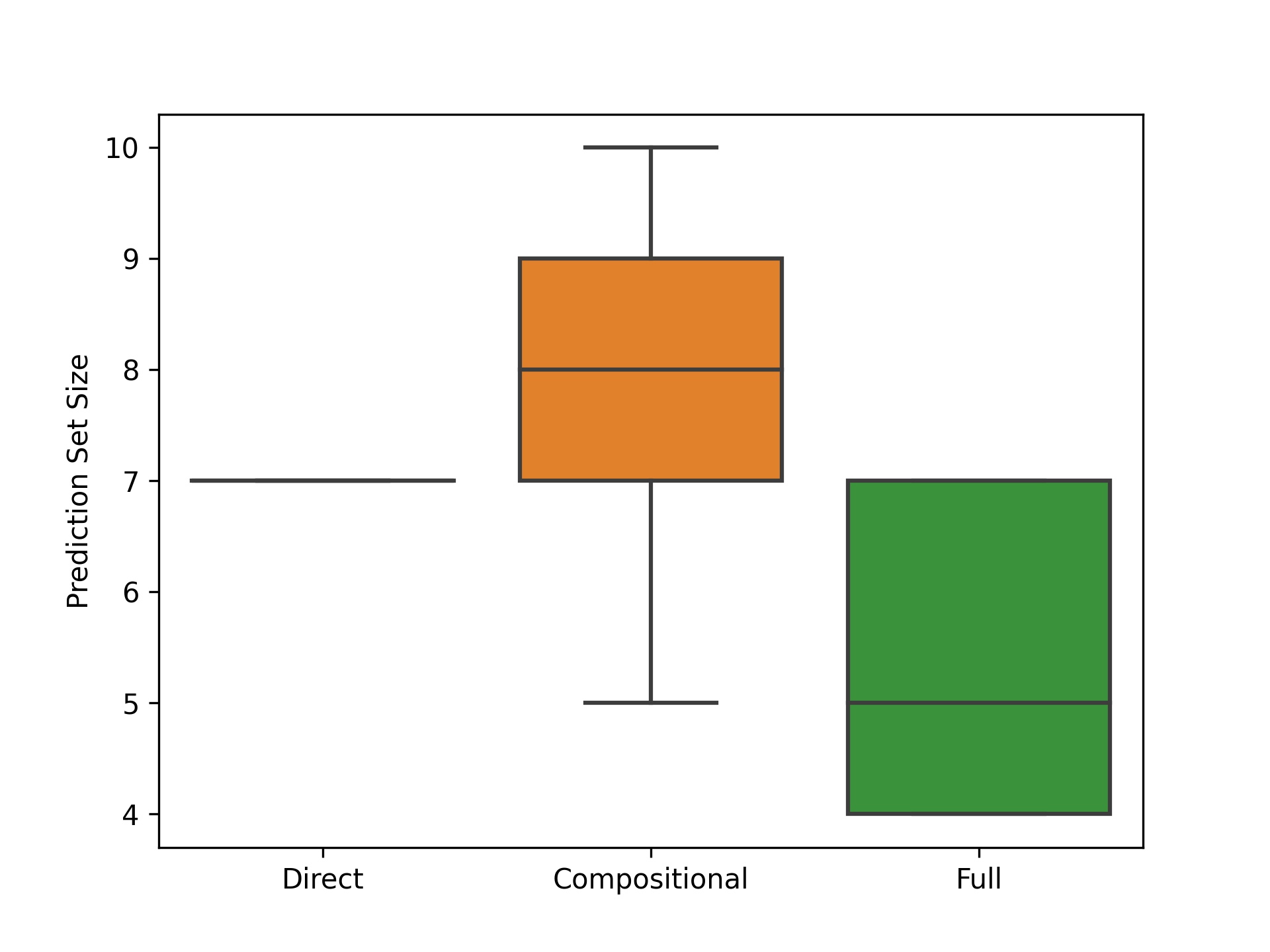}
\caption{Program 3}
\label{fig:prog-3}
\end{subfigure}
\vfill \vfill \vfill
\begin{subfigure}[b]{0.3\textwidth}
\centering
\includegraphics[width=\textwidth]{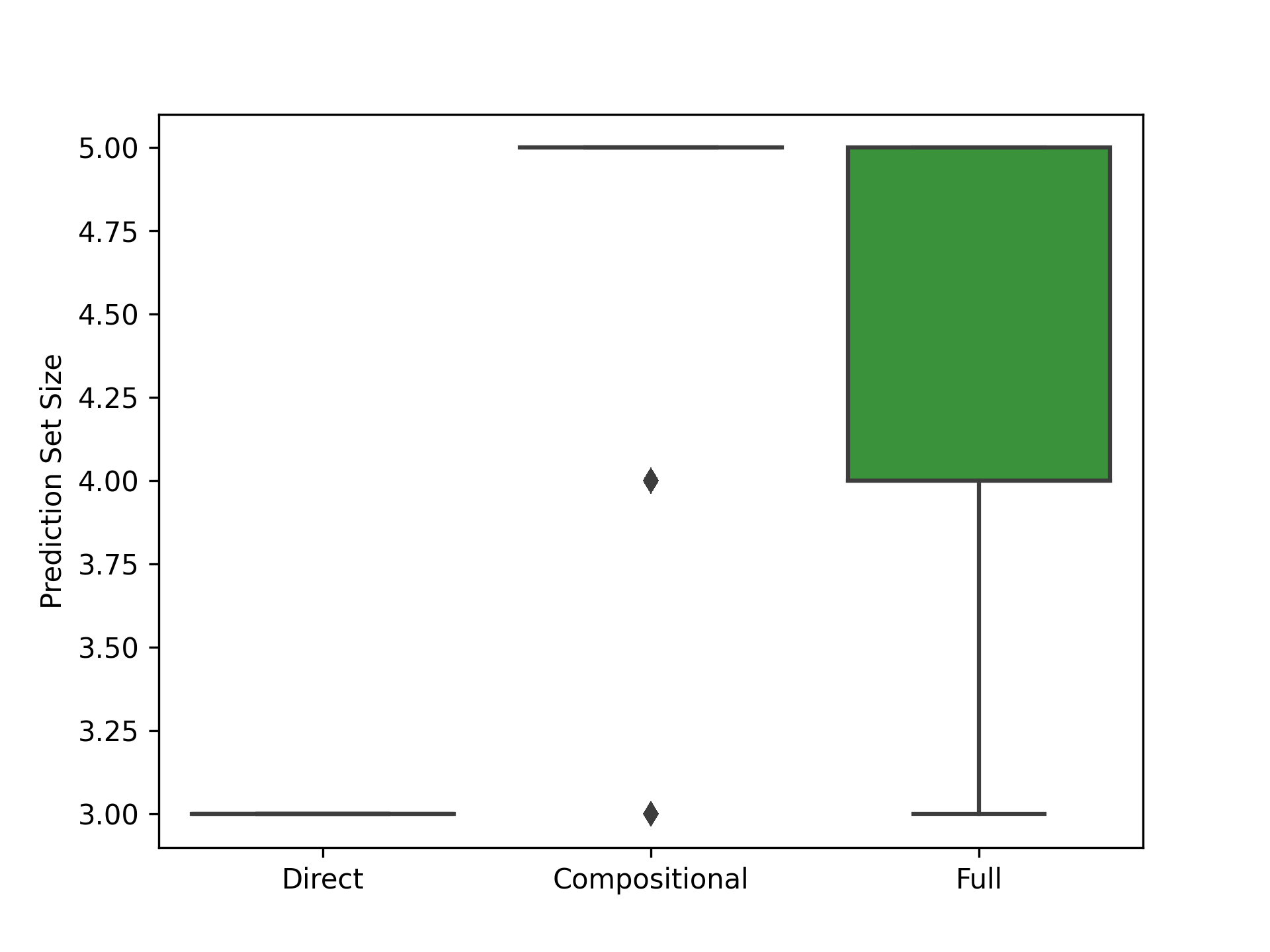}
\caption{Program 4}
\label{fig:prog-4}
\end{subfigure}
\hfill
\begin{subfigure}[b]{0.3\textwidth}
\centering
\includegraphics[width=\textwidth]{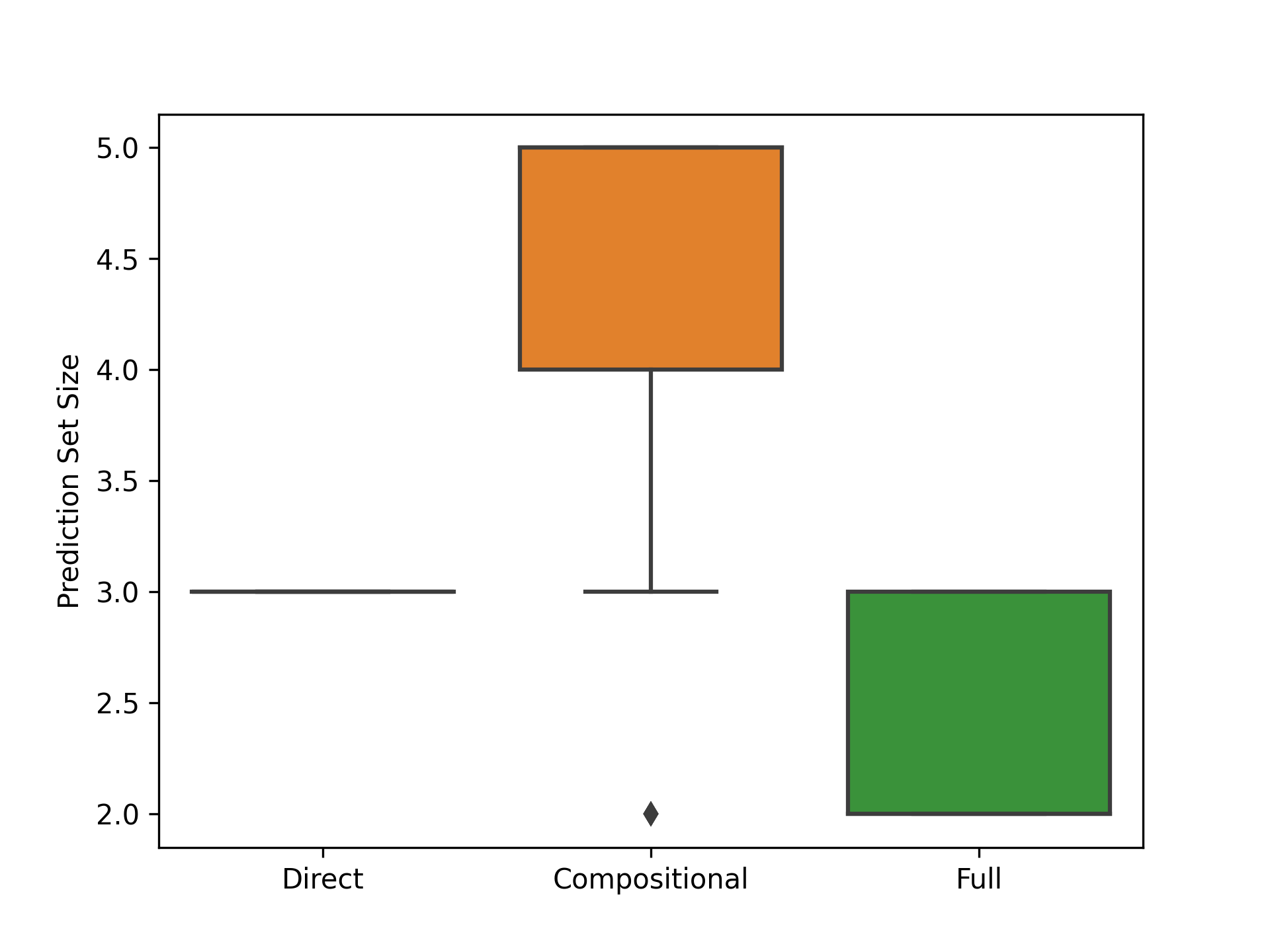}
\caption{Program 5}
\label{fig:prog-5}
\end{subfigure}
\hfill
\begin{subfigure}[b]{0.3\textwidth}
\centering
\includegraphics[width=\textwidth]{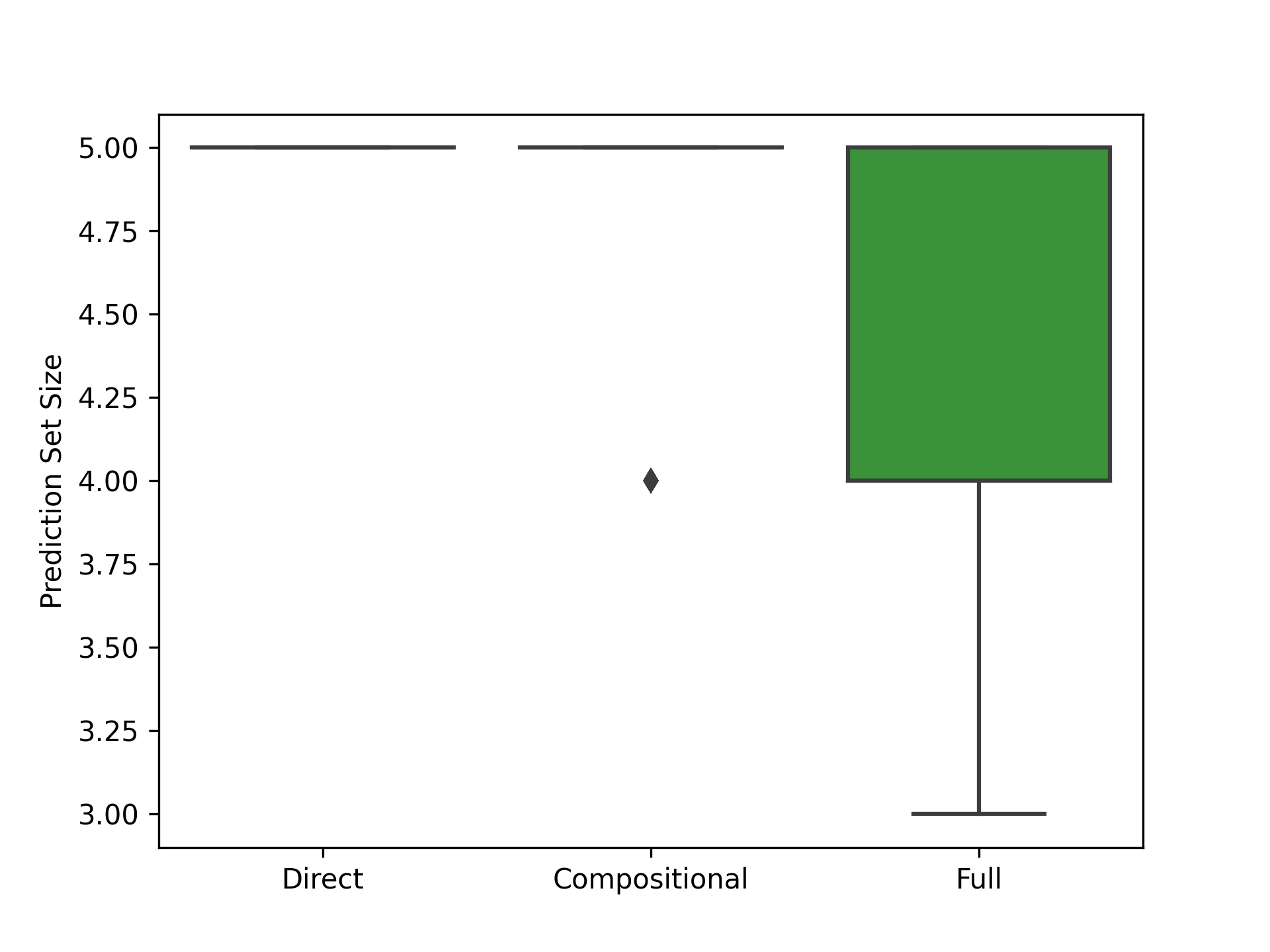}
\caption{Program 6}
\label{fig:prog-6}
\end{subfigure}
\vfill \vfill \vfill
\begin{subfigure}[b]{0.3\textwidth}
\centering
\includegraphics[width=\textwidth]{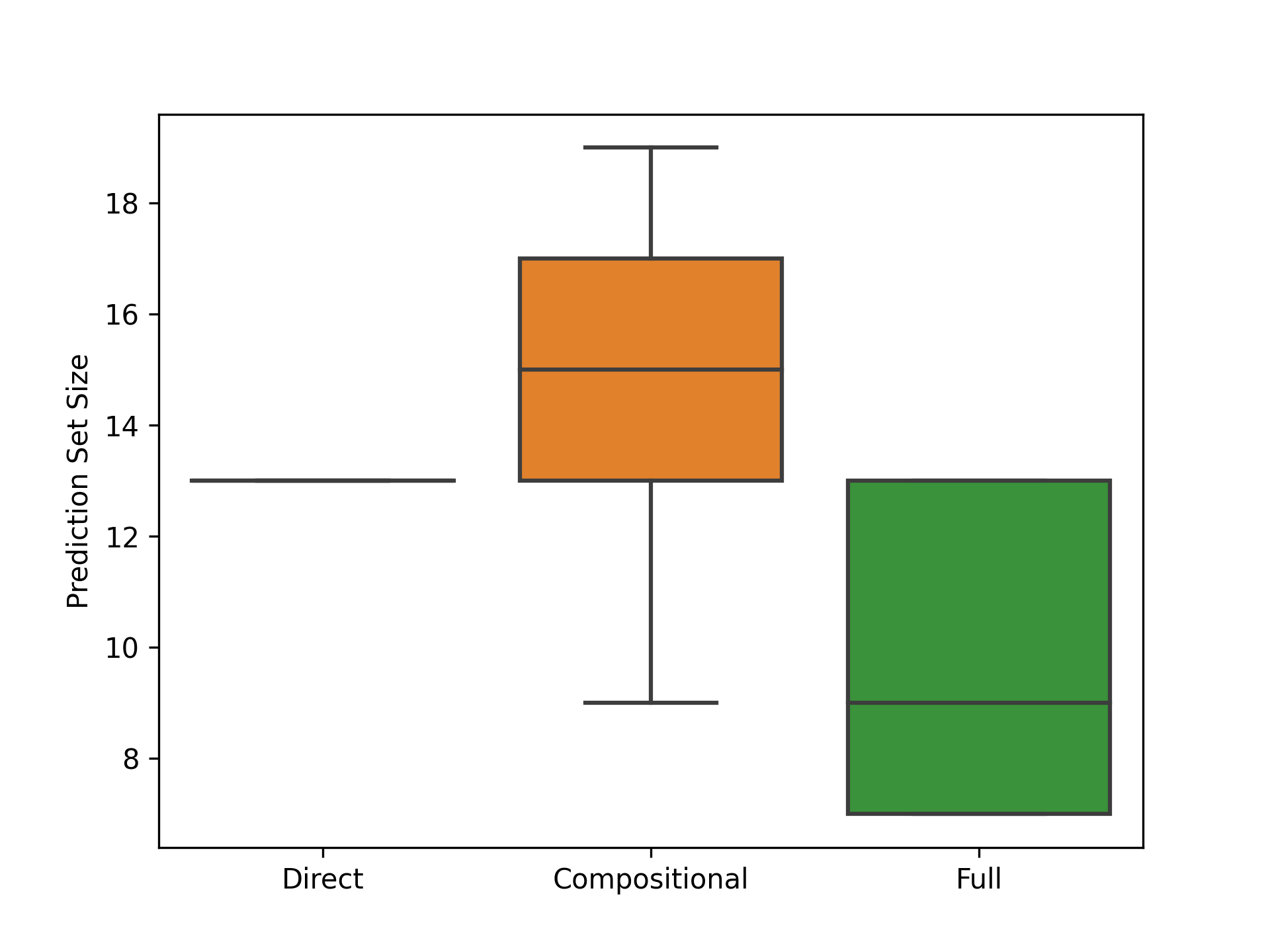}
\caption{Program 7}
\label{fig:prog-7}
\end{subfigure}
\hfill
\begin{subfigure}[b]{0.3\textwidth}
\centering
\includegraphics[width=\textwidth]{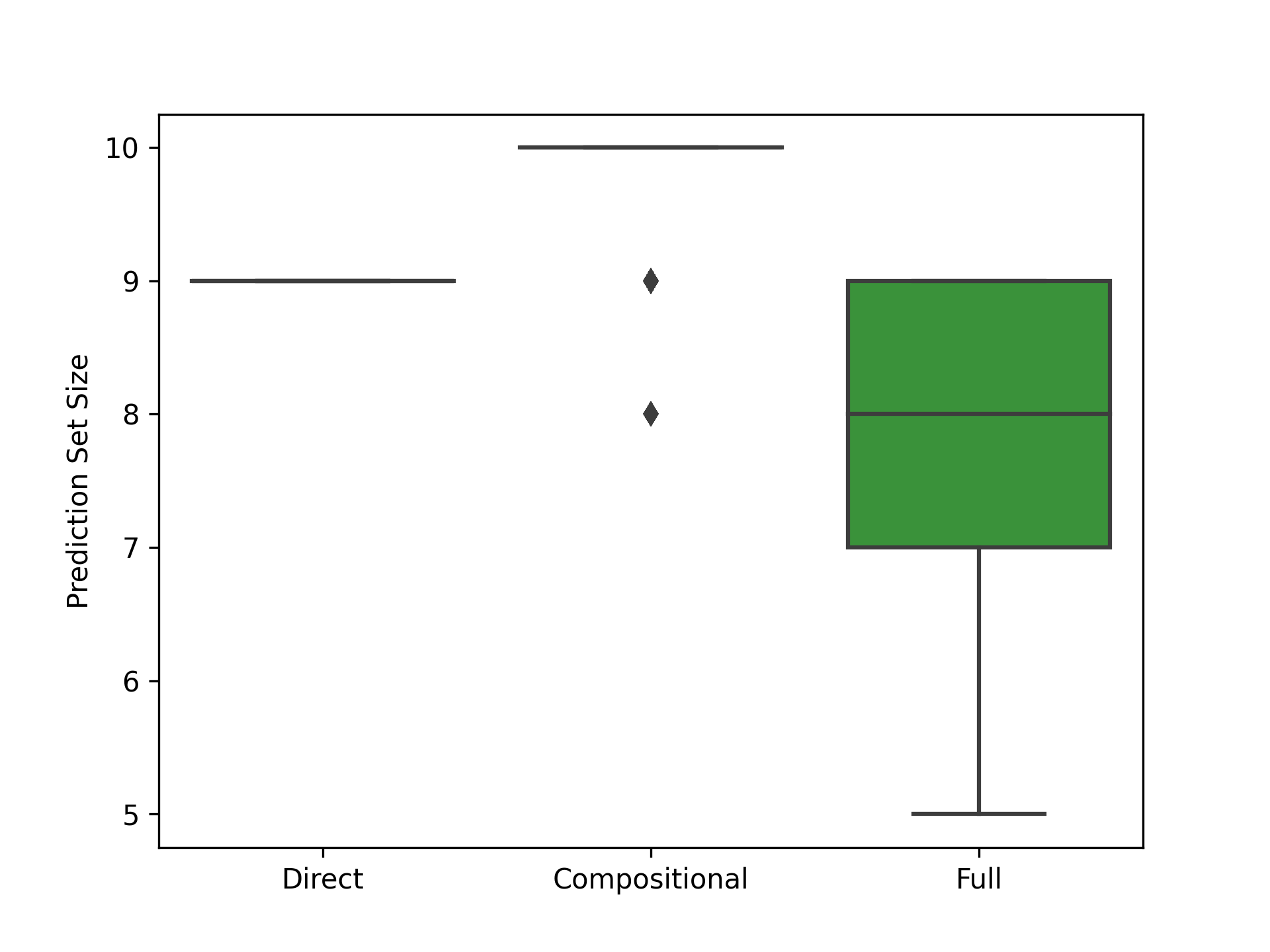}
\caption{Program 8}
\label{fig:prog-8}
\end{subfigure}
\hfill
\begin{subfigure}[b]{0.3\textwidth}
\centering
\includegraphics[width=\textwidth]{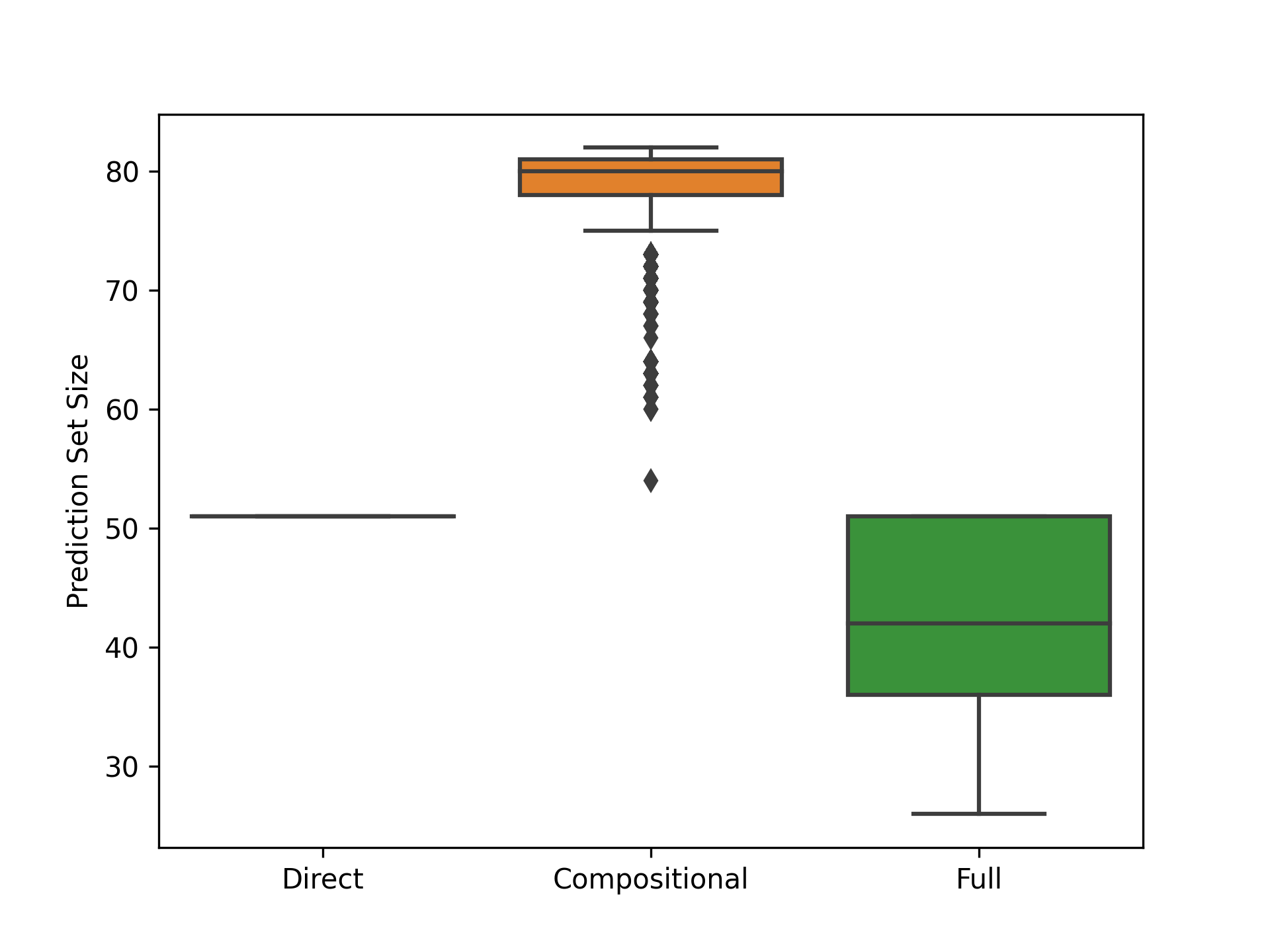}
\caption{Program 9}
\label{fig:prog-9}
\end{subfigure}
\vfill \vfill \vfill
\begin{subfigure}[b]{0.3\textwidth}
\centering
\includegraphics[width=\textwidth]{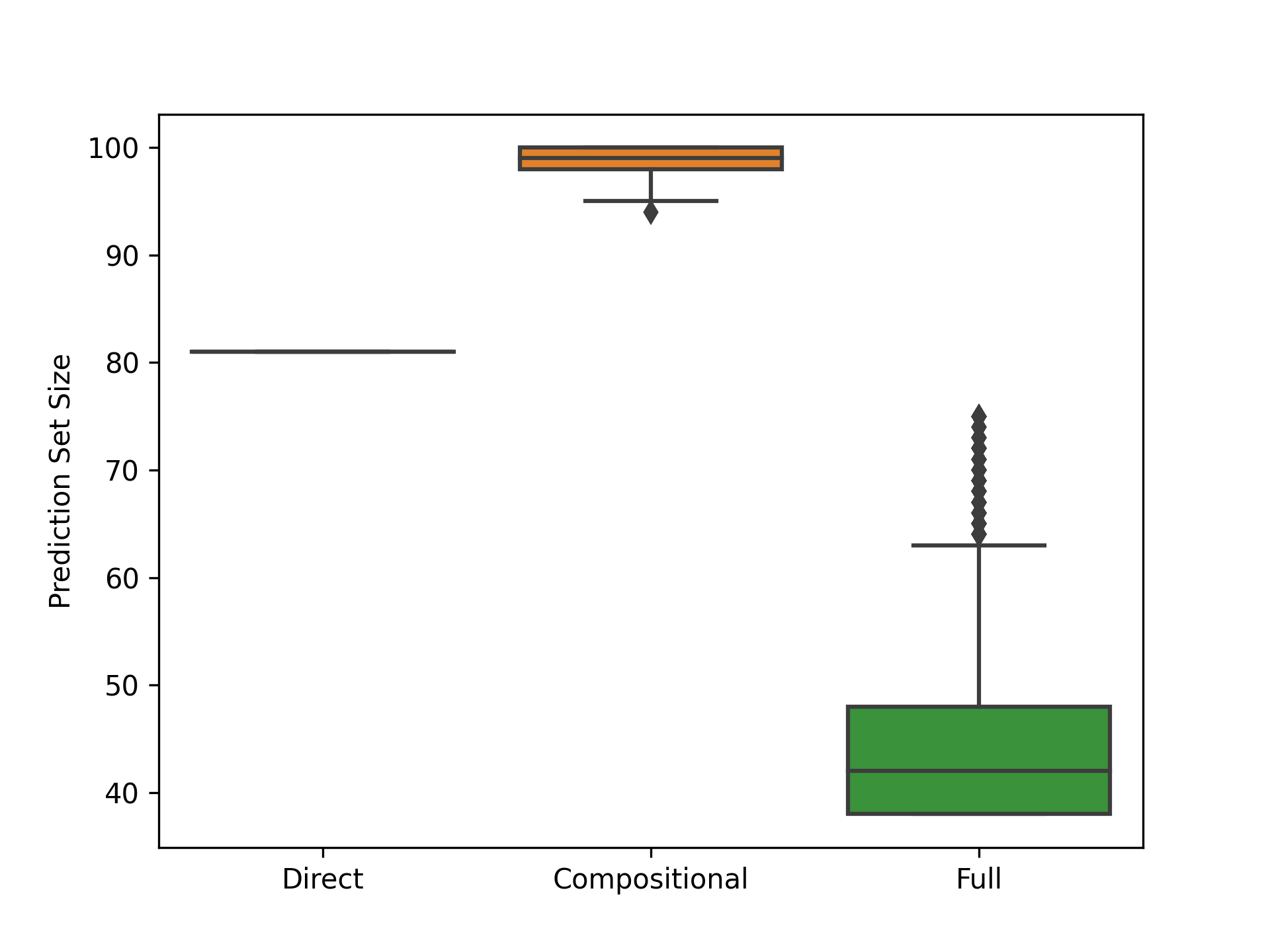}
\caption{Program 10}
\label{fig:prog-10}
\end{subfigure}
\caption{Box plots of prediction set sizes across all test examples for each program in Section~\ref{sec:mnist}. Note that for a fixed calibration set, the direct approach always produces prediction sets of the same size.}
\label{fig:MNIST-Set-Size}
\end{figure}

\begin{figure}[t]
\centering
\begin{subfigure}[b]{0.3\textwidth}
\centering
\includegraphics[width=\textwidth]{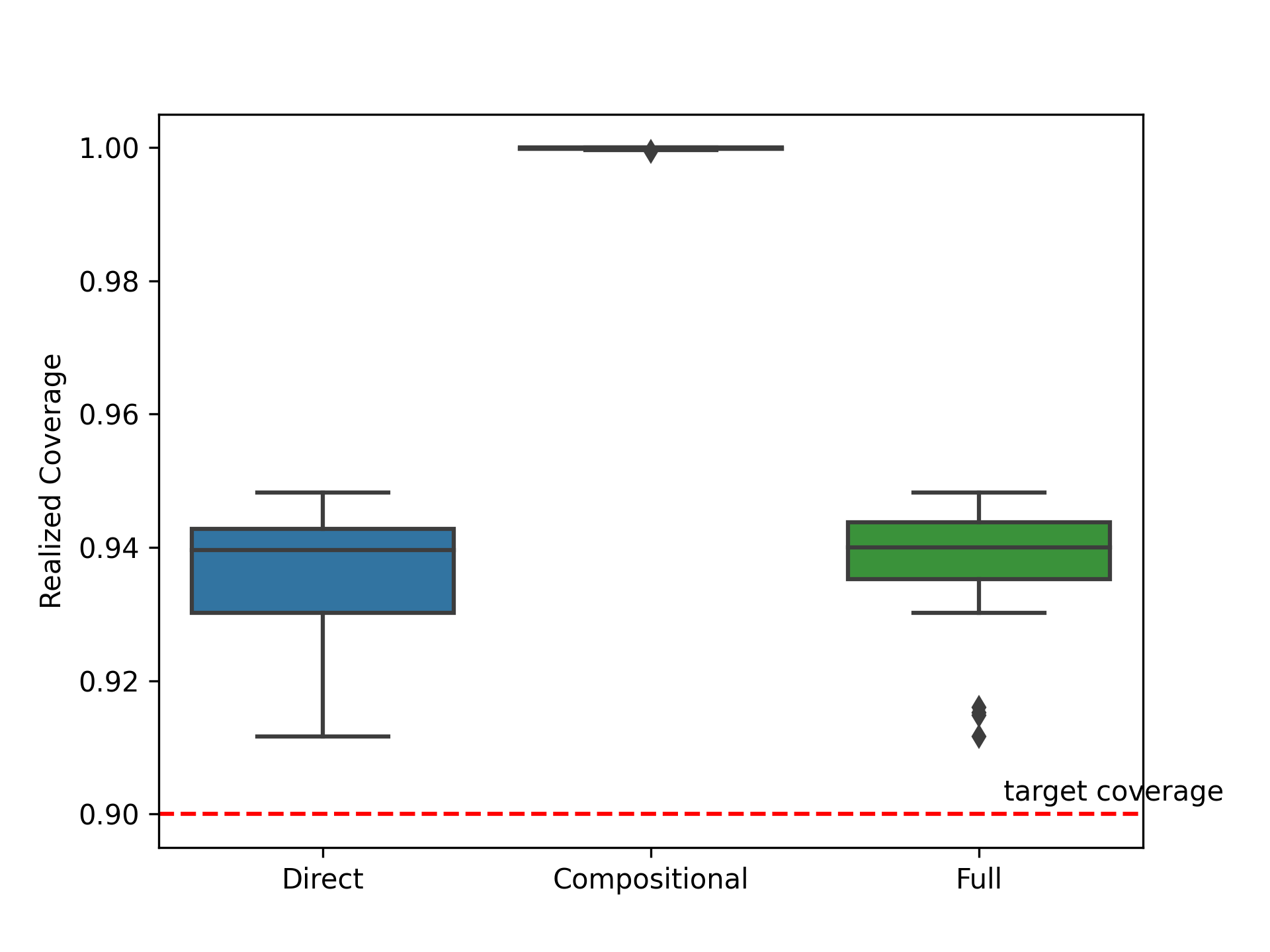}
\caption{Program 1}
\label{fig:prog-1}
\end{subfigure}
\hfill
\begin{subfigure}[b]{0.3\textwidth}
\centering
\includegraphics[width=\textwidth]{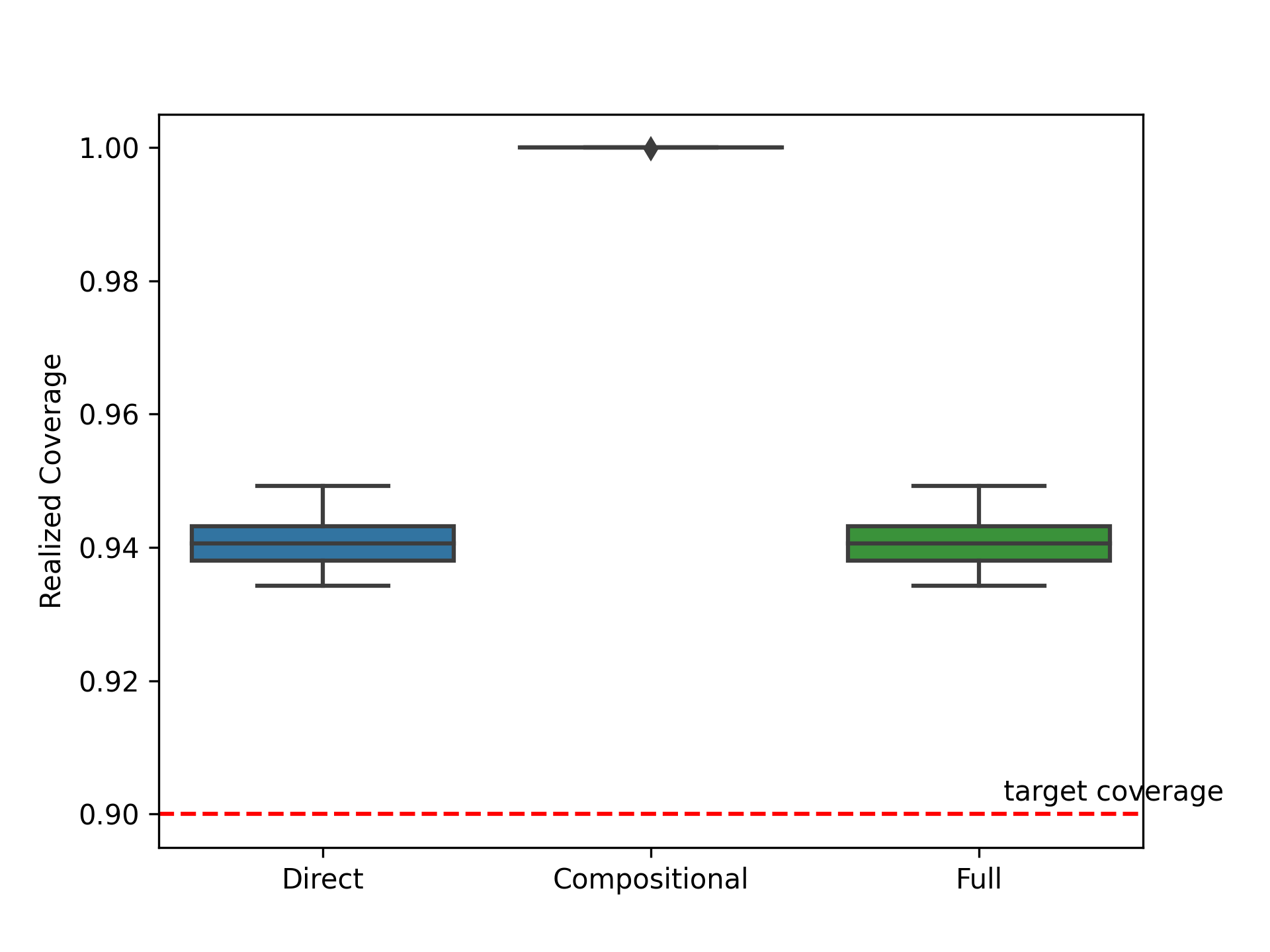}
\caption{Program 2}
\label{fig:prog-2}
\end{subfigure}
\hfill
\begin{subfigure}[b]{0.3\textwidth}
\centering
\includegraphics[width=\textwidth]{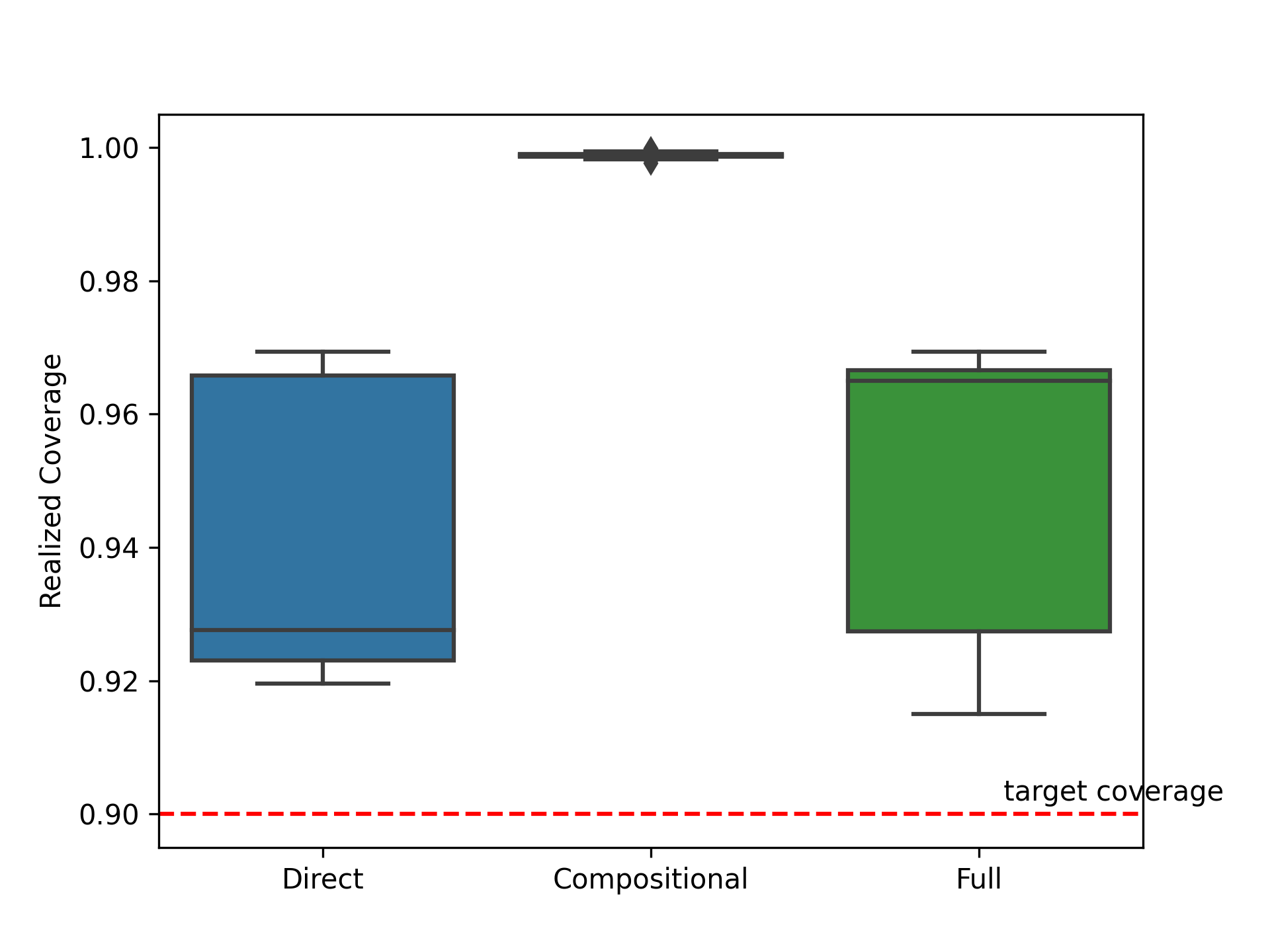}
\caption{Program 3}
\label{fig:prog-3}
\end{subfigure}
\vfill \vfill \vfill
\begin{subfigure}[b]{0.3\textwidth}
\centering
\includegraphics[width=\textwidth]{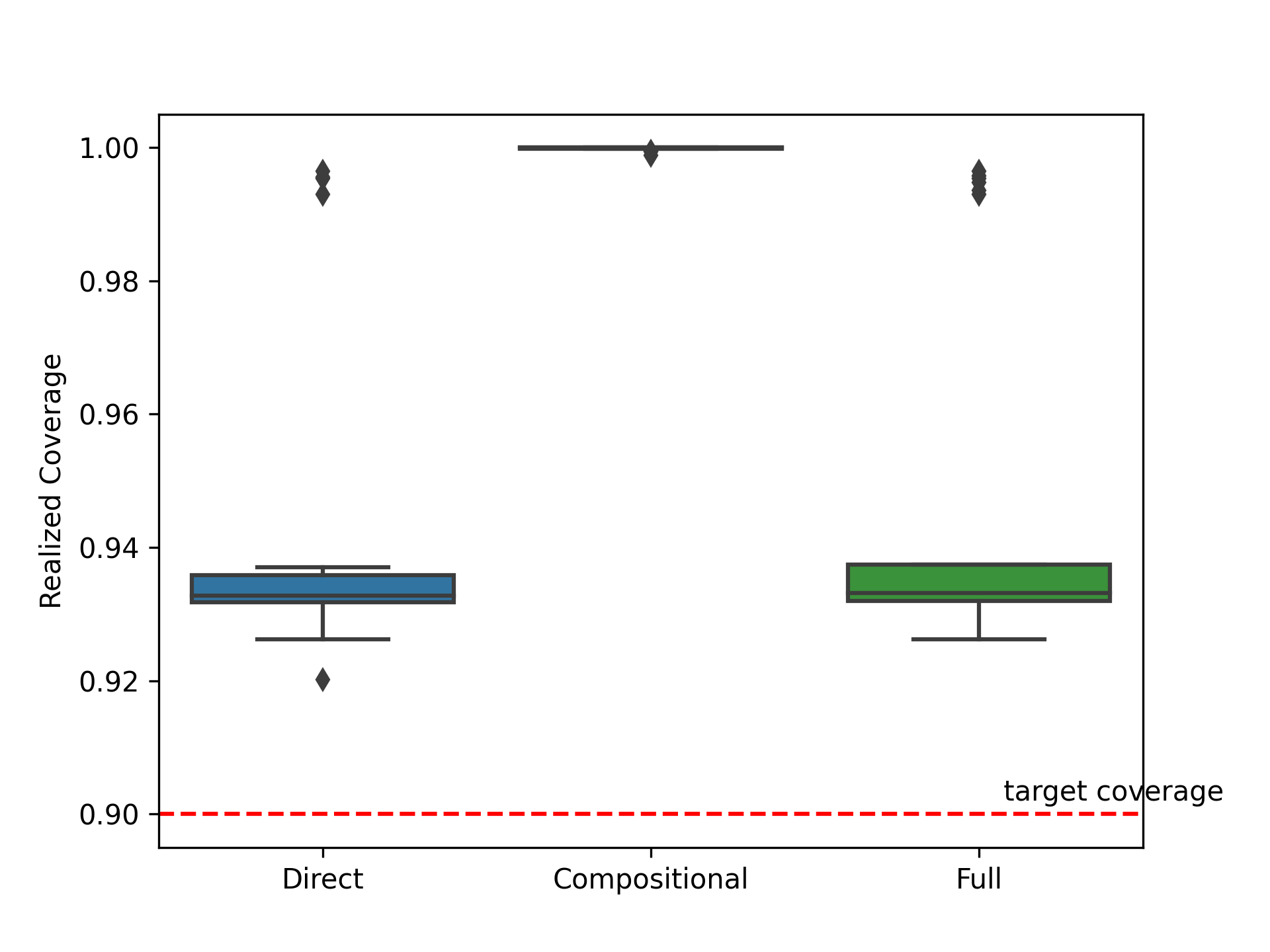}
\caption{Program 4}
\label{fig:prog-4}
\end{subfigure}
\hfill
\begin{subfigure}[b]{0.3\textwidth}
\centering
\includegraphics[width=\textwidth]{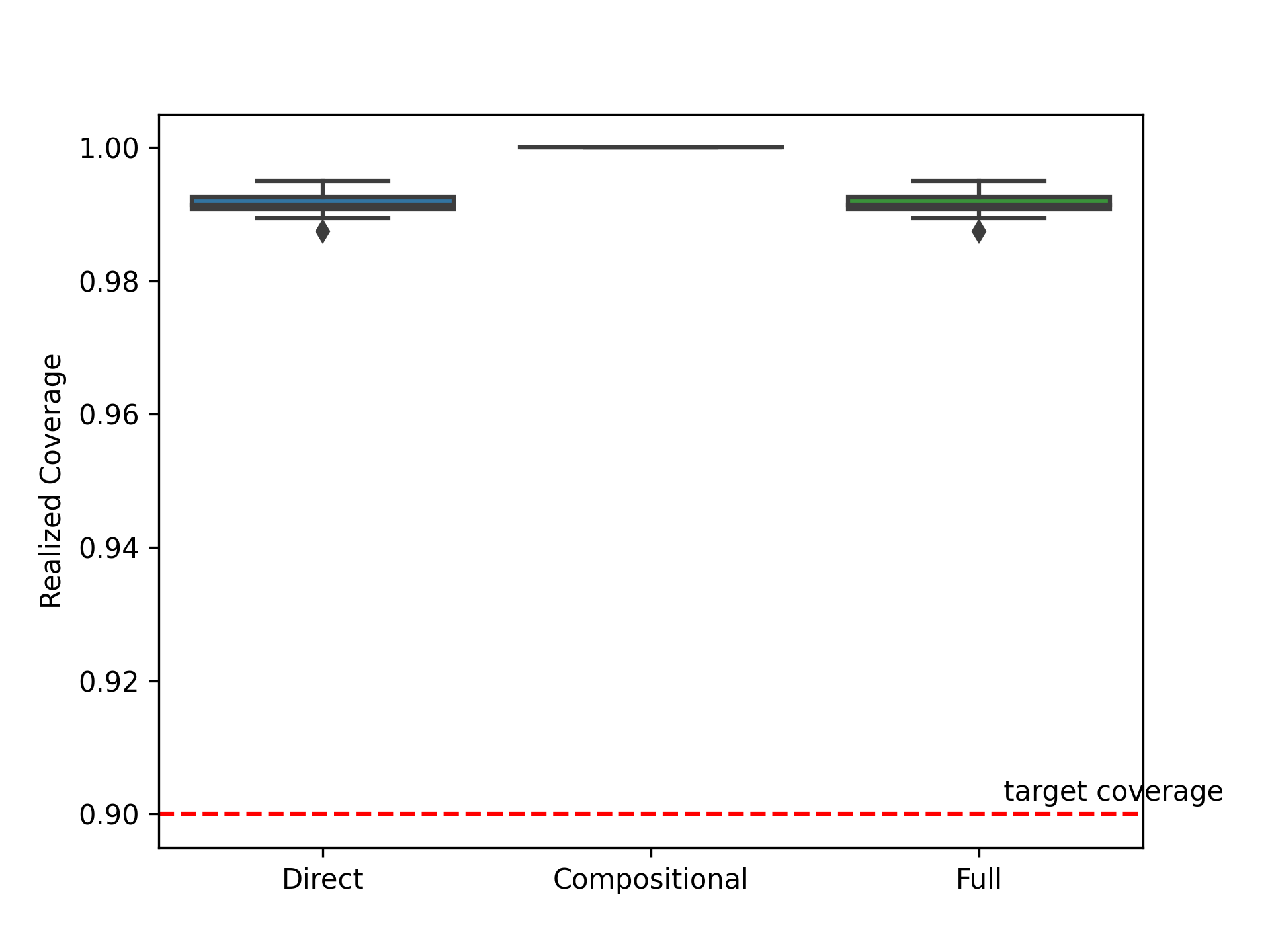}
\caption{Program 5}
\label{fig:prog-5}
\end{subfigure}
\hfill
\begin{subfigure}[b]{0.3\textwidth}
\centering
\includegraphics[width=\textwidth]{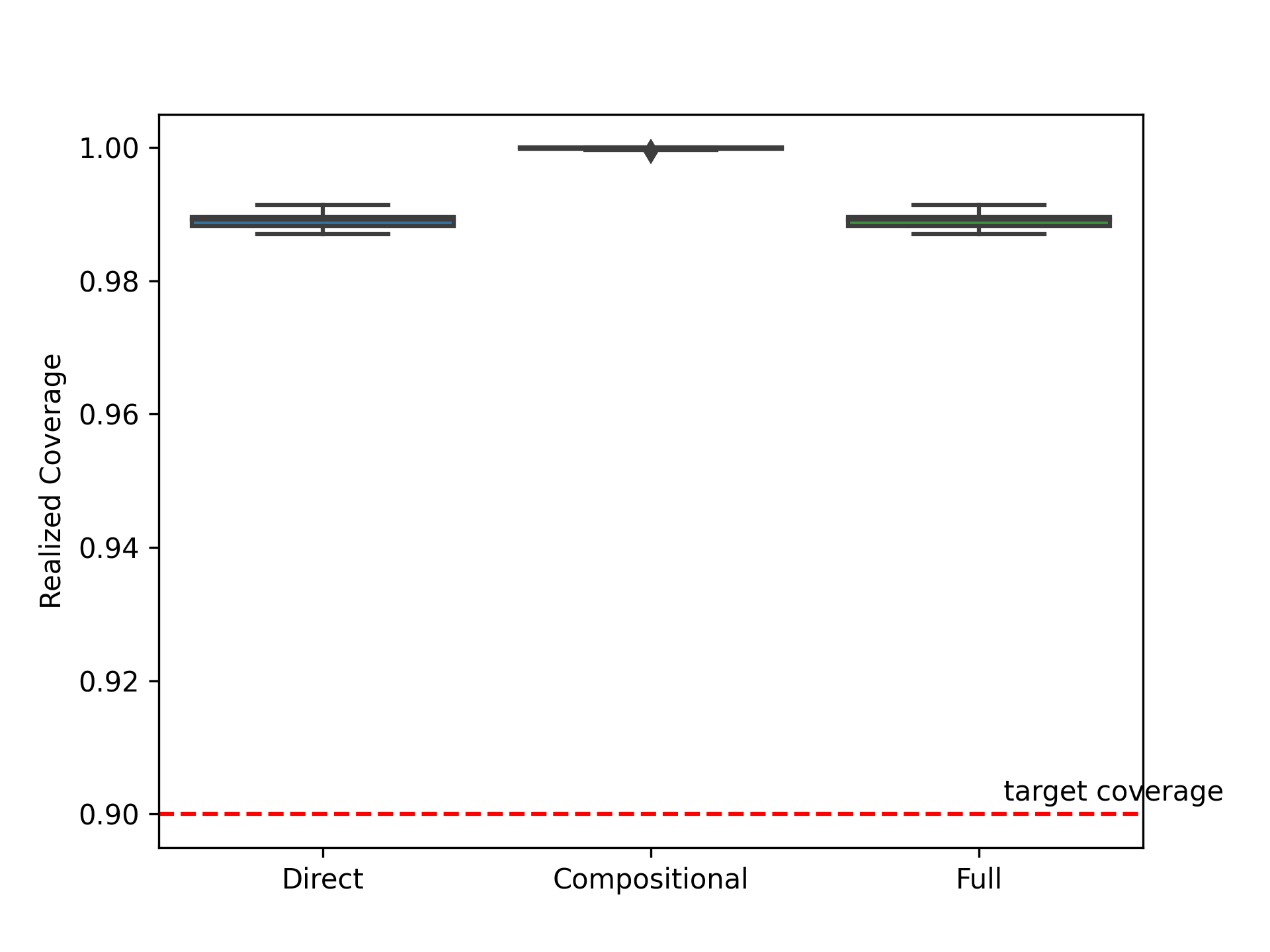}
\caption{Program 6}
\label{fig:prog-6}
\end{subfigure}
\vfill \vfill \vfill
\begin{subfigure}[b]{0.3\textwidth}
\centering
\includegraphics[width=\textwidth]{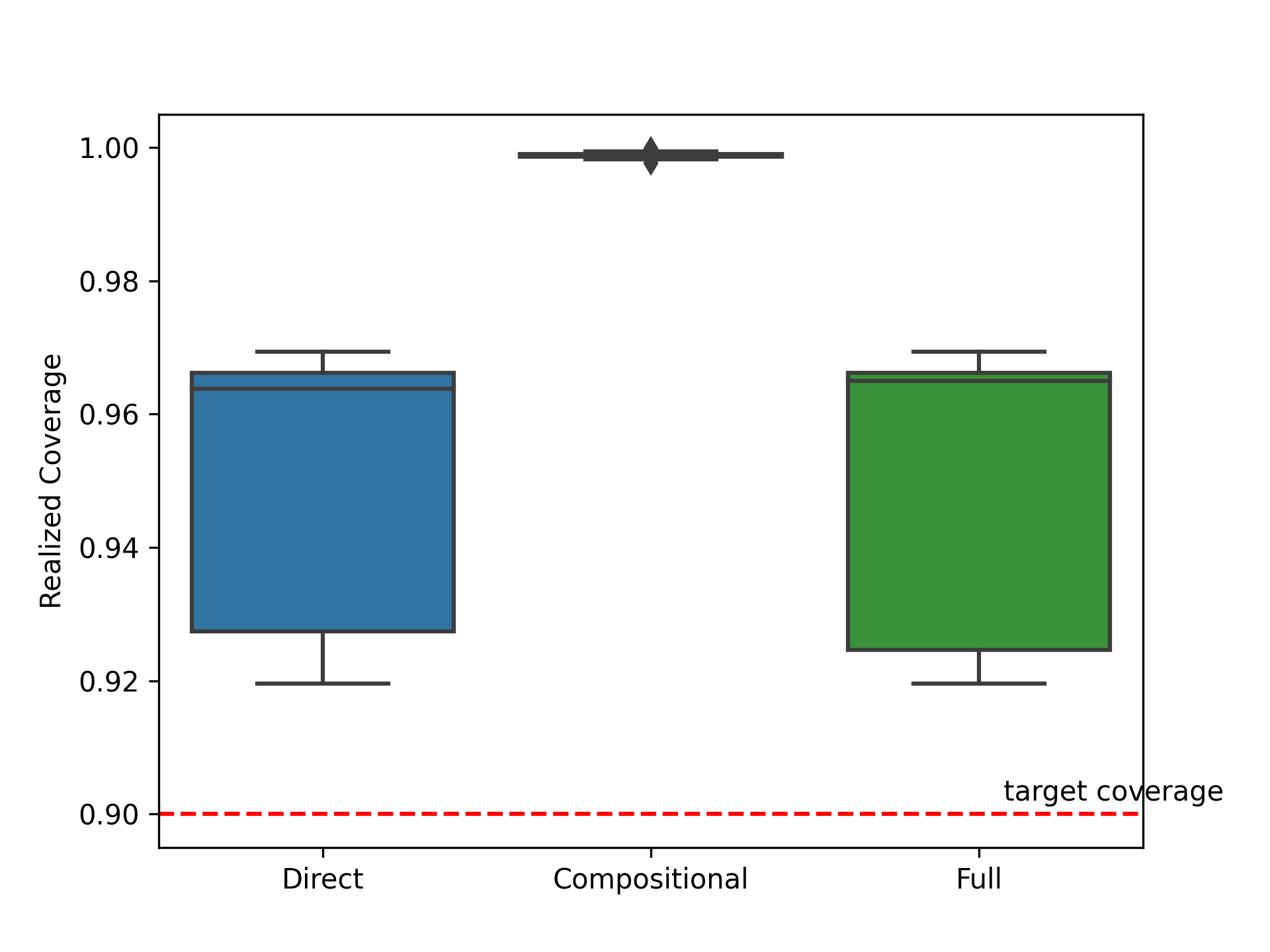}
\caption{Program 7}
\label{fig:prog-7}
\end{subfigure}
\hfill
\begin{subfigure}[b]{0.3\textwidth}
\centering
\includegraphics[width=\textwidth]{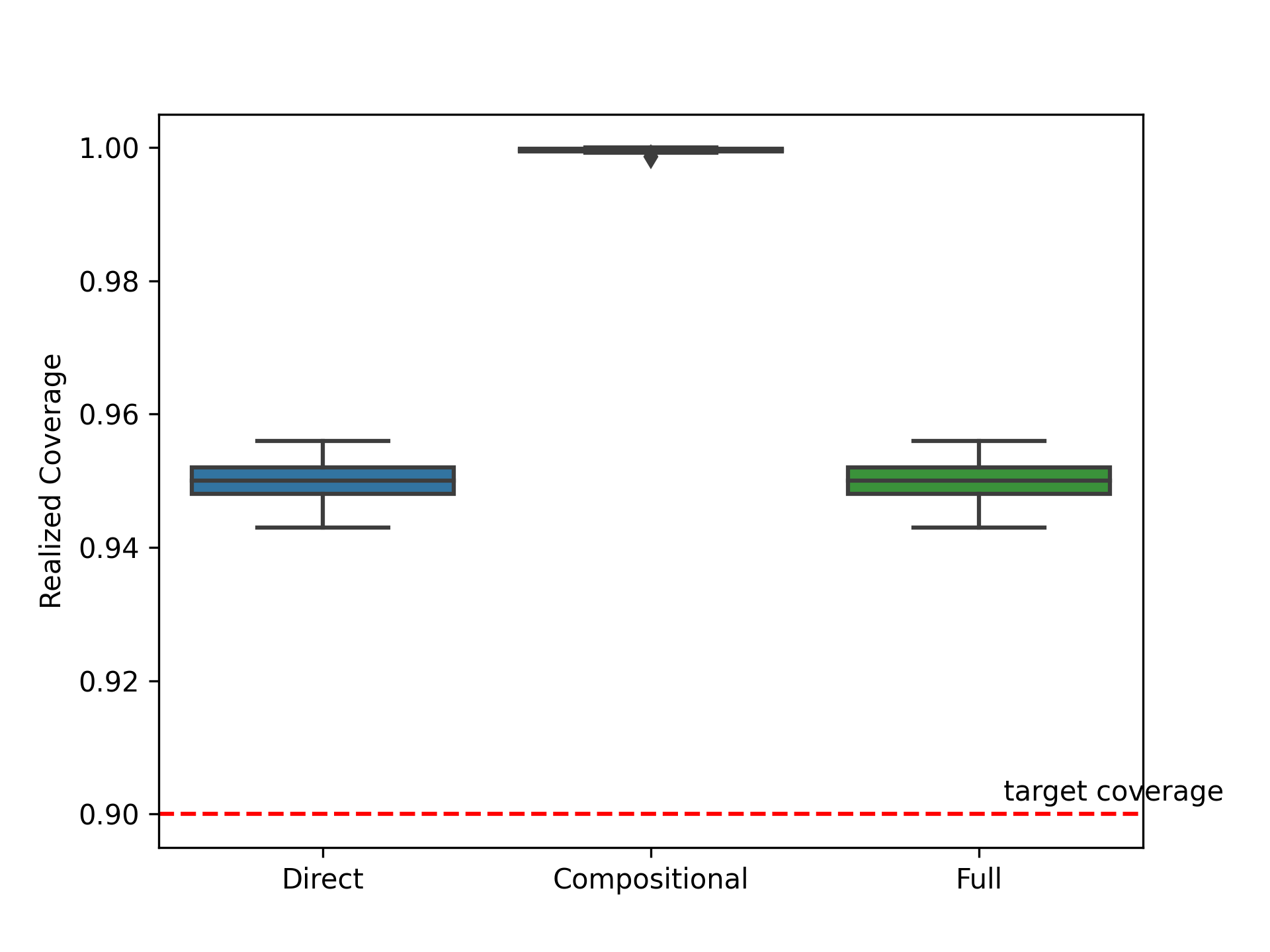}
\caption{Program 8}
\label{fig:prog-8}
\end{subfigure}
\hfill
\begin{subfigure}[b]{0.3\textwidth}
\centering
\includegraphics[width=\textwidth]{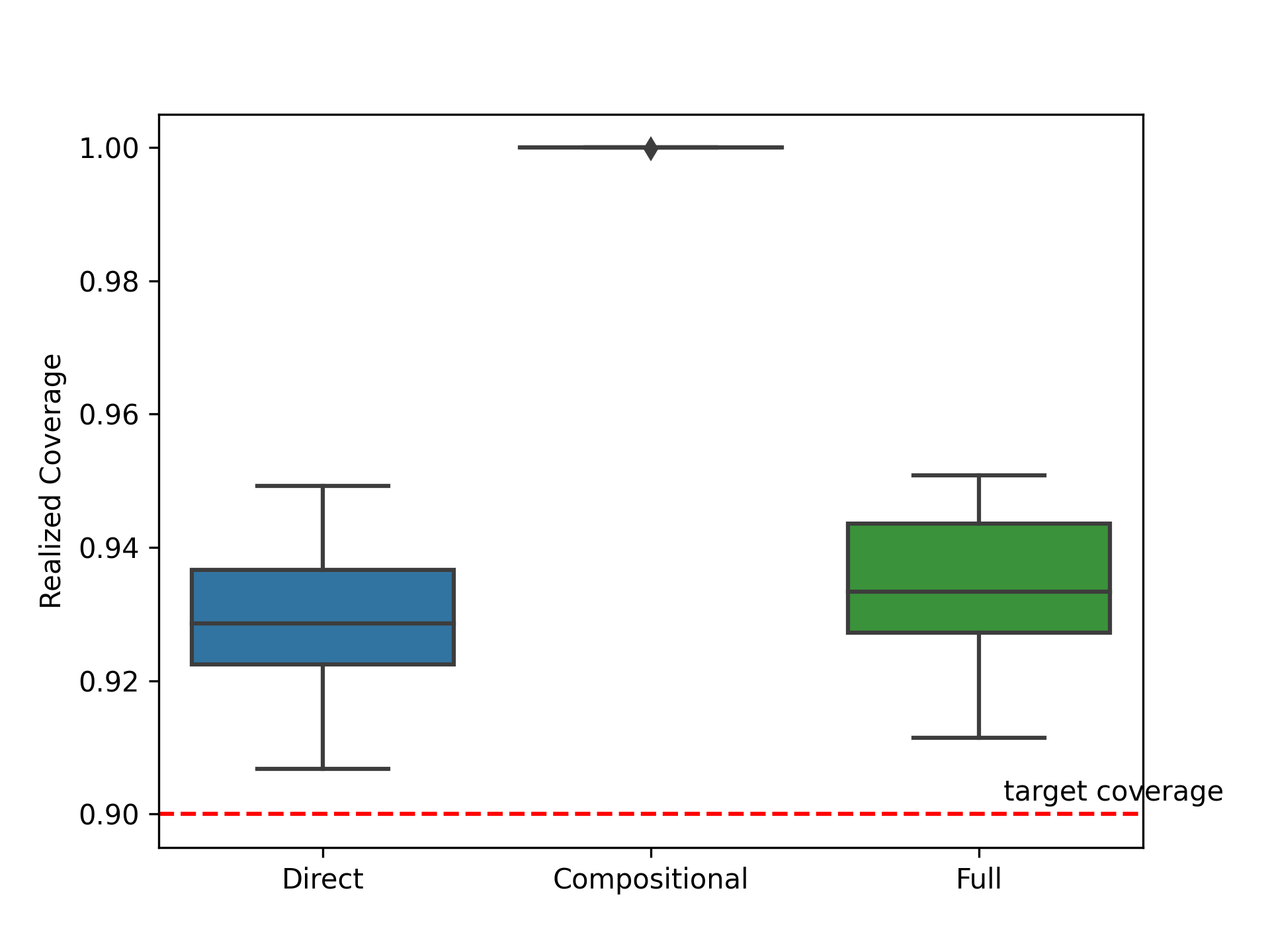}
\caption{Program 9}
\label{fig:prog-9}
\end{subfigure}
\vfill \vfill \vfill
\begin{subfigure}[b]{0.3\textwidth}
\centering
\includegraphics[width=\textwidth]{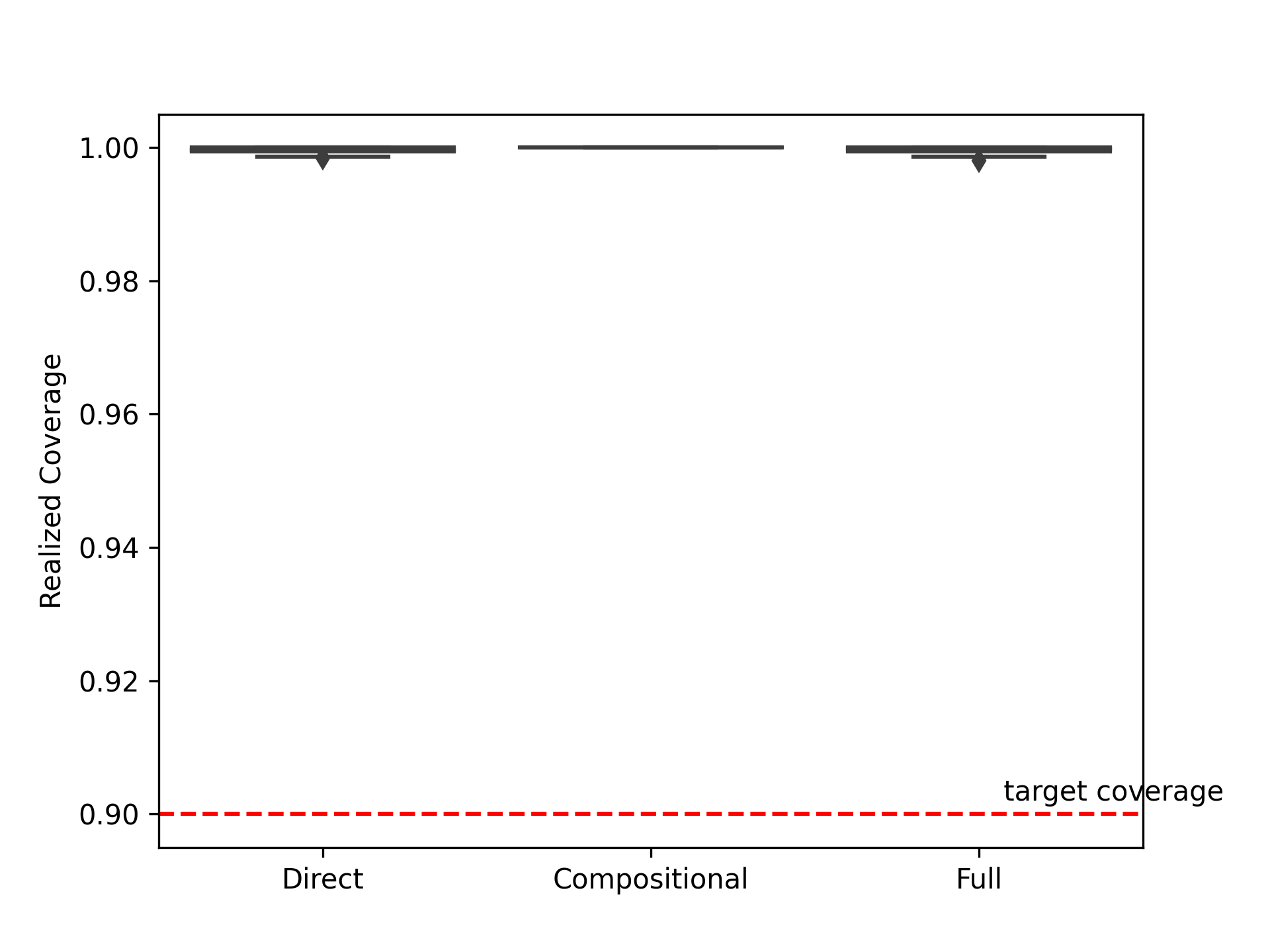}
\caption{Program 10}
\label{fig:prog-10}
\end{subfigure}
\caption{Box plots of coverages for all programs in Section~\ref{sec:mnist}, over 25 random splits of data into calibration and test sets. The dotted red line shows the desired coverage rate $1 - \epsilon = 0.9$.}
\label{fig:MNIST-coverage}
\end{figure}

\begin{figure}[t]
\centering
\begin{subfigure}[b]{0.3\textwidth}
\centering
\includegraphics[width=\textwidth]{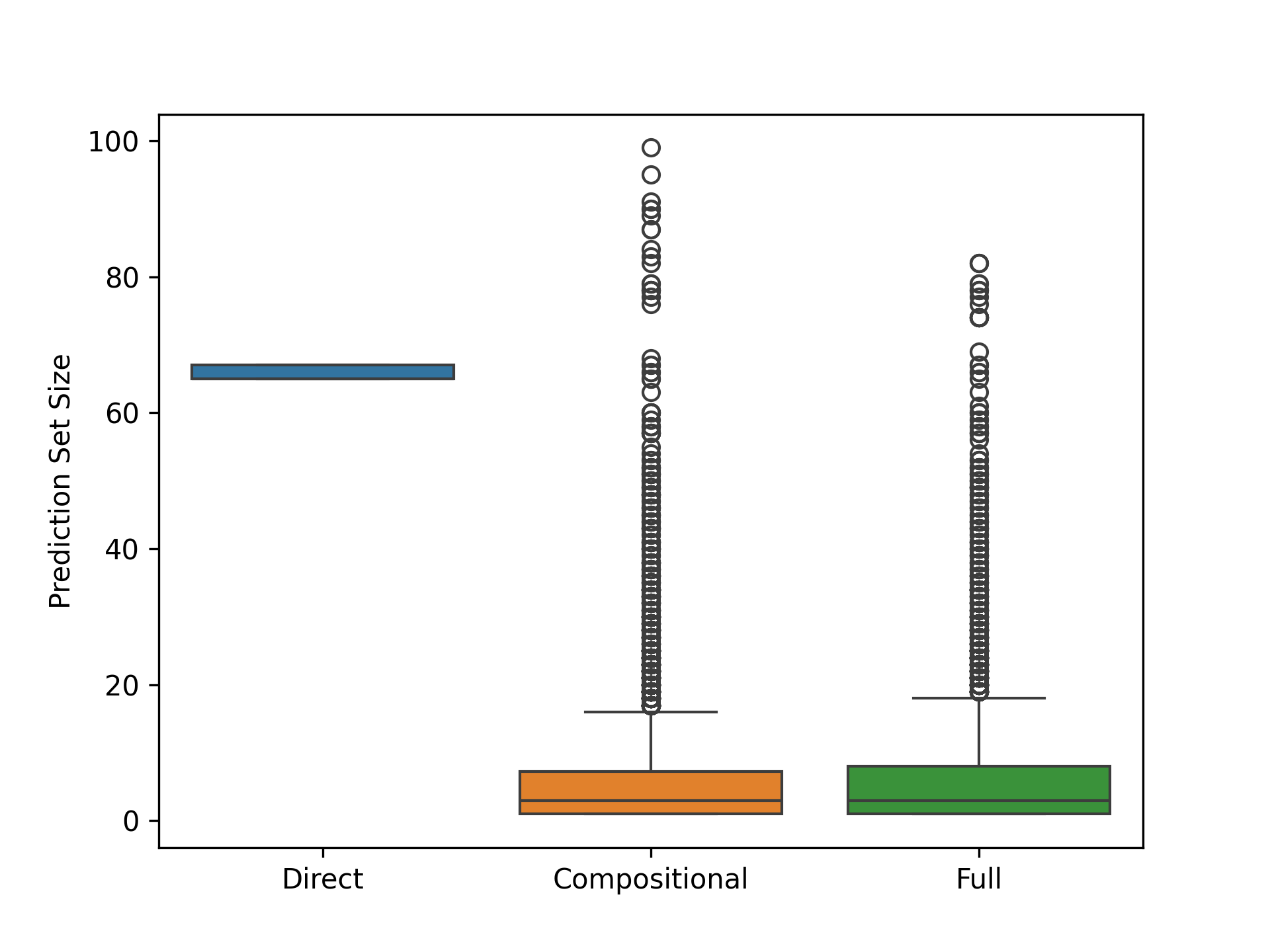}
\caption{Program 1}
\label{fig:prog-1}
\end{subfigure}
\hfill
\begin{subfigure}[b]{0.3\textwidth}
\centering
\includegraphics[width=\textwidth]{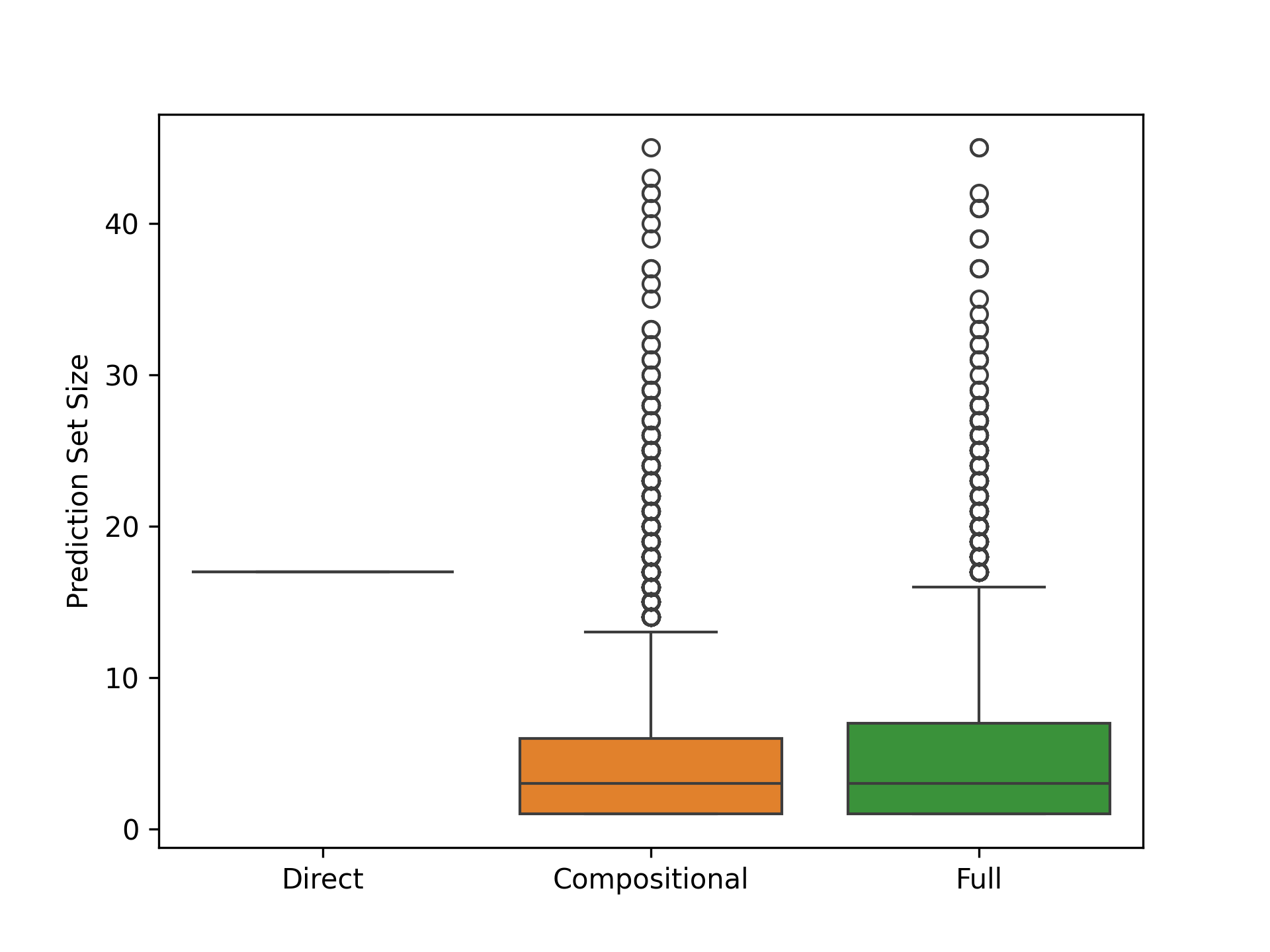}
\caption{Program 2}
\label{fig:prog-2}
\end{subfigure}
\hfill
\begin{subfigure}[b]{0.3\textwidth}
\centering
\includegraphics[width=\textwidth]{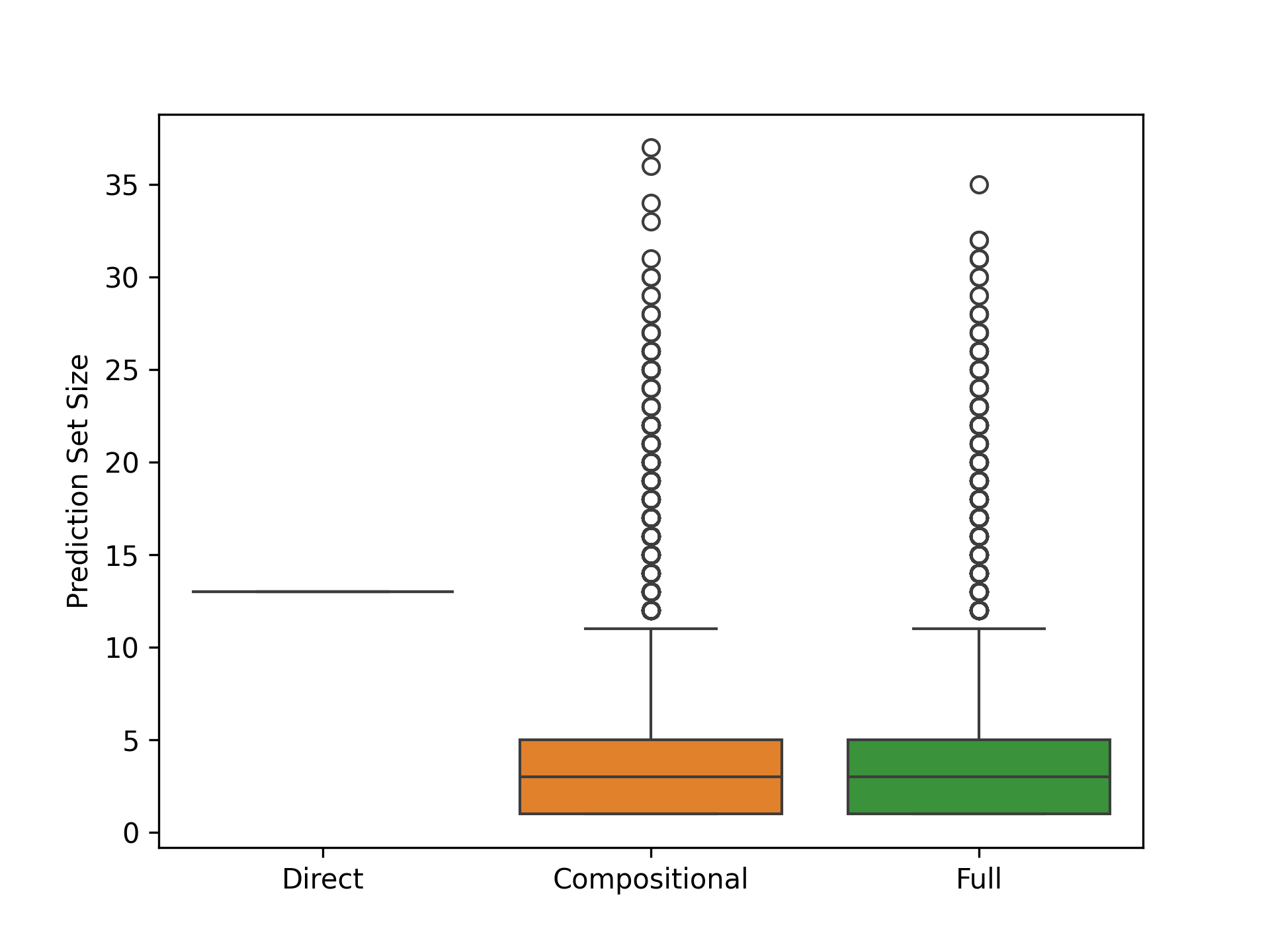}
\caption{Program 3}
\label{fig:prog-3}
\end{subfigure}
\vfill \vfill \vfill
\begin{subfigure}[b]{0.3\textwidth}
\centering
\includegraphics[width=\textwidth]{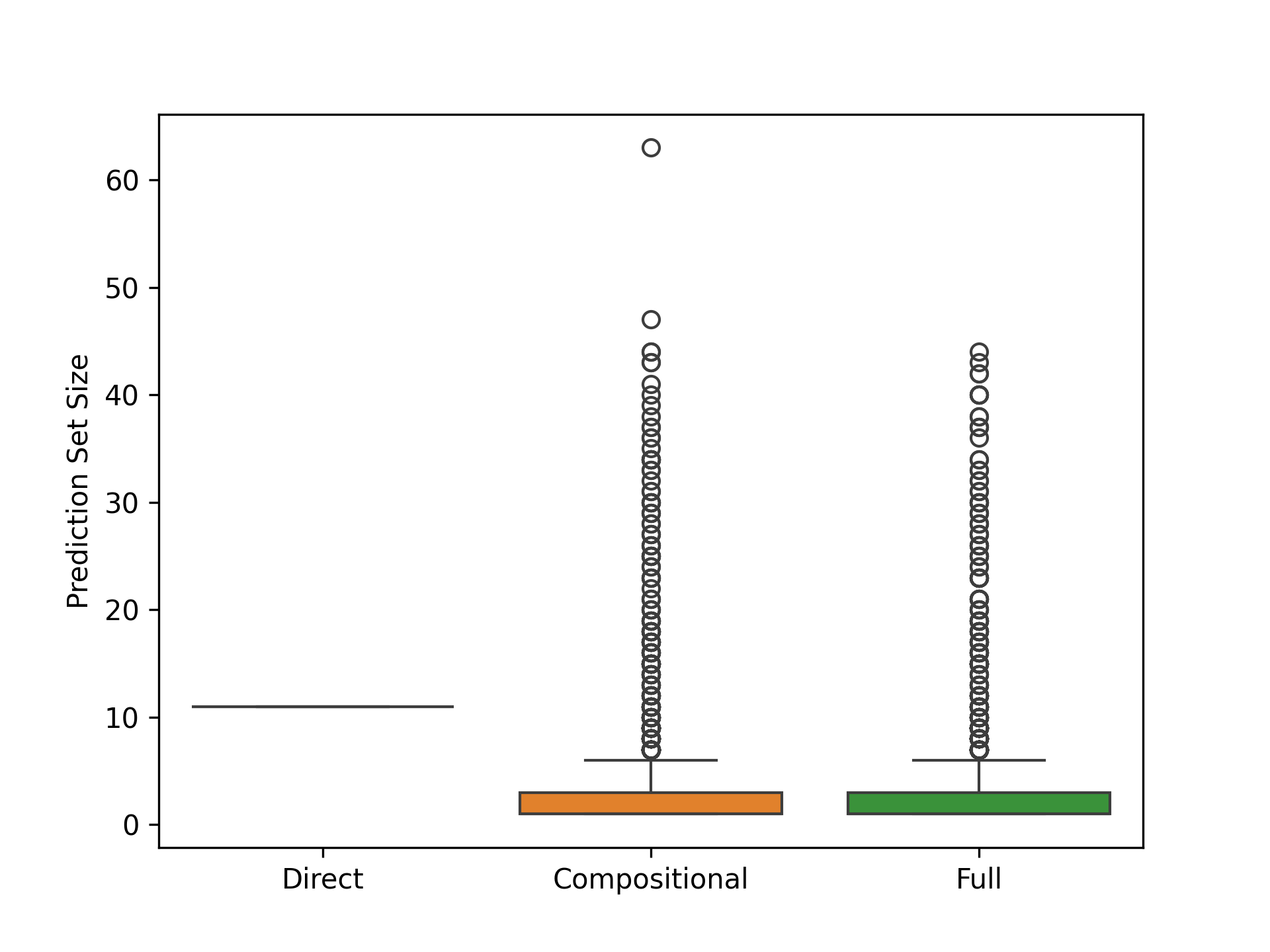}
\caption{Program 4}
\label{fig:prog-4}
\end{subfigure}
\hfill
\begin{subfigure}[b]{0.3\textwidth}
\centering
\includegraphics[width=\textwidth]{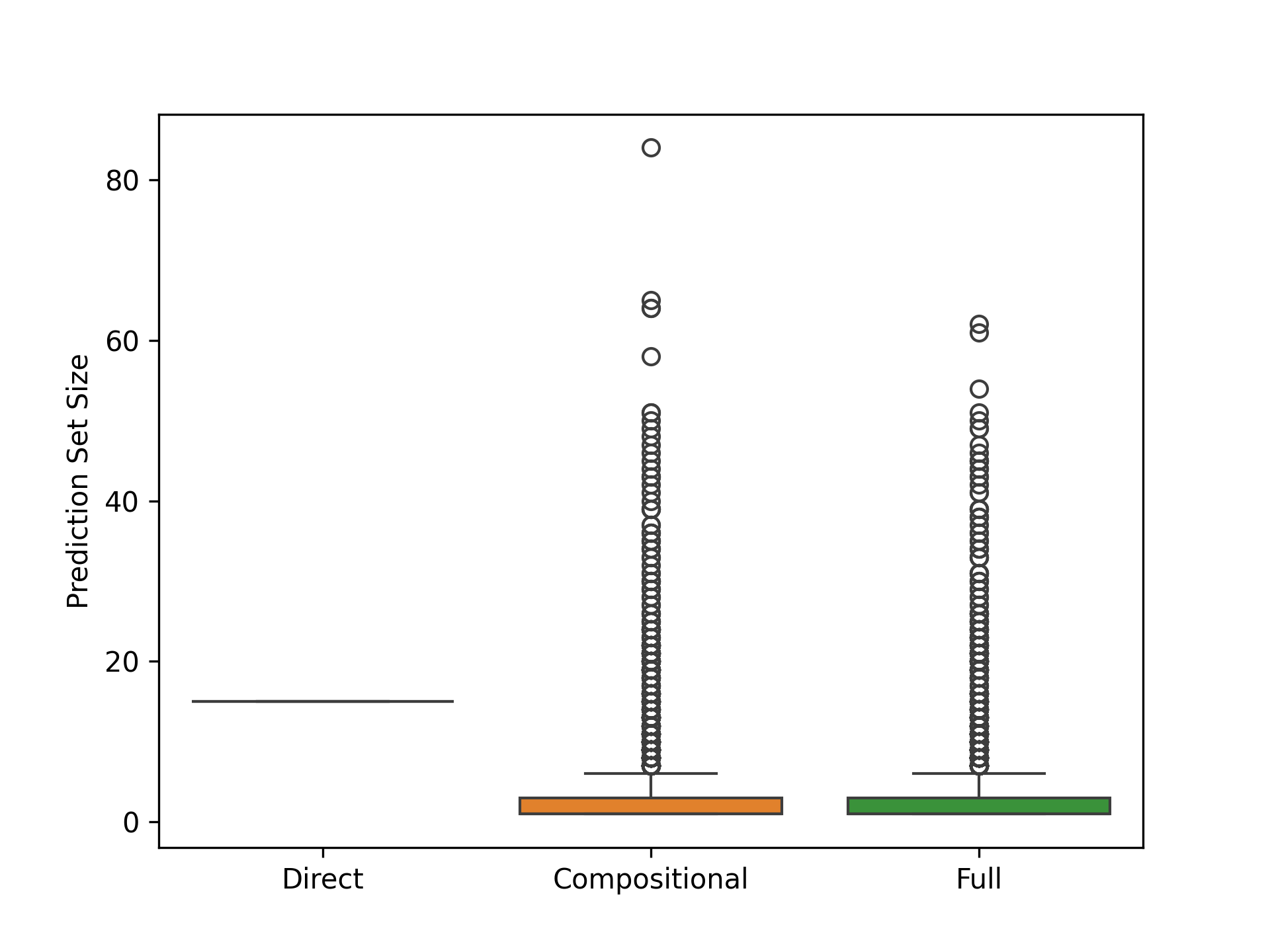}
\caption{Program 5}
\label{fig:prog-5}
\end{subfigure}
\hfill
\begin{subfigure}[b]{0.3\textwidth}
\centering
\includegraphics[width=\textwidth]{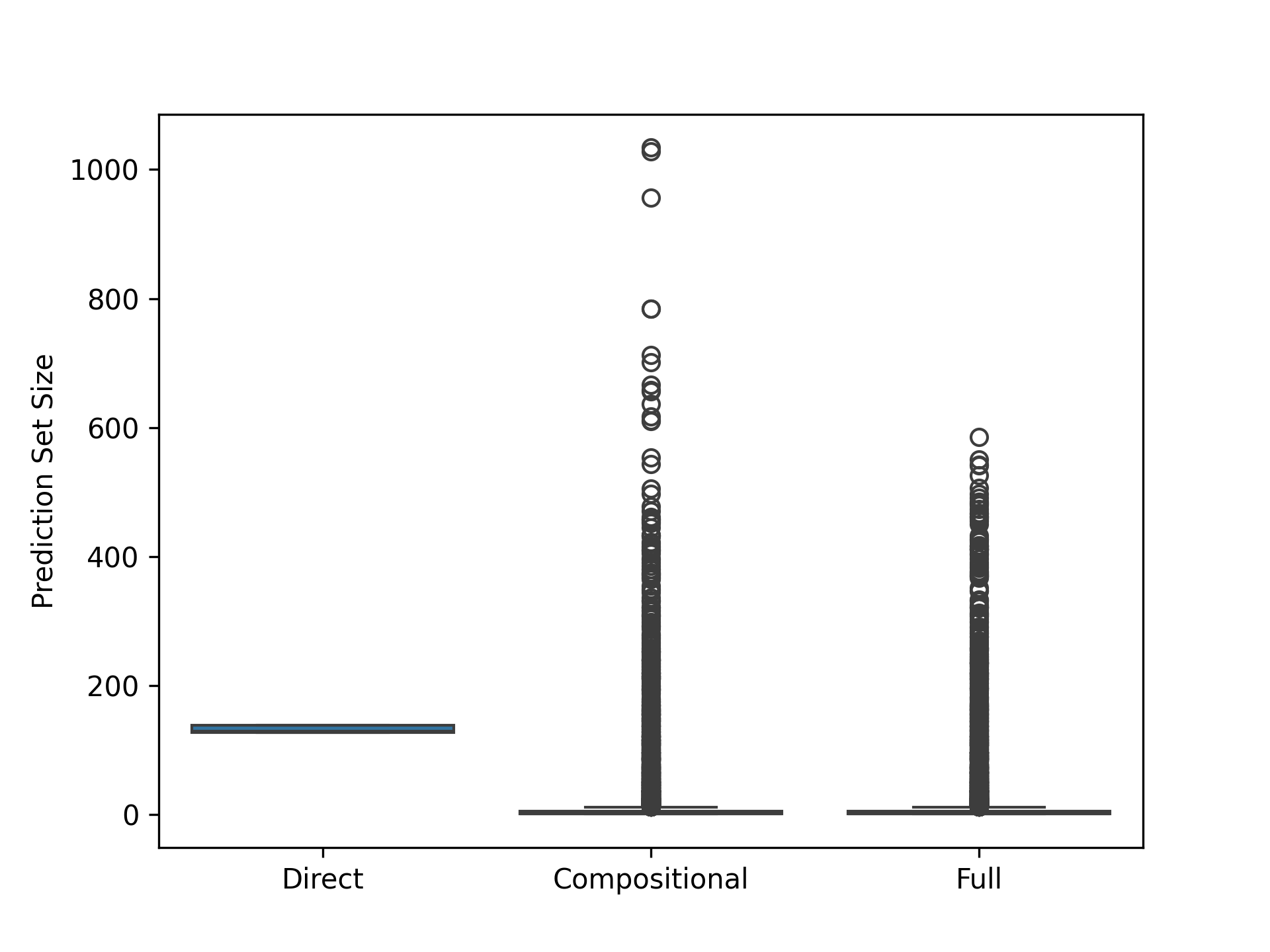}
\caption{Program 6}
\label{fig:prog-6}
\end{subfigure}
\vfill \vfill \vfill
\begin{subfigure}[b]{0.3\textwidth}
\centering
\includegraphics[width=\textwidth]{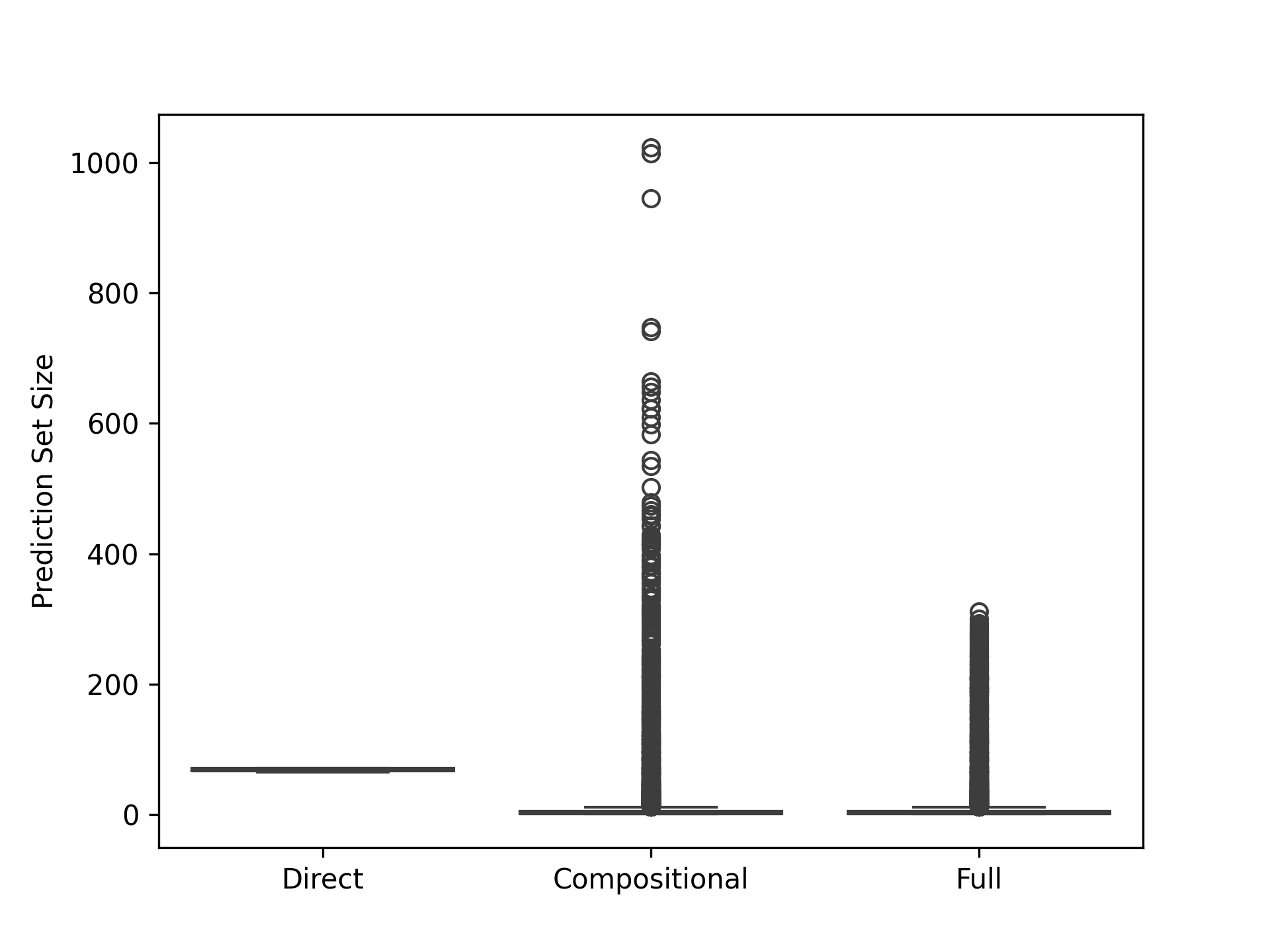}
\caption{Program 7}
\label{fig:prog-7}
\end{subfigure}
\hfill
\begin{subfigure}[b]{0.3\textwidth}
\centering
\includegraphics[width=\textwidth]{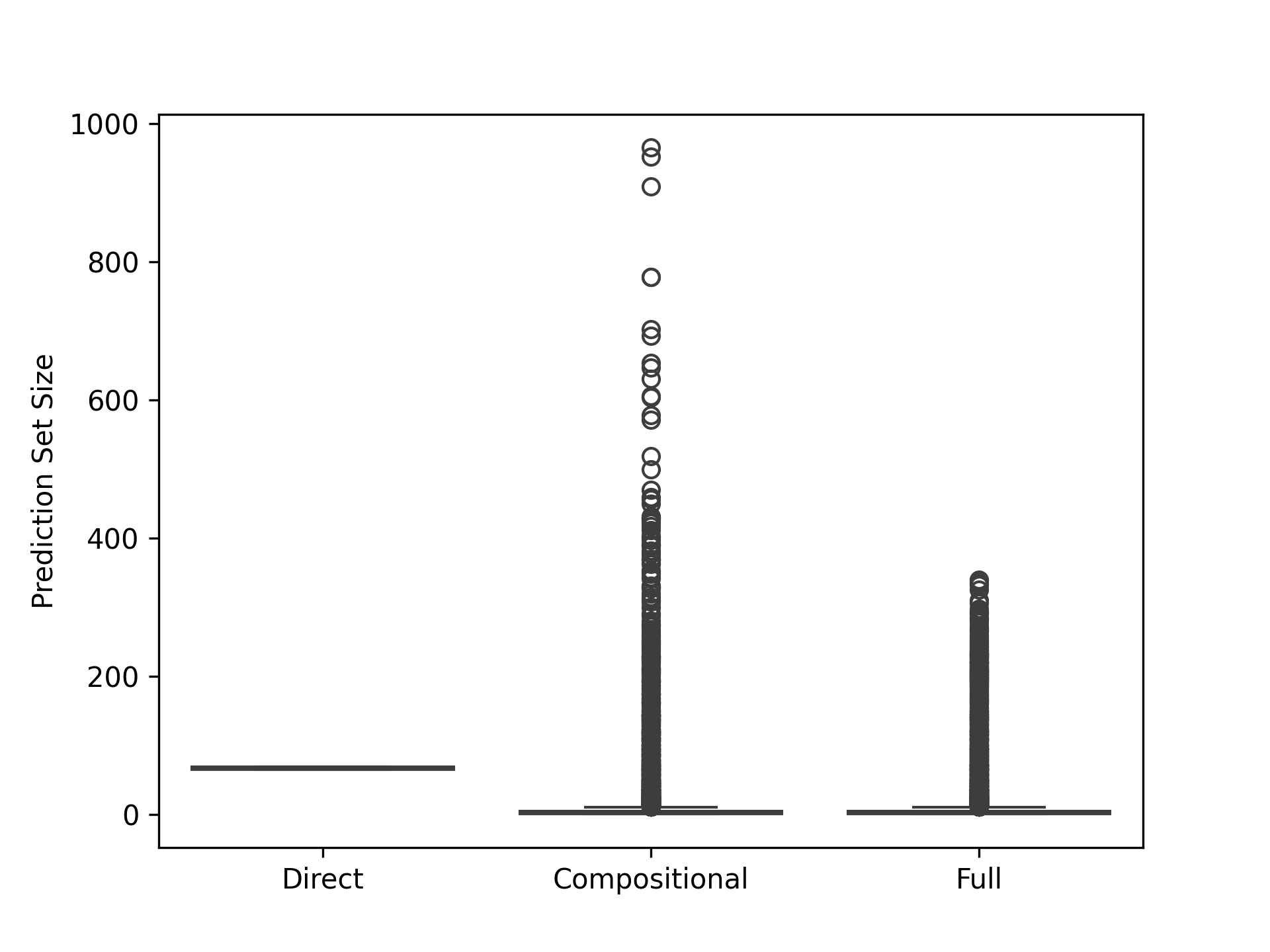}
\caption{Program 8}
\label{fig:prog-8}
\end{subfigure}
\hfill
\begin{subfigure}[b]{0.3\textwidth}
\centering
\includegraphics[width=\textwidth]{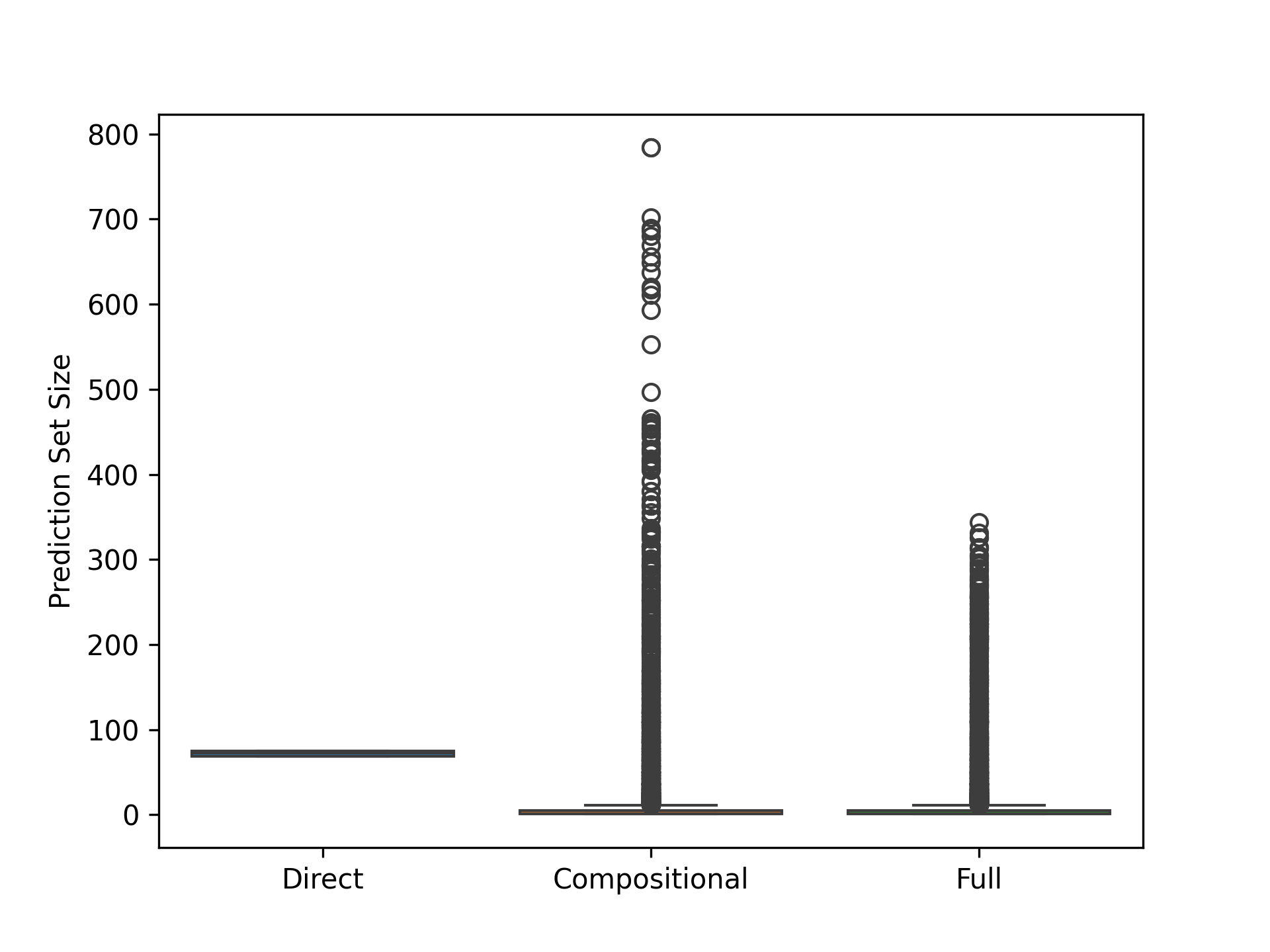}
\caption{Program9}
\label{fig:prog-9}
\end{subfigure}
\vfill \vfill \vfill
\begin{subfigure}[b]{0.3\textwidth}
\centering
\includegraphics[width=\textwidth]{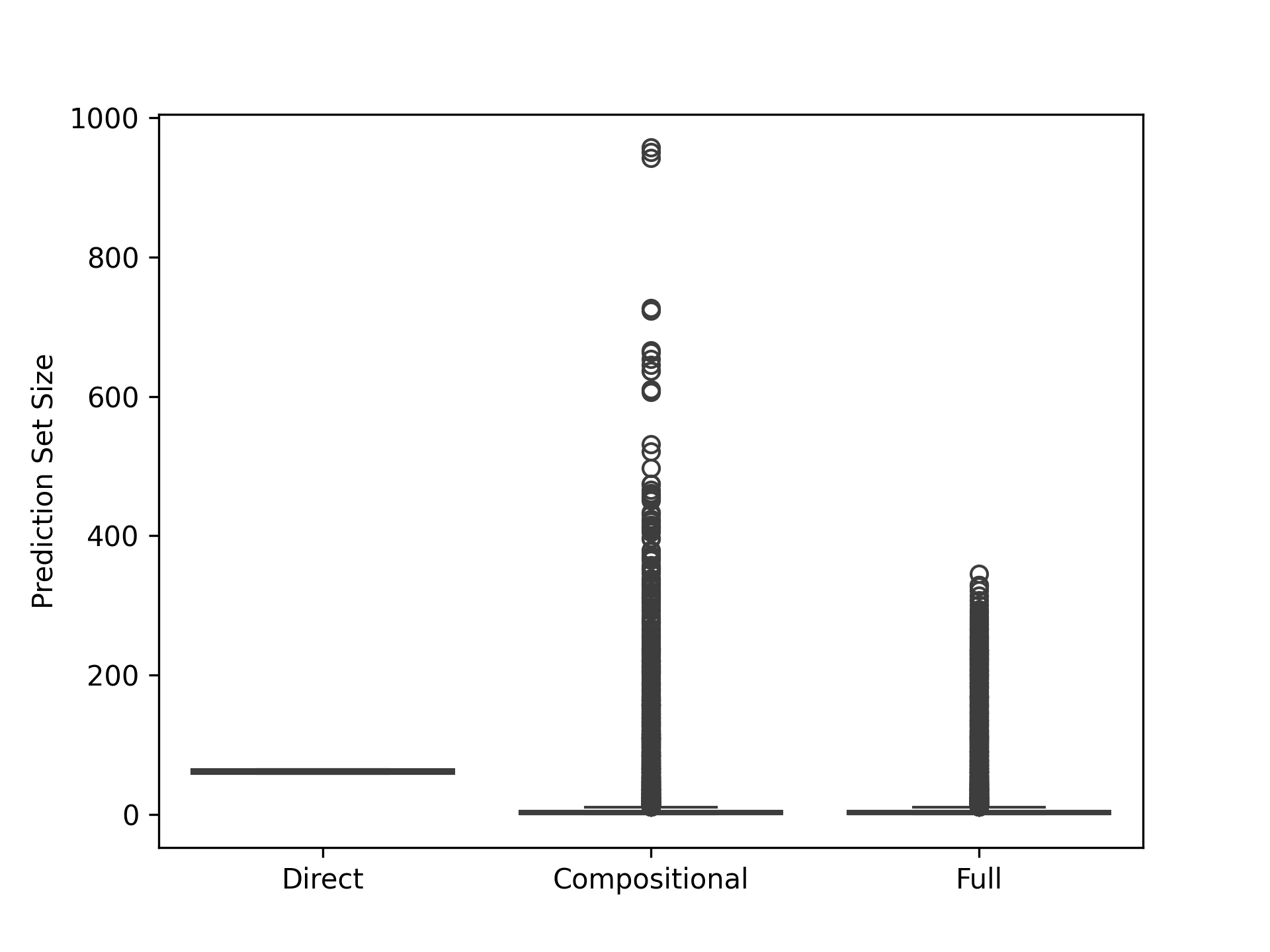}
\caption{Program 10}
\label{fig:prog-10}
\end{subfigure}
\hfill
\begin{subfigure}[b]{0.3\textwidth}
\centering
\includegraphics[width=\textwidth]{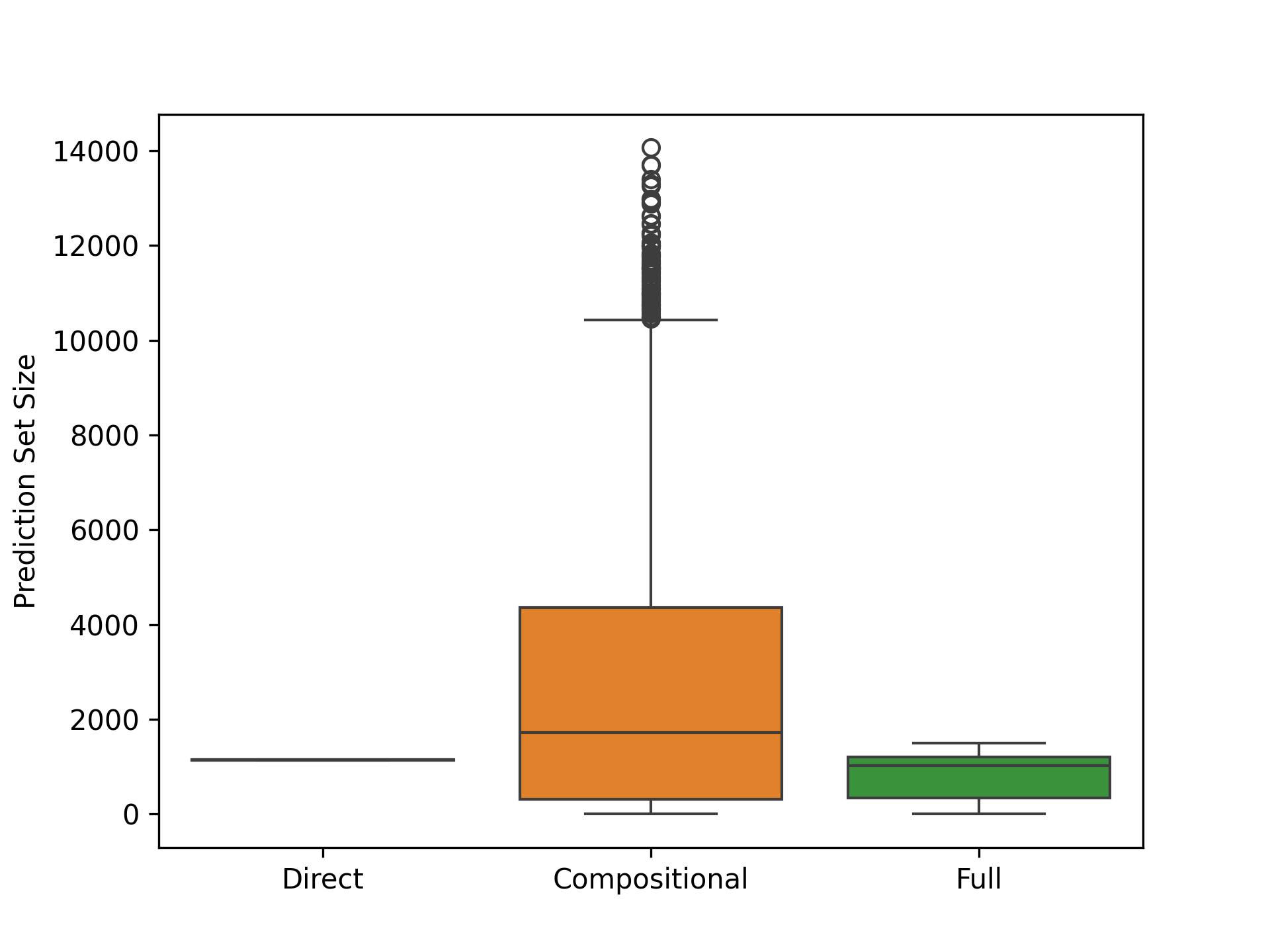}
\caption{Program 11}
\label{fig:prog-11}
\end{subfigure}
\hfill
\begin{subfigure}[b]{0.3\textwidth}
\centering
\includegraphics[width=\textwidth]{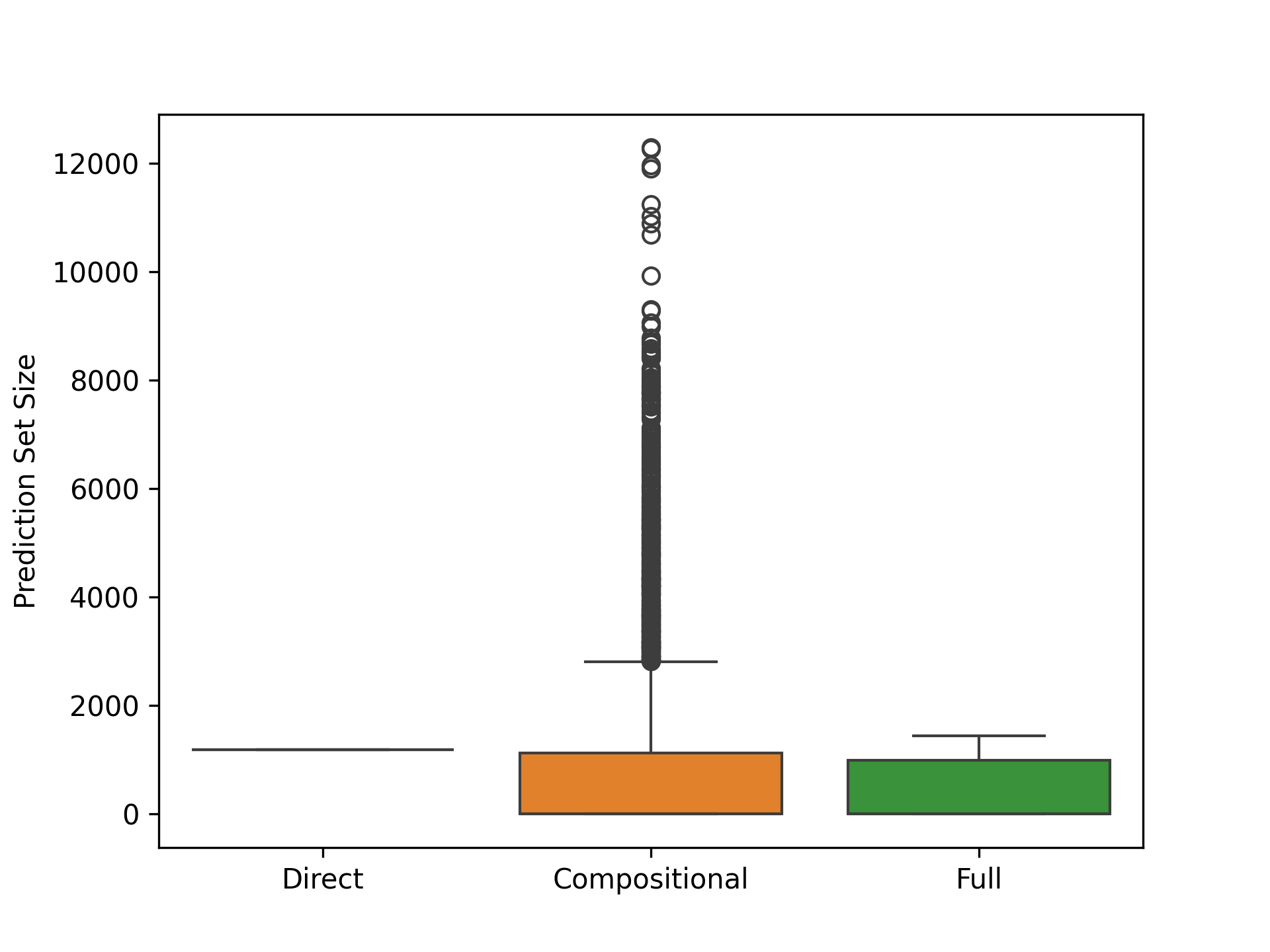}
\caption{Program 12}
\label{fig:prog-12}
\end{subfigure}
\caption{
Box plots of prediction set sizes across all test examples for each program in Section~\ref{sec:obj-detect}. Note that for a fixed calibration set, the direct approach always produces prediction sets of the same size.}
\label{fig:obj-det-set-size}
\end{figure}

\begin{figure}[t]
\centering
\begin{subfigure}[b]{0.3\textwidth}
\centering
\includegraphics[width=\textwidth]{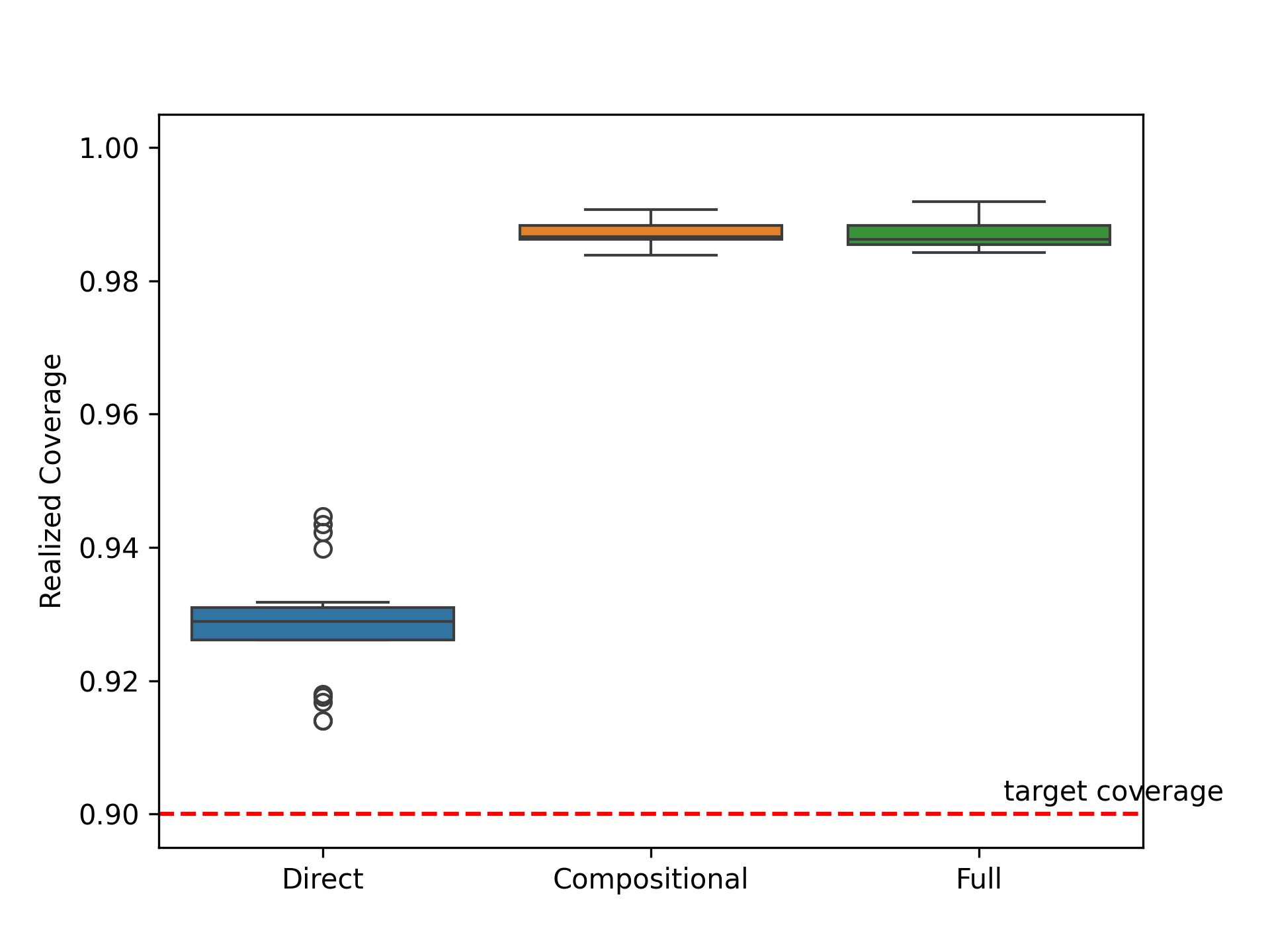}
\caption{Program 1}
\label{fig:prog-1}
\end{subfigure}
\hfill
\begin{subfigure}[b]{0.3\textwidth}
\centering
\includegraphics[width=\textwidth]{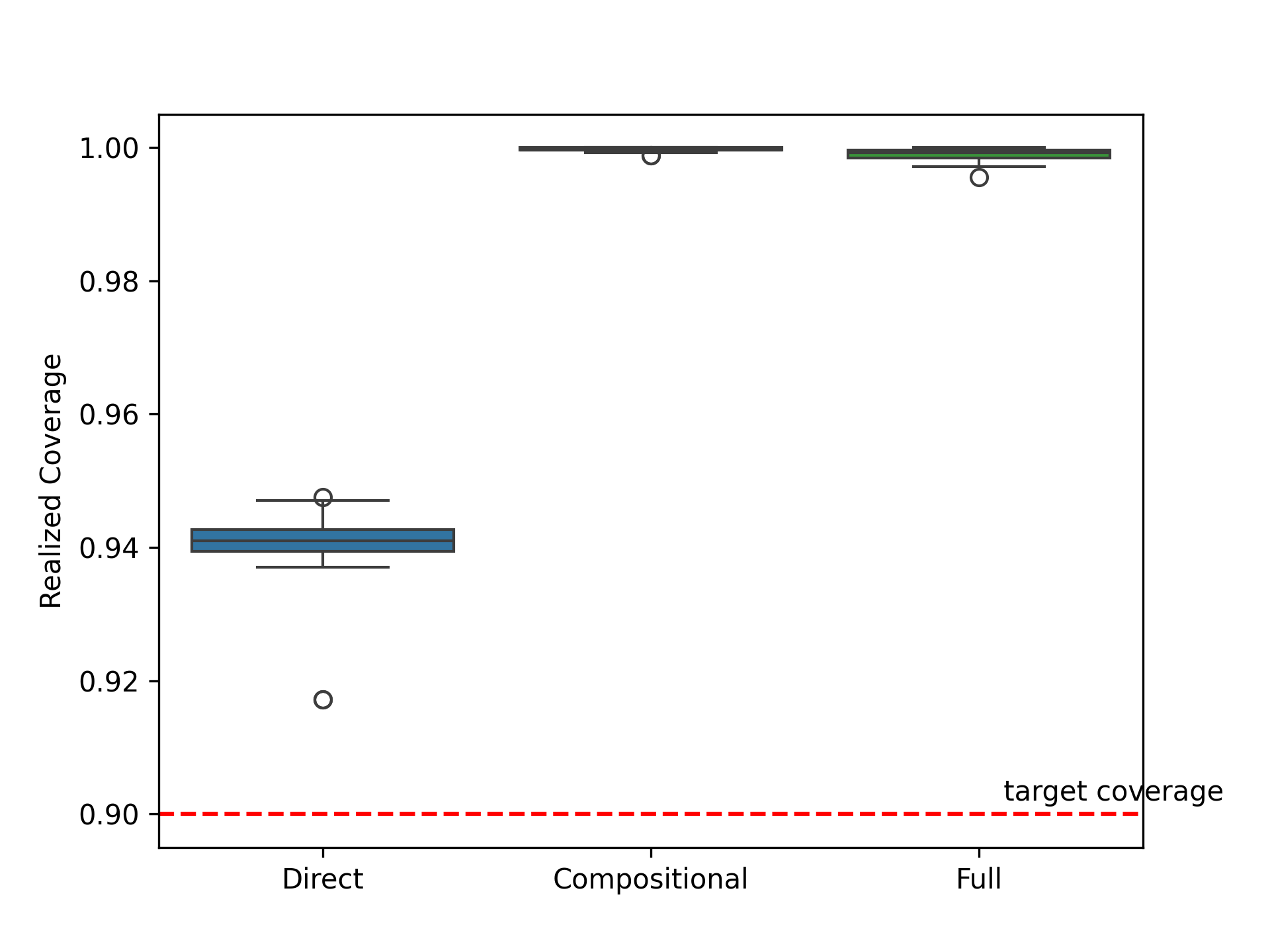}
\caption{Program 2}
\label{fig:prog-2}
\end{subfigure}
\hfill
\begin{subfigure}[b]{0.3\textwidth}
\centering
\includegraphics[width=\textwidth]{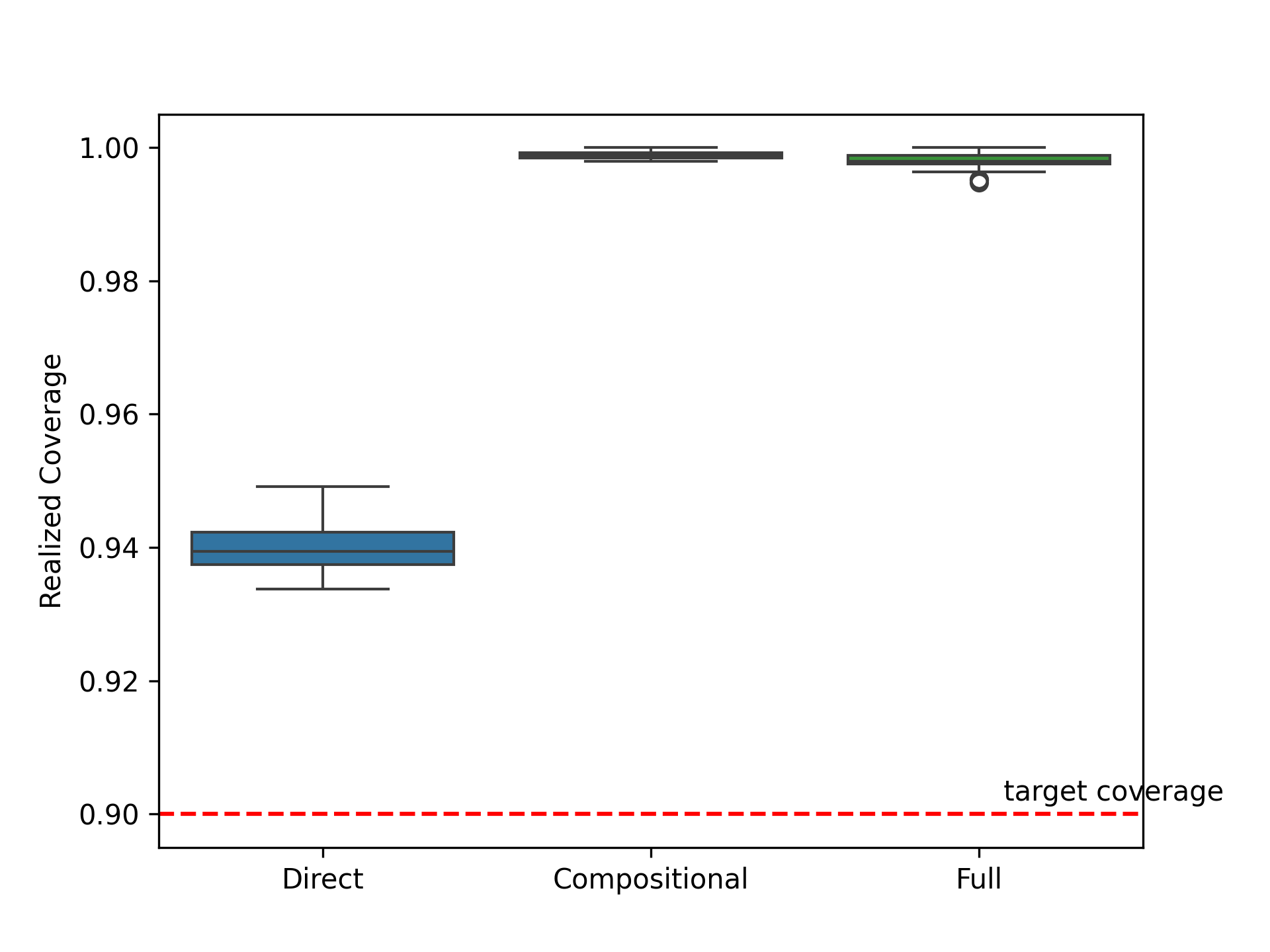}
\caption{Program 3}
\label{fig:prog-3}
\end{subfigure}
\vfill \vfill \vfill
\begin{subfigure}[b]{0.3\textwidth}
\centering
\includegraphics[width=\textwidth]{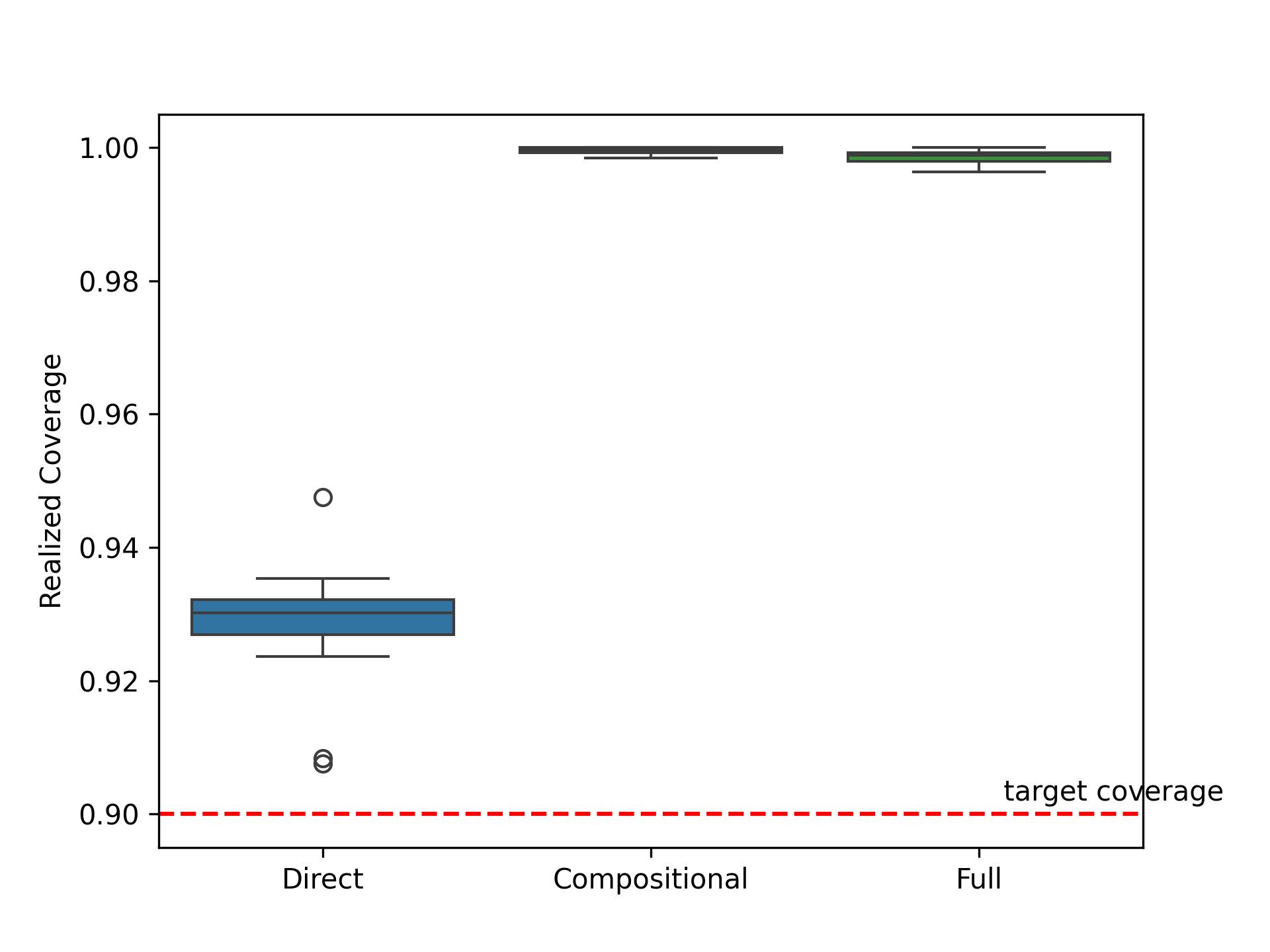}
\caption{Program 4}
\label{fig:prog-4}
\end{subfigure}
\hfill
\begin{subfigure}[b]{0.3\textwidth}
\centering
\includegraphics[width=\textwidth]{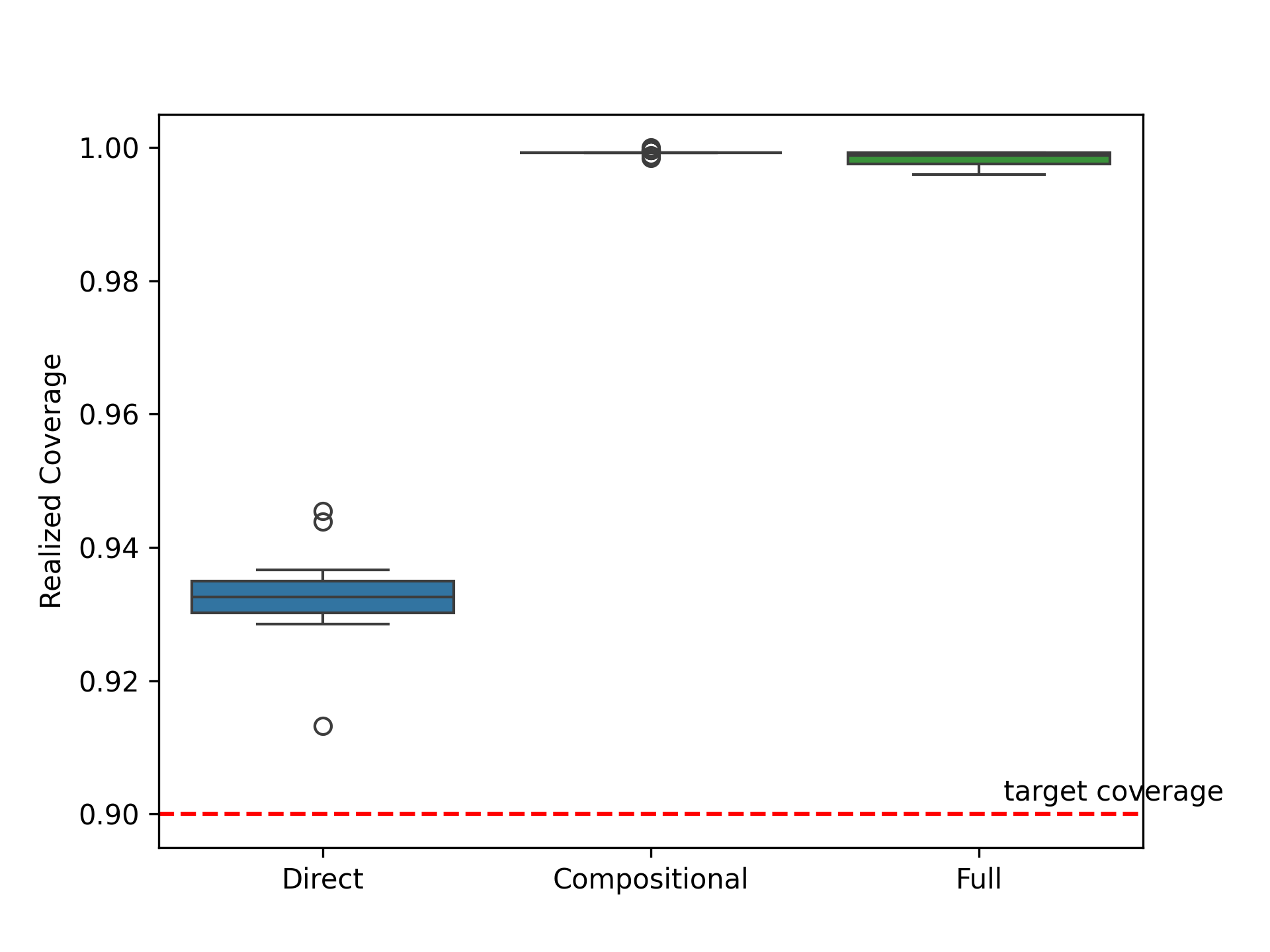}
\caption{Program 5}
\label{fig:prog-5}
\end{subfigure}
\hfill
\begin{subfigure}[b]{0.3\textwidth}
\centering
\includegraphics[width=\textwidth]{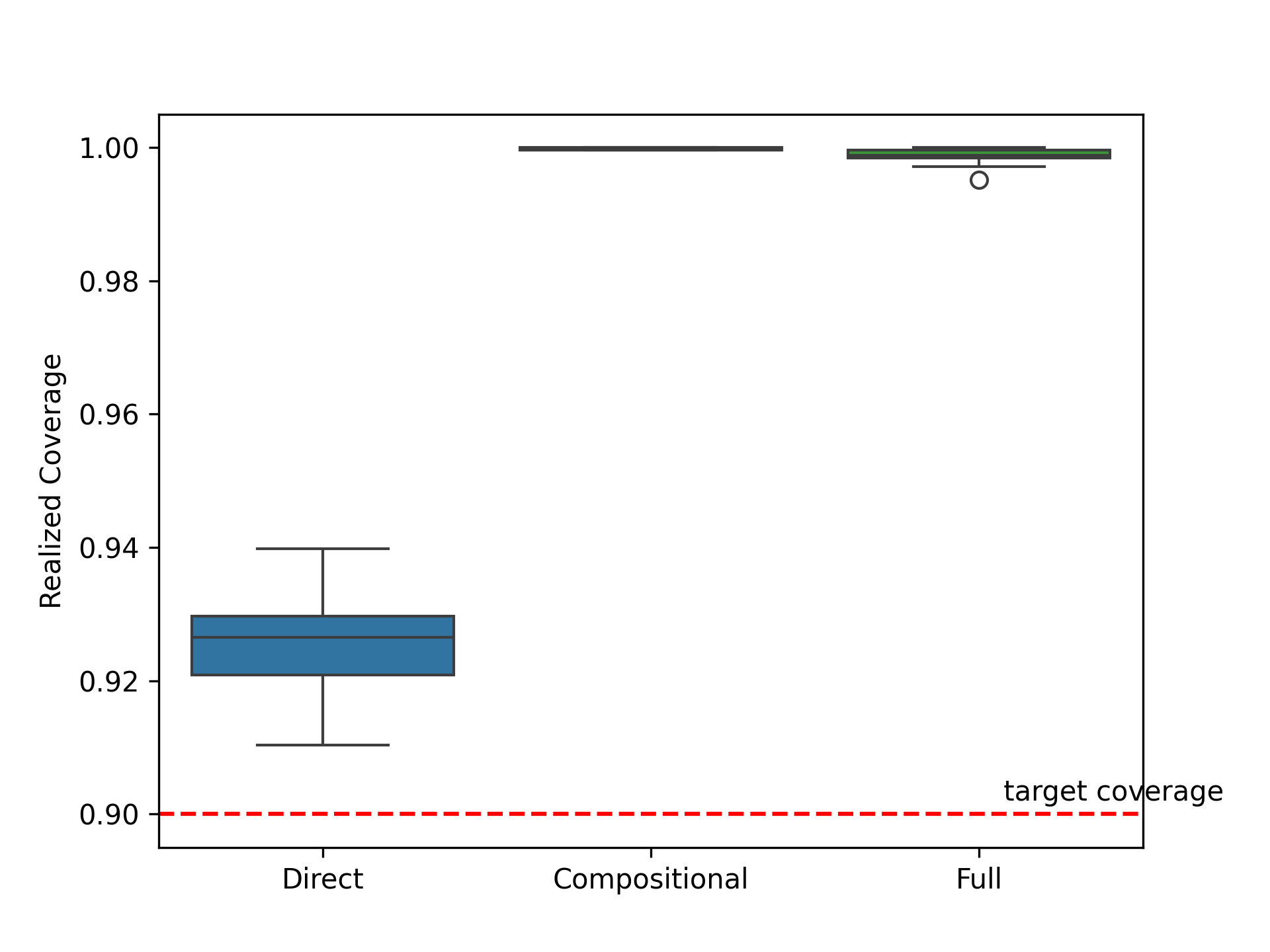}
\caption{Program 6}
\label{fig:prog-6}
\end{subfigure}
\vfill \vfill \vfill
\begin{subfigure}[b]{0.3\textwidth}
\centering
\includegraphics[width=\textwidth]{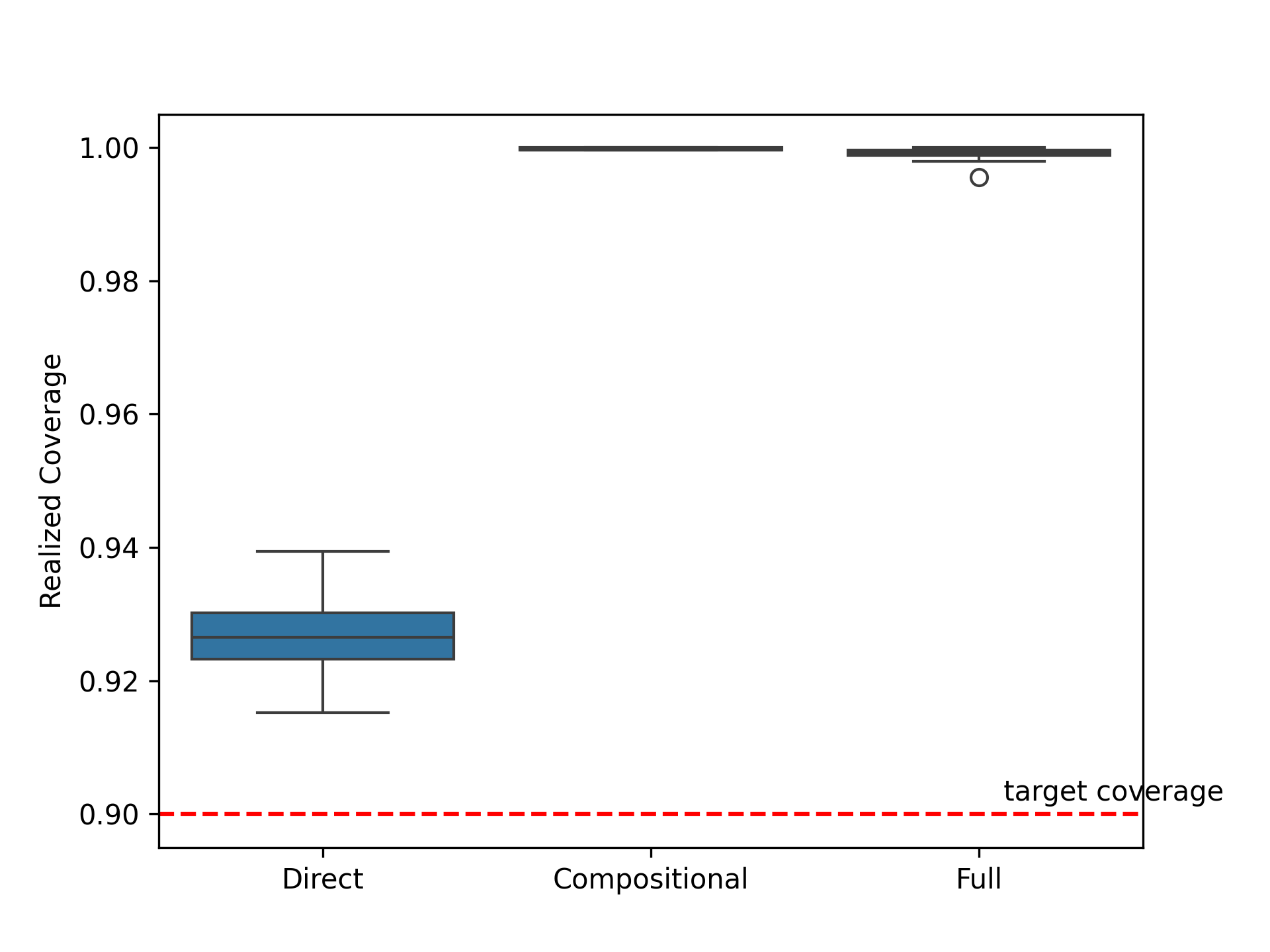}
\caption{Program 7}
\label{fig:prog-7}
\end{subfigure}
\hfill
\begin{subfigure}[b]{0.3\textwidth}
\centering
\includegraphics[width=\textwidth]{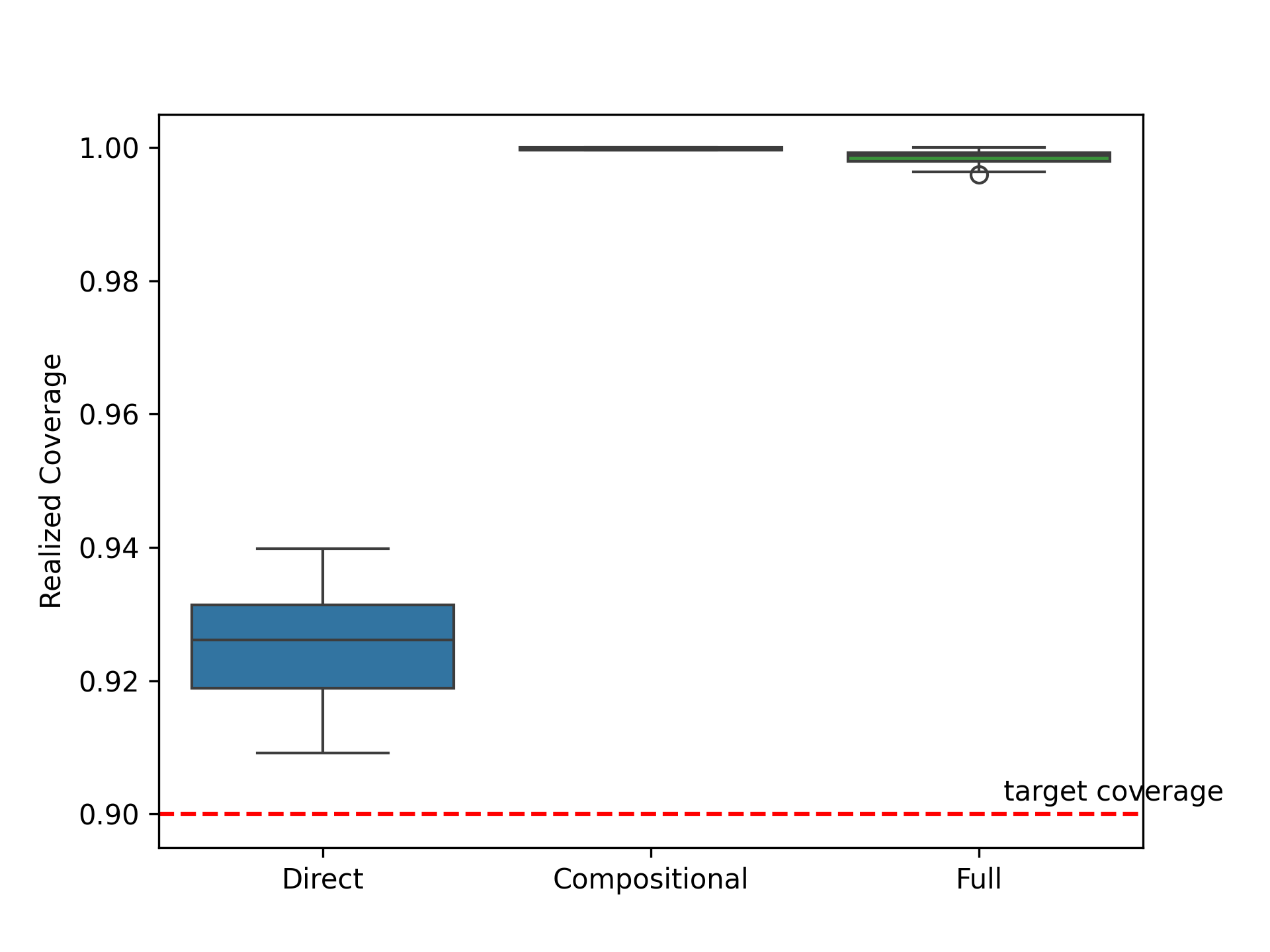}
\caption{Program 8}
\label{fig:prog-8}
\end{subfigure}
\hfill
\begin{subfigure}[b]{0.3\textwidth}
\centering
\includegraphics[width=\textwidth]{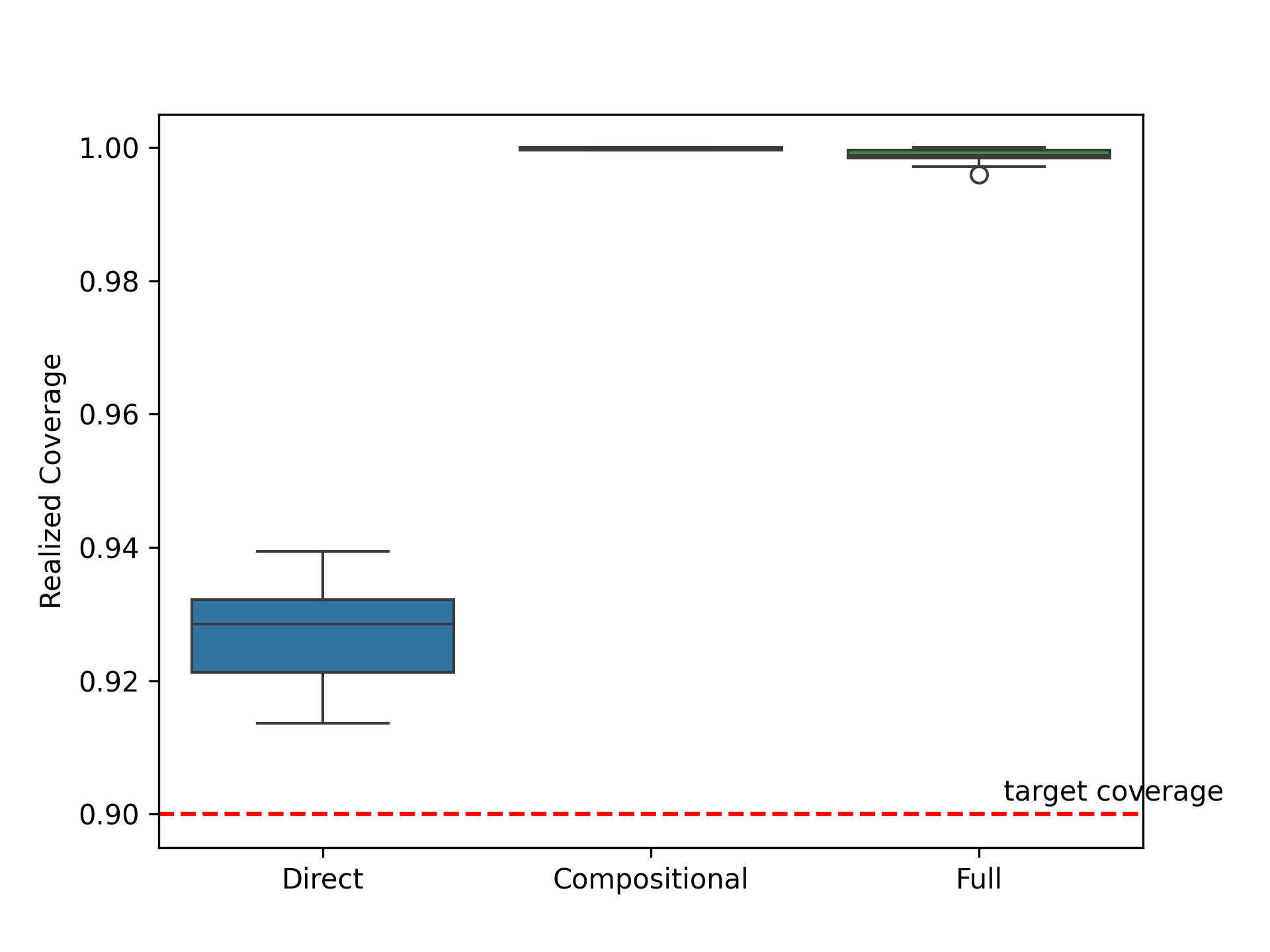}
\caption{Program9}
\label{fig:prog-9}
\end{subfigure}
\vfill \vfill \vfill
\begin{subfigure}[b]{0.3\textwidth}
\centering
\includegraphics[width=\textwidth]{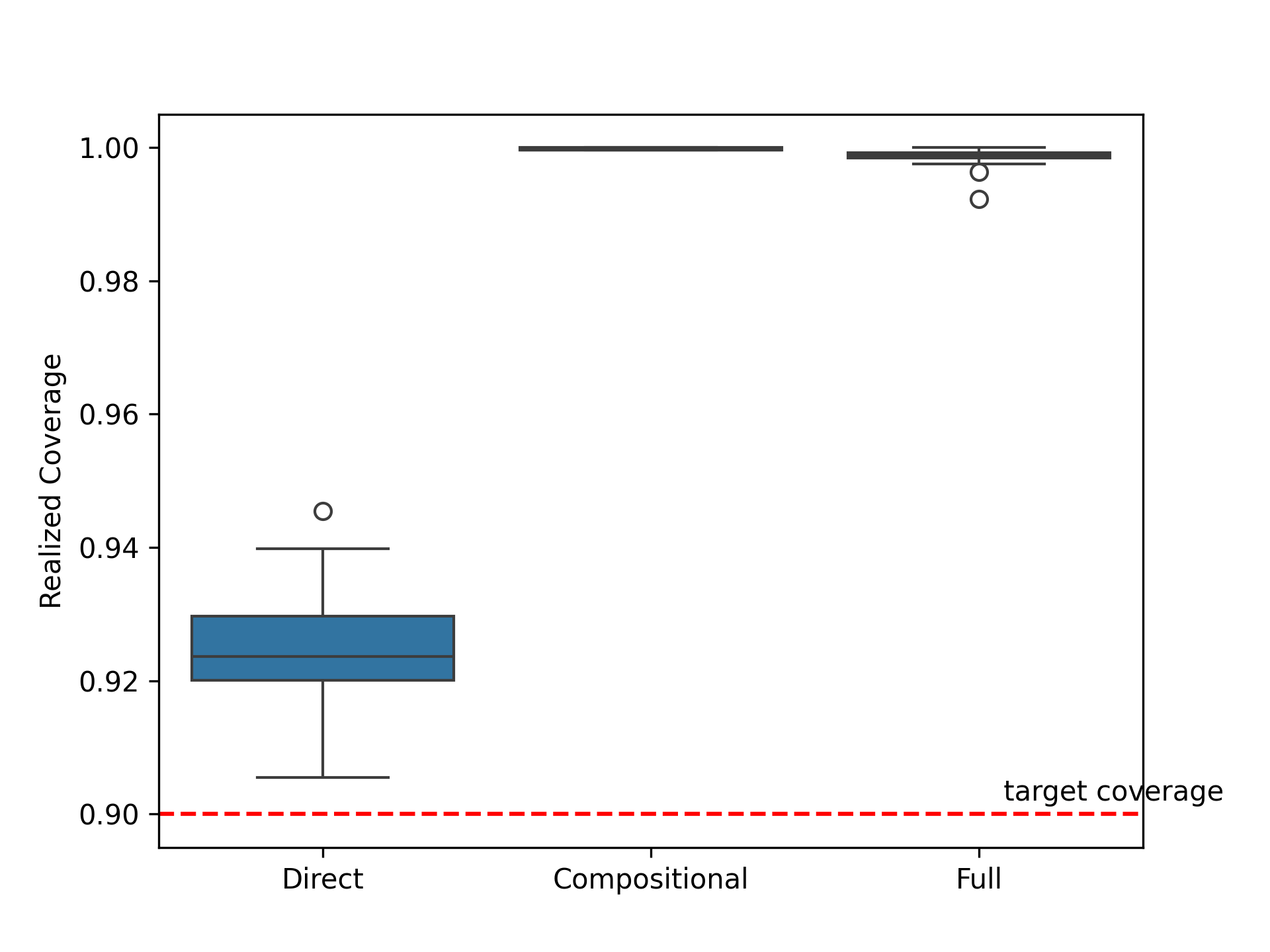}
\caption{Program 10}
\label{fig:prog-10}
\end{subfigure}
\hfill
\begin{subfigure}[b]{0.3\textwidth}
\centering
\includegraphics[width=\textwidth]{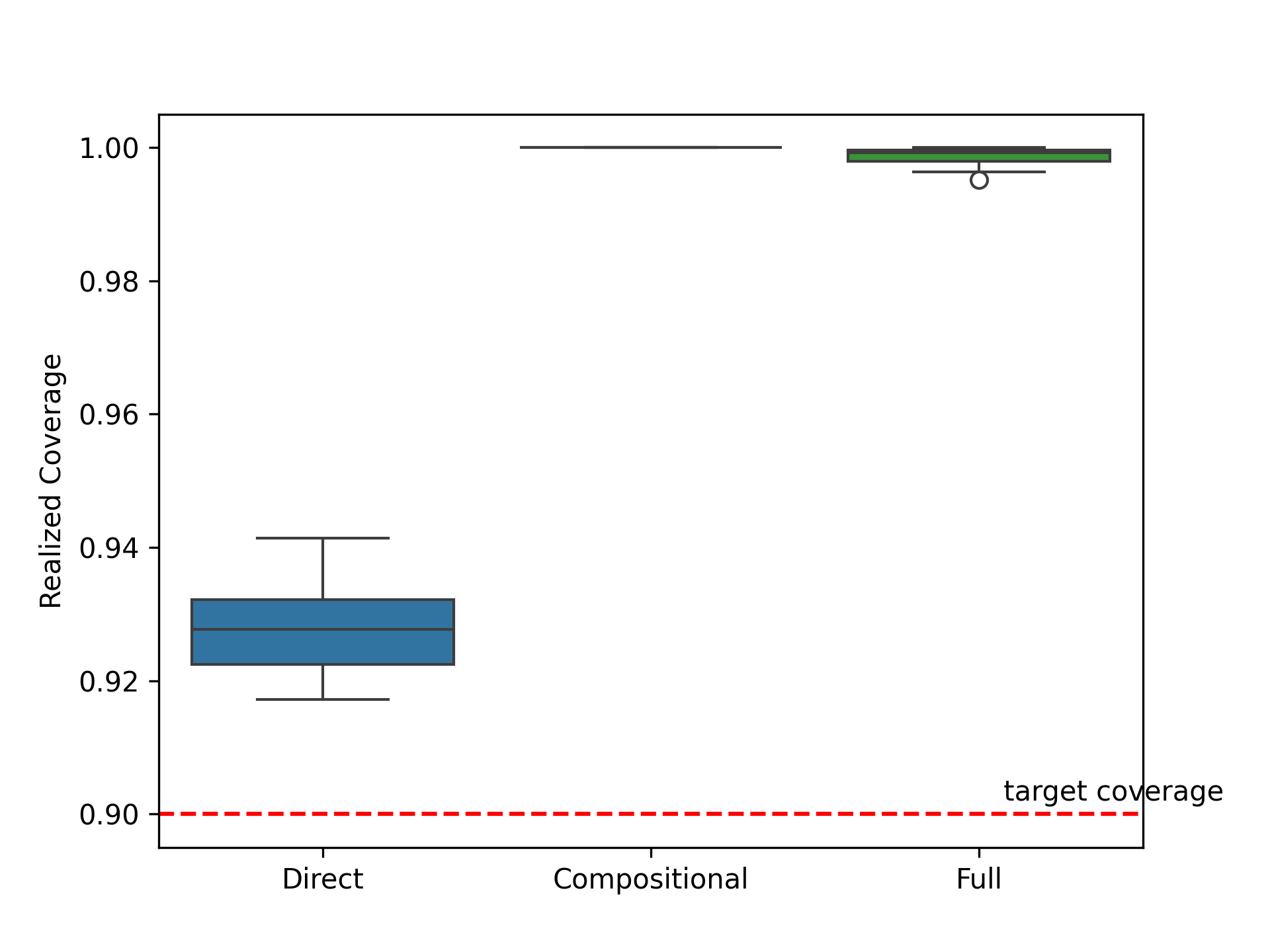}
\caption{Program 11}
\label{fig:prog-11}
\end{subfigure}
\hfill
\begin{subfigure}[b]{0.3\textwidth}
\centering
\includegraphics[width=\textwidth]{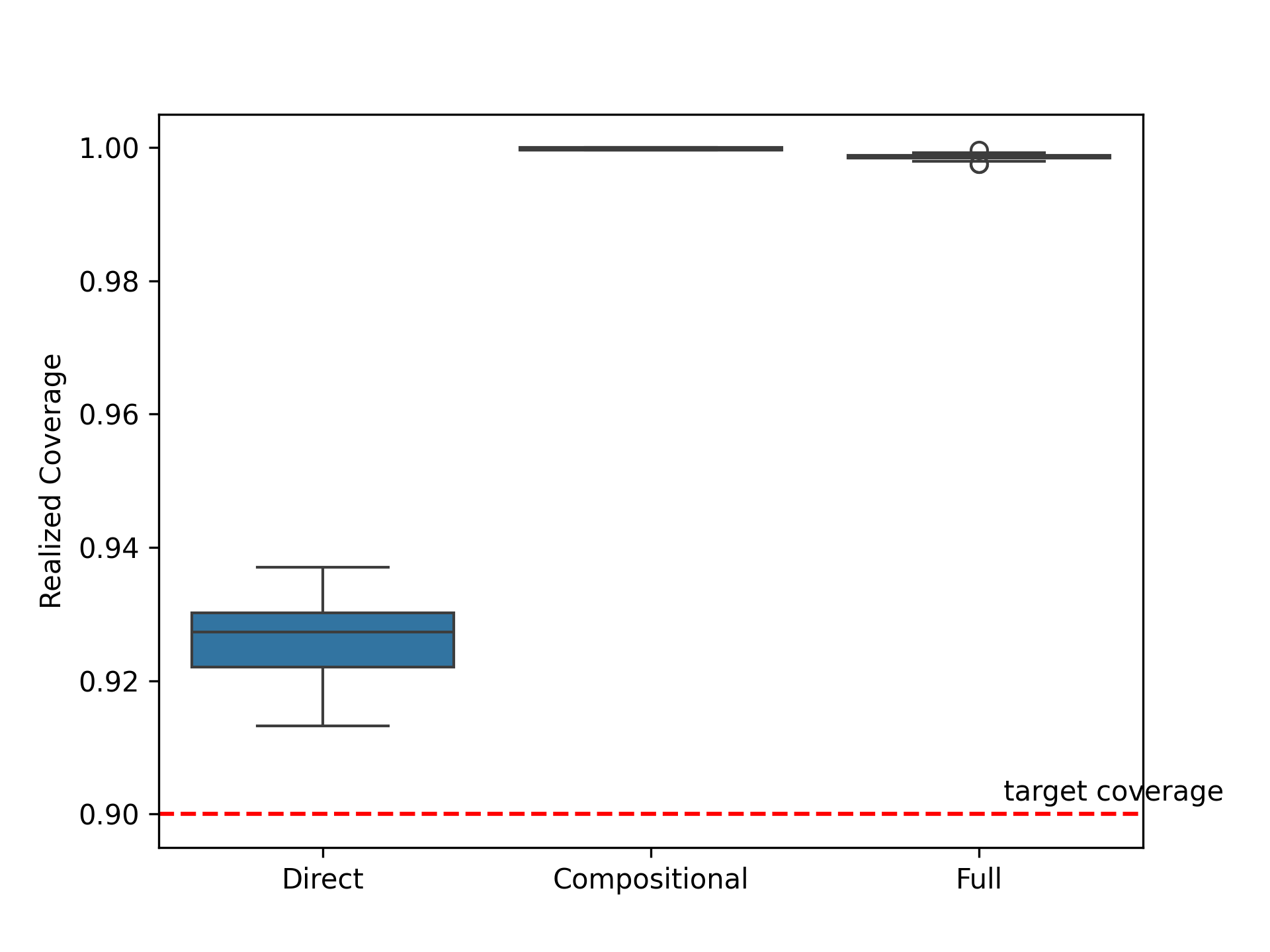}
\caption{Program 12}
\label{fig:prog-12}
\end{subfigure}
\caption{
Box plots of realized coverage for all programs in Section~\ref{sec:obj-detect}. The dotted red line shows the desired coverage rate $1 - \epsilon = 0.9$.}
\label{fig:obj-det-coverage}
\end{figure}

Here we present additional plots for the experiments. Figures~\ref{fig:MNIST-Set-Size} \&~\ref{fig:MNIST-coverage} show results for experiments in the MNIST domain (Section~\ref{sec:mnist}), and Figures~\ref{fig:obj-det-set-size} \&~\ref{fig:obj-det-coverage} show results for experiments in the MS-COCO object detection domain (Section~\ref{sec:obj-detect}). 

\end{document}